\documentclass[12pt]{article}
\usepackage{slashed}
\usepackage{amsmath,amssymb}
\usepackage{bbm}
\pdfoutput=1
\usepackage{hyperref}
\usepackage{graphicx}
\usepackage{amsbsy}

\begin{document}
\setlength{\unitlength}{1mm}
\begin{titlepage}
\vspace*{0.8cm}
\begin{center}
 {\huge \bf  CP violation and FCNC in a warped A$_4$ flavor model  }
\end{center}
\vskip0.2cm

\begin{center}
 {\bf Avihay Kadosh$^a$ and Elisabetta Pallante$^a$}
\end{center}
\vskip 4pt

\begin{center}
$^a$ {\it Centre for Theoretical Physics, University of Groningen, 9747 AG, Netherlands}

\vspace*{0.1cm}

{\tt a.kadosh@rug.nl, e.pallante@rug.nl }
\end{center}
\vglue 0.3truecm

\begin{abstract}
\vskip 3pt \noindent

We recently proposed a spontaneous A${}_4$
flavor symmetry breaking scheme implemented in a warped extra
dimensional setup to explain the observed pattern of quark and
lepton masses and mixings. The quark mixing is induced by bulk
A${}_4$ flavons mediating ``cross-brane"
interactions and a ``cross-talk" between the quark and neutrino
sectors.
In this work we explore the phenomenology of RS-A${}_4$ and systematically obtain bounds on the
Kaluza-Klein mass scale implied by flavor changing neutral current (FCNC)
processes. In particular, we study the constraints arising from
$Re(\epsilon'/\epsilon_K)$, $b\rightarrow s\gamma$, the neutron
EDM and Higgs mediated FCNCs, while the tree level contribution
 to $\epsilon_K$ through a KK gluon exchange vanishes.
We find an overall lower bound on the Kaluza-Klein mass scale
$M_{KK}\gtrsim 1.3$ TeV from FCNCs, induced by $b\to s\gamma$
differently from flavor anarchic models. This bound is still
weaker than the bound $M_{KK}\gtrsim 4.6$ TeV induced by
$Zb_L\bar{b}_L$ in RS-A${}_4$. The little CP problem, related to
the largely enhanced new physics contributions to the neutron EDM
in flavor anarchic models, is absent. The subtleties of having the
Higgs and flavons in the bulk are taken into account and final
predictions are derived in the complete three-generation case.
\end{abstract}


\end{titlepage}

\section{Introduction}

In a recent
paper \cite{A4Warped} we have proposed a model based on a bulk A${}_4$ flavor
symmetry \cite{a4} in warped geometry \cite{RS}, in an attempt to
describe masses and mixing patterns of Standard Model (SM) quarks and leptons.
As in previous models based on A${}_{4}$  \cite{Csaki:2008qq}, the
three generations of left-handed quarks transform as triplets of A${}_4$; this assignment
forbids tree level gauge mediated FCNCs and allows to obtain realistic masses and
almost realistic mixing angles in the quark sector. The scalar sector of the RS-A${}_4$
model contains two bulk flavon fields, in addition to a bulk Higgs field.
The bulk flavons transform as triplets of A${}_{4}$, and allow for a complete
 ``cross-talk" \cite{Volkas} between the A${}_{4}\to Z_{2}$
spontaneous symmetry breaking (SSB) pattern associated with the
heavy neutrino sector -- with scalar mediator  peaked towards the
UV brane -- and the A${}_{4}\to Z_{3}$ SSB pattern associated with
the quark and charged lepton sectors -- with scalar mediator peaked
towards the IR brane. A bulk custodial symmetry, broken
differently at the two branes \cite{Agashe:2003zs}, guarantees the
suppression of large contributions to electroweak precision
observables \cite{Carena:2007}, such as the Peskin-Takeuchi $S$, $T$ parameters.
However, the mixing  between zero modes of the 5D theory and their
Kaluza-Klein (KK) excitations -- after 4D reduction -- may still
cause significant new physics (NP) contributions to SM suppressed
flavor changing neutral current (FCNC) processes.

\noindent In the most general case, without imposing any additional flavor symmetry and assuming
anarchical 5D Yukawa couplings, new physics contributions can already be generated at
tree level through a KK gauge boson exchange.
Even if a RS-GIM suppression mechanism \cite{rsgim1,Cacciapaglia:2007fw} is at
 work,
stringent constraints on the KK scale come from the $K^{0}-\overline{K^{0}}$ oscillation
 parameter $\epsilon_{K}$ and the radiative decays $b\to s(d)\gamma$ \cite{Agashe:2004cp,Azatov},
the direct CP violation parameter $\epsilon^\prime/\epsilon_K$ \cite{IsidoriPLB}, and
 especially the neutron electric dipole moment
\cite{Agashe:2004cp}, where a KK mass of $\mathcal{O}$(3 TeV)
gives rise to a NP contribution which is roughly forty times
larger than the current experimental bound -- a CP problem in
itself, referred to as little CP problem. The bounds become
increasingly stringent by IR localizing the Higgs field.

\noindent Conclusions may differ
if a flavor pattern of the Yukawa couplings is assumed to hold in
the 5D theory due to bulk flavor symmetries. They typically imply
an increased alignment between the 4D fermion mass matrix and the
Yukawa and gauge couplings, thus suppressing the amount of flavor
violation induced by the interactions with KK states.
One example that removes or suppresses all tree level contributions
is the generalization to 5D of minimal flavor violation in the quark sector
\cite{Fitzpatrick:2007sa} and in the lepton
sector \cite{Chen:2008qg, Perez:2008ee}.
In these settings, the bulk mass matrices are aligned with the 5D
Yukawa matrices as a result of a bulk $[U(3)]^{6}$ flavor symmetry that is broken in a
controlled manner.
In \cite{Csaki:2009wc} a shining mechanism is proposed, where the suppression of flavor
violation in the effective 4D theory on
the IR brane
is obtained by confining the sources of flavor violation to the UV brane, and communicating
its effects through gauge bosons of the gauged bulk flavor symmetry.

\noindent In our case,
the most relevant consequence of imposing an A${}_4$ flavor symmetry is  the
degeneracy of the left-handed fermion bulk profiles $f_Q$, i.e.
$diag(f_{Q_1,Q_2,Q_3})=f_Q\times \mathbbm{1}$. In addition, the
distribution of phases, CKM and Majorana-like, in the mixing
matrices might induce zeros in the imaginary components of the
Wilson coefficients contributing to CP violating quantities.
 In \cite{A4Warped} we already observed a few consequences of the A${}_4$ flavor symmetry.
First, the new physics contribution to $\epsilon_K$ coming from a
KK gluon exchange at tree level vanishes \cite{A4Warped}, thus
relaxing the most stringent bound on the KK scale induced by
$\epsilon_K$ in flavor anarchic models \cite{IsidoriPLB}.
 This leaves  $b\to s(d)\gamma$, $\epsilon^\prime /\epsilon_K$,
the neutron EDM and Higgs mediated FCNCs as possible candidates to
produce the most stringent lower bounds on the KK scale. In
addition, a milder lower bound from the EDM and $\epsilon^\prime
/\epsilon_K$ should be expected in our model due to the vanishing
of down-type dipole contributions in the naive spurion analysis
and mass insertion approximation.
It should also be interesting to compare this
pattern to the case of larger realizations
of the flavor symmetry, like $T'$ \cite{TPrime}, usually associated
with a rather richer flavon sector.

\noindent In this paper we analyze the above processes, $b\to
s(d)\gamma$, $\epsilon^\prime /\epsilon_K$, the neutron EDM and
Higgs mediated FCNC (HMFCNC) processes
\cite{HiggsContino,HiggsFCNC}, in the context of RS-A${}_4$.
Differently from flavor anarchy, it is particularly relevant in
this case to properly describe the flavor pattern of Yukawa
interactions and the mixing among generations. For this reason, we
predict all quantities at various levels of approximation,
starting with the generalization of the spurion analysis in the
mass insertion approximation to include bulk effects parameterized
by overlap factors. The latter quantities measure the deviation from the case of a IR
localized Higgs. We then proceed beyond the mass insertion
approximation, for each generation separately: this means that KK
mass eigenstates for each separate generation are obtained by
disregarding generational mixing, while the latter is
approximately described by the flavor structure of the spurion
analysis. Finally, we compare with the {\em exact} three-generation case,
where all contributions are obtained in terms of the KK mass
eigenstates, after the complete mass matrix for the zero modes and
KK modes is diagonalized numerically, or by means of an approximate
analytical procedure.

\noindent The paper is organized as follows. In
Sec.~\ref{sec:Quark} we recall the important components of the RS-A${}_4$
model proposed in \cite{A4Warped}, focusing on the Yukawa sector of the theory. In
Sec.~\ref{sec:Dipole} we derive new physics contributions to the
Wilson coefficients of magnetic and chromo-magnetic dipole
operators, relevant for the estimate of the neutron EDM,
$b\rightarrow s\gamma$ and $Re(\epsilon'/\epsilon_K)$. In
particular, we describe the various degrees of approximation, in
which the KK mixing within each generation and the mixing among
generations can be incorporated.  The analysis is then performed
separately for each observable in Sec.~\ref{Sec:NumericalResults}
and predictions are studied by varying the model input parameters.
Sec.~\ref{sec:higgs} describes Higgs mediated FCNC processes.
We conclude in Sec.~\ref{sec:conclusion}. A few appendices are included. The overlap factors are defined and computed in Appendix~\ref{app:A}. Appendix~\ref{app:B} contains details of the diagonalization of the KK mass matrices in the one-generation approximation and for three-generations.

\section{Quark sector of the A${}_4$ warped model}
\label{sec:Quark}
We start by reviewing some useful results and definitions for the quark
sector in RS-A${}_4$. In this model \cite{A4Warped} we adopt a
custodial RS setup without an additional $P_{LR}$ symmetry \cite{Agashe:2006at}. We then assign the three generations of left-handed fermion weak doublets to
triplets of the discrete non-abelian
flavor symmetry, A${}_4$. The right-handed charged fermions are
assigned to the 3 distinct one-dimensional representations of
A${}_4$. The SSB pattern A${}_4\rightarrow nothing$ is driven by the
VEVs of two flavons $\Phi$ and $\chi$, which are assigned to be
triplets of A${}_4$ peaked towards the IR and UV branes,
respectively, and it is responsible for the generation of fermion
masses and mixings in good agreement with the experimental results
\cite{PDG}.
\subsection{The  4D Yukawa Lagrangian}
Since the Higgs field and the A${}_4$
flavons $\Phi$ and $\chi$ live in the bulk, it will be instructive
to generalize \cite{IsidoriPLB} and write the 4D Yukawa lagrangian
in terms of overlap correction factors $r$'s, which quantify the deviation from the IR localized case. All overlap factors,
 defined as the ratio between the bulk wave function
overlaps and the approximate coupling on the IR brane, are derived in Appendix \ref{app:A}.

\noindent
The leading order (LO) 4D Yukawa lagrangian, generated by the LO 5D and A${}_4$-invariant Yukawa lagrangian in \cite{A4Warped}, and including all the effective interactions in the KK tower, carries similar structure in
the up- and down-quark sector. In particular, the leading order interactions with the neutral Higgs can be written as follows
\begin{eqnarray} \label{lagrangian} \mathcal{L}^{4D} \supset
 & \hat{Y}^{u,d}_{ij}h_{0(4D)}^{(*)} \left[\,\psi^{0 \dagger}_{Q_i}
f^{-1}_{Q_i} \psi^0_{u_j,d_j}f^{-1}_{u_j,d_j} r^{H\Phi}_{00}
(c_{Q_i},c_{u_j,d_j},\beta)+ \sum_{n} \psi^{0 \dagger}_{Q_i}
f^{-1}_{Q_i} \psi^n_{u_j,d_j} r^{H\Phi}_{0n} (c_{Q_i},c_{u_j,d_j},\beta) \right.\nonumber \\
& \left. + \sum_{n} \psi^{0 \dagger}_{Q_i} f^{-1}_{Q_i}
\psi^{n^{-+}}_{u_j,d_j} r^{H\Phi}_{0n^{-+}} (c_{Q_i},c_{d_j,u_j},\beta)+
 \sum_{n} \psi^{n \dagger}_{Q_i} \psi^0_{u_j,d_j}
f^{-1}_{u_j,d_j} r^{H\Phi}_{n0} (c_{Q_i},c_{u_j,d_j},\beta)
\right.\nonumber
\\  & \left.+  \sum_{n,m} \psi^{n \dagger}_{Q_i} \psi^m_{u_j,d_j}
 r^{H\Phi}_{nm} (c_{Q_i},c_{u_j,d_j},\beta)  +
\sum_{n,m} \psi^{n^{--}}_{Q_{i}} (\psi^{m^{--}}_{u_j,d_j})^\dagger
r^{H\Phi}_{n^{-}m^{-}} (c_{Q_i},c_{u_j,d_j},\beta) \right. \nonumber\\
& \left.+ \sum_{n,m} \psi^{n \dagger}_{Q_i}
\psi^{m^{-+}}_{u_j,d_j}
 r^{H\Phi}_{nm^{-+}} (c_{Q_i},c_{d_j,u_j},\beta)  +
\sum_{n,m} \psi^{n^{--}}_{Q_i} (\psi^{m^{+-}}_{u_j,d_j})^\dagger
r^{H\Phi}_{n^{-}m^{+-}} (c_{Q_i},c_{d_j,u_j},\beta)
\right] \,,\nonumber\\
\end{eqnarray}
where $h_{0(4D)}^{(*)}$ couples to the down (up) sector, corresponding to the first
(second) label in $c_{q_i,q'_i}$, and
$f_{Q_i,u_i,d_i}=\sqrt{2k}/\hat{\chi}_{0_{Q_i,u_i,d_i}}$, with
$\hat{\chi}_{0_{Q_i,u_i,d_i}}$ the canonically normalized zero
mode profile of the corresponding fermion at the IR
brane -- see Appendix \ref{app:A}. With the same
convention, all KK wave functions on the IR brane are
approximately equal to $\sqrt{2k}$. The $\psi$'s denote the 4D wave
functions of the fermion fields in the KK tower.  The boundary
condition (BC) for each KK mode is also specified.
 Unless stated otherwise, the BC are of the type $(++)$ on the UV
and IR brane, respectively. A single $(-)$ in the overlap subscript stands for $(--)$, and all
other BC are fully specified.  In the custodial case, each fermion zero mode, with $(++)$ boundary conditions, is accompanied by three first level KK modes, with $(++)$, $(--)$ and $(+-)$ (or $(-+)$) boundary conditions.
The quantities
$r^{H\Phi}_{nm}$ are the overlap correction factors for the states
$n$ and $m$ calculated in appendix \ref{app:A}. They are functions of the left-handed (LH) fermion bulk mass parameters $c_{Q_i}$, the right-handed (RH) ones $c_{d_i,u_i}$ and the scalar bulk mass parameter $\beta$ (see appendix \ref{app:A}).
The  Higgs field transforms as a bidoublet under ${\mbox SU}(2)_L\times
SU(2)_R$, but contains only two degrees of freedom, $h_0$ and $h_+$
\begin{equation}
{\cal H}=\left( \begin{array}{cc} H & \tilde{H}
\end{array}\right)=\left(\begin{array}{ccc}h_0^*&h_+\\-h_+^*&h_0\\\end{array}\right)\qquad
h_0(x,y)=v_H(\beta_H,y)+\sum_{n}h_0^{(n)}(x)\phi_n(y)\, .
\label{HiggsRep.}
\end{equation}
The profile of the Higgs VEV along
the fifth dimension (see also \cite{WiseScalar}) is
\begin{equation}
v_H=H_0e^{(2+\beta_H)k(|y|-\pi
R)}\,,\label{VEVprofile}
\end{equation}
with $\beta_H=\sqrt{4+\mu_H^2}$, and $\mu_H$ the bulk mass of the
5D Higgs field in units of the $AdS_5$ curvature scale $k\approx
M_{Pl}$. As in \cite{A4Warped}, we assume $\beta_H\simeq 2$,
which yields $H_0\simeq 0.39M_{Pl}^{3/2}$, for $k\pi R\simeq 34.8$ and matching with the
measured $W$ boson mass. In
addition, the profile of the physical higgs $h_0^{(1)}(y)$ is
almost identical to the VEV profile for $m_h<<M_{KK}$. The VEV
profile for the A${}_4$ flavon $\Phi$, peaked towards the IR brane, is
of similar structure to the one of the Higgs, with
$\beta_\phi\simeq 2$ and $\Phi_0\simeq 0.577M_{Pl}^{3/2}$. The VEV
profile of the UV peaked A${}_4$ flavon, $\chi$, will only enter
through the subdominant Yukawa interactions and is approximately
\begin{equation}v_\chi=\chi_0e^{(2-\beta_\chi)k|y|}(1-e^{(2\beta_\chi)k(|y|-\pi R)})
\label{VEVchi},\end{equation} with $\beta_\chi\simeq2$ and
$\chi_0\simeq 0.155M_{Pl}^{3/2}$.
The leading order 5D and A${}_4$-invariant Yukawa
lagrangian in \cite{A4Warped}, consisting of operators of the form
$(y_{u_i,d_i,e_i}/M_{Pl}^2)\bar{Q}_L(\bar{L}_L)\Phi H
u_{R_i}(d_{R_i}, e_{R_i})$, was shown to induce the same pattern of masses and mixings
in the up, down and charged lepton sectors. After spontaneous symmetry breaking of A${}_4$, by
the VEVs of the $\Phi$ triplet, the leading order 4D Yukawa matrices in these sectors
take the form
\begin{equation}
(\hat{Y}^{u,d,e}_{ij})_{LO}=\frac{2v_\Phi^{4D}e^{k\pi
R}}{k}\left(\begin{array}{ccc}y_{u,d,e}&y_{c,s,\mu}&y_{t,b,\tau}\\y_{u,d,e}&
\omega y_{c,s,\mu}&\omega^2 y_{t,b,\tau}\\y_{u,d,e}& \omega^2
y_{c,s,\mu}& \omega y_{t,b,\tau}
\end{array}\right)\equiv
\frac{2v_\Phi^{4D}e^{k\pi R}}{k}(\hat{y}^{u,d,e}_{ij})_{LO}  \,,
\label{LOYukawa}
\end{equation}
where $y_{u,d,e}$ are the dimensionless 5D Yukawa couplings and
 $\omega=e^{2\pi i/3}$. The parameter $v_{\Phi}^{4D}$ denotes the 4D VEV of
$\Phi$ in the IR localized case, and it is given by $\Phi_0 \simeq
\sqrt{2k\,(1+\beta_\Phi)}v_{\Phi}^{4D}e^{k\pi R}$, up to
exponentially suppressed contributions -- see Appendix \ref{app:A}
for its exact expression.
Notice that, differently from the flavor anarchic case, the
overlap factors in Eq.~(\ref{lagrangian}) are now functions of the
VEV profiles of all scalar fields, $H$ and $\Phi$ at leading order,
with $\beta =\beta_H +\beta_\Phi$. The other crucial ingredient of
the RS-A${}_4$ model is the degeneracy of the LH fermion bulk mass parameters,
since the corresponding fermions are unified in triplets of A$_4$;
consequently $f_{Q_i}\equiv f_Q$ and
$\hat{\chi}_{0_{Q_i}}\equiv\hat{\chi}_{0_{Q}}$ in
Eq.~(\ref{lagrangian}).

\noindent The Yukawa texture in Eq.~(\ref{LOYukawa}) was shown \cite{A4Warped} to induce the
same left-diagonalization matrix
\begin{equation}
V_L^{u,d,e}=U(\omega)=\frac{1}{\sqrt{3}}\left(\begin{array}{ccc}1&1&1\\1&\omega&\omega^2\\1&\omega^2&\omega
\end{array}\right)\label{UOmega}\end{equation}
 for all charged fermions and in
particular for the zero modes, identified with the SM
fermion content.
At leading order, the right diagonalization matrix for all charged
fermions is simply the identity. This pattern of the
diagonalization matrices, independent of the leading order 5D
Yukawa couplings, will be shown not to
induce any of the flavor violating interactions we wish to
inspect.

\noindent The deviation from unity of the CKM matrix, and thus quark
mixing, is induced by cross-talk effects \cite{Volkas} in
RS-A${}_4$ \cite{A4Warped}. They mediate between the IR and UV
branes and between the SSB patterns of the neutrino and quark
sectors, in the form of higher dimensional operators
$(1/M_{Pl}^{7/2})\,\bar{Q}_L\Phi\chi H (u_R, u_R', u_R'', d_R,
d_R', d_R'')$ and breaking completely the A${}_4$ flavor symmetry. Each
of these operators turned out to yield two independent
contributions to the up- and down-quark mass matrices, for which we
label the dimensionless 5D coefficients as $\tilde{x}_i^{u,d}$ and
$\tilde{y}_i^{u,d}$. Since the leading order diagonalization
matrices are independent of the corresponding Yukawa couplings,
the perturbed diagonalization matrices are governed by
$\tilde{x}_i^{u,d}$, $\tilde{y}_i^{u,d}$ only. Although we need a
specific assignment of these parameters to match the CKM matrix
while maintaining the magnitude of all parameters naturally of
order one, we will explore the full parameter space of the model
to account for the largest possible contributions of new physics
to FCNC processes. They will provide the most stringent
constraints on the KK mass scale in the RS-A${}_4$ setup. The
 4D Yukawa matrix induced by the above higher
order effects can thus be parameterized as follows
\begin{equation} (\hat{Y}_{ij}^{u,d})_{NLO}=\frac{2
v_\Phi^{4D}v_{\chi}^{4D}e^{2k\pi R}}{k^{2}}
\left(\begin{array}{ccc}\tilde{x}_1^{u,d}&\tilde{x}_2^{u,d}
&\tilde{x}_3^{u,d}\\0&0&0\\\tilde{y}_1^{u,d}&\tilde{y}_2^{u,d}
&\tilde{y}_3^{u,d}\end{array}\right)\equiv \frac{
2v_\Phi^{4D}v_{\chi}^{4D}e^{2k\pi
R}}{k^2}(\hat{y}_{ij}^{u,d})_{NLO}\,.\label{NLOYukawa}\end{equation}
This time the VEV profile of the UV peaked flavon field $\chi$
will also enter all the corresponding overlap correction factors,
leading to the NLO 4D lagrangian analogous to
Eq.~(\ref{lagrangian}), with overlaps $r^{H\Phi\chi}_{nm}$ as
defined in appendix \ref{app:A}. The modified left- and right-diagonalization
matrices for the up and down mass matrix
have a simple structure, up to and including linear terms in
$\tilde{x}_i^{u,d}$, $\tilde{y}_i^{u,d}$ and working in the zero mode
approximation (ZMA)\cite{A4Warped}. The left-handed matrix is
given by 
\begin{equation}V_L^q=U(\omega)\left( \begin{array}{ccc} 1 &
f_\chi^{q_2}(\tilde{x}_2^q+\tilde{y}_2^q) &
f_\chi^{q_3}(\tilde{x}_3^q+\tilde{y}_3^q)\\
-f_\chi^{q_2}[(\tilde{x}_2^{q})^* + (\tilde{y}_2^{q})^*] & 1 &
f_\chi^{q_3}(\tilde{x}_3^{q} + \omega \tilde{y}_3^{q})\\
-f_\chi^{q_3}[(\tilde{x}_3^{q})^* + (\tilde{y}_3^{q})^*]&
-f_\chi^{q_3}[(\tilde{x}_3^{q})^* + \omega^2(\tilde{y}_3^{q})^*]&
1
\end{array}\right)\label{VLQ},
\end{equation}
with $q=u,d$ and $f_\chi^{q_i}=4C_\chi/(12-c_q^L-c_{q_i})$, with $C_\chi=\chi_0/k^{3/2}\simeq 0.155$ and $\omega=e^{2\pi
i/3}$. In \cite{A4Warped}, we assigned the degenerate left-handed bulk parameter $c_q^L=0.4$, and the right-handed parameters
$c_u=0.78$, $c_d=0.76$, $c_s=0.683$, $c_c=0.606$, $c_b=0.557$ and
$c_t=-0.17$, to yield the physical running quark masses at the KK scale of 1.8
TeV and satisfy the stringent constraints coming from $Zb_L\bar{b}_L$.
The CKM matrix elements to first order in $f_\chi^{q_i}
(\tilde{x}_i^{u,d}, \tilde{y}_i^{u,d})$ are easily obtained from
$V_{CKM}=(V_L^u)^\dagger V_L^d$, leading to
\begin{equation}
V_{us}=-V_{cd}^{*}\simeq\left((\tilde{x}_2^d+\tilde{y}_2^d)f_{\chi}^s-(\tilde{x}_2^u+\tilde{y}_2^u)
f_{\chi}^c \right),\label{Vus}
\end{equation}
\begin{equation}
V_{cb}=-V_{ts}^{*}\simeq\left((\tilde{x}_3^d+\omega\tilde{y}_3^d)f_{\chi}^b-(\tilde{x}_3^u+
\omega\tilde{y}_3^u)f_{\chi}^t\right),  \label{Vcb}
\end{equation}
\begin{equation}
V_{ub}=-V_{td}^{*}\simeq\left((\tilde{x}_3^d+\tilde{y}_3^d)f_{\chi}^b-(\tilde{x}_3^u+
\tilde{y}_3^u)f_{\chi}^t\right). \label{Vub}
\end{equation}
An almost realistic CKM matrix can be obtained with minimal deviations
from the universality assumption that all magnitudes of
$\tilde{x}_i^{u,d}$, $\tilde{y}_i^{u,d}$ are of $\mathcal{O}(1)$; in particular
\begin{equation}
\tilde{x}_2^{u}=\tilde{y}_2^{u}=-\tilde{x}_2^{d}=-\tilde{y}_2^{d}=
e^{i\delta_2^u}\,,\,\,\,\,\tilde{x}_3^u \simeq
0.67-0.19i\,,\,\,\,\, \tilde{y}_3^d \simeq
0.59-0.23i\,.\label{CKMAssignment}
\end{equation}
Considering the global fit of the parameters of the Wolfenstein
parametrization~\cite{PDG}, we can obtain real $V_{us}$ and consequently
real $V_{cd}$ with the choices $\delta_2^u=0, \pi$.
 All other
$\tilde{x}_i^{u,d}$, $\tilde{y}_i^{u,d}$ parameters are simply set
to unity, as explained in \cite{A4Warped}. The CKM matrix obtained
by this choice has $|V_{us}|=|V_{cd}|=0.2257$,
$|V_{cb}|=|V_{ts}|=0.0415$, $|V_{ub}|=|V_{td}|=0.00359$ and
$V_{ii}=1$.  The phase of $V_{ub}$ is matched by the same
assignments to its experimental value, $\delta\simeq 1.2$, while
the other off-diagonal elements are real. This provides an almost realistic CKM matrix.
The main deviation from the global fit~\cite{PDG}
amounts to the difference in magnitude of $V_{ub}$ and $V_{td}$.
In addition, one still has to account for the
$\mathcal{O}(\lambda_{CKM}^2)$ deviations from unity of the
diagonal elements and match the phases of the CKM elements to the
9 constraints implied by the Jarlskog invariant. All deviations
 have to come from higher order corrections in the RS-A${}_4$ model, rendering the corresponding
parameter assignments less appealing.

\noindent The right diagonalization matrices do not enter the CKM matrix, however, they are crucial in the
evaluation of the Wilson coefficients contributing to the FCNC processes we are interested in. To first order in
$f_\chi^{q_i} (\tilde{x}_i^{u,d}, \tilde{y}_i^{u,d})$ one obtains
\begin{equation}V_R^q=\left( \begin{array}{ccc} 1 &
\Delta_1^q &
\Delta_2^q\\
-(\Delta_1^q)^*  & 1 &
\Delta_3^q\\
-(\Delta_2^q)^* & -(\Delta_3^q)^* & 1
\end{array}\right)\label{VRQ},
\end{equation}
where $q=u,d$ and the $\Delta_i^q$ are  given by:
\begin{equation}
\Delta_1^q
=\frac{m_{q_1}}{m_{q_2}}\left[f_\chi^{q_1}\left((\tilde{x}_1^q)^*
+\omega^2 (\tilde{y}_1^q)^*\right)
+f_\chi^{q_2}\left(\tilde{x}_2^q+\tilde{y}_2^q\right)\right]
\label{DeltaR1},\end{equation}
\begin{equation}
\Delta_2^q
=\frac{m_{q_1}}{m_{q_3}}\left[f_\chi^{q_1}\left((\tilde{x}_1^q)^*
+\omega (\tilde{y}_1^q)^*
\right)+f_\chi^{q_3}\left(\tilde{x}_3^q+\tilde{y}_3^q\right)\right]
\label{DeltaR2},\end{equation}
\begin{equation}
\Delta_3^q
=\frac{m_{q_2}}{m_{q_3}}\left[f_\chi^{q_2}\left((\tilde{x}_2^q)^*
+\omega (\tilde{y}_2^q)^*\right)
+f_\chi^{q_3}\left(\tilde{x}_3^q+\omega\tilde{y}_3^q\right)\right]
\label{DeltaR3}.\end{equation}
The suppression by quark mass ratios of the off-diagonal elements
in $V_R^{u,d}$ will turn out to play an important role in relaxing
the flavor violation bounds on the KK mass scale, as compared to
flavor anarchic frameworks.

\subsection{Parameter counting and physical phases}
In order to estimate the new physics contributions associated with
the imaginary parts of amplitudes, we need to know how many real
and imaginary physical parameters are in our model. We start
with the 6 leading order Yukawa couplings $y_{q_i}$ and the 12
$\tilde{x}_i^q$ and $\tilde{y}_i^q$ couplings of the cross-talk
operators, $\bar{Q}_L\Phi\chi H (u_R, u_R', u_R'', d_R, d_R',
d_R'')$. Besides the Yukawas, we have 6 real and 3 imaginary
parameters in the spurions $F_{u,d}=diag(f^{-1}_{u_j,d_j})$, and 1 real parameter $F_Q=f_Q^{-1}\mathbbm{1}$. Hence,
in total, we have 31 real and 24 imaginary parameters in the
most general case.

 \noindent We now consider the flavor symmetry breaking pattern before the SSB
of A${}_4$, $U(3)_Q\times U(3)_u\times U(3)_d\longrightarrow A_4$, induced by
the leading order Yukawa lagrangian and the cross-talk operators
in charge of quark mixing. We realize that we can
eliminate 17 phases -- the baryon number should still be conserved -- and
6 real parameters. This leaves us with
25 physical real parameters, that is the 12 mixing angles in
$V_{L,R}^{u,d}$, 6 quark masses and the 7 eigenvalues of
$F_{Q,u,d}$. In the imaginary sector, we are left with
7 phases, 4 of which are CKM-like phases, one in each of the
$V_{L,R}^{u,d}$ matrices, while the other 3 are Majorana-like
phases which can be rotated between the left and right
diagonalization matrices of both the up and down sectors. We
should take these phases into account when evaluating the imaginary
parts of amplitudes and we will do so by parametrizing the phase of each
element of $V_{L,R}^{u,d}$ in terms of phases of the parameters
$\tilde{x}_i^{u,d}$ and $\tilde{y}_i^{u,d}$, which govern the
structure of the diagonalization matrices.

\section{ Dipole Operators and helicity flipping FCNCs }
\label{sec:Dipole} FCNC processes are known to provide among the
stringest constraints for physics beyond the standard model. This
is also the case for flavor anarchic models in warped extra
dimensions \cite{Agashe:2004cp,Azatov,IsidoriPLB}. In the quark
sector, significant bounds on the KK mass scale may typically come
from the neutron electric dipole moment (EDM), the CP violation
parameters $\epsilon_K$ and $Re(\epsilon^\prime/\epsilon_K)$, and
radiative $B$ decays such as $b\to s\gamma$. All these processes
are mediated by effective dipole operators. It is also well known
\cite{Buras} that SM interactions only induce, to leading order,
the dipole operators $O_{7\gamma}$ and $O_{8g}$ 
\begin{equation}
O_{7\gamma(8g)}=\bar{d}_R^i \sigma^{\mu\nu}d_L^j
F_{\mu\nu}(G_{\mu\nu}),\label{SM-O7}\end{equation}
where $F_{\mu\nu}$ and $G_{\mu\nu}$ are the field strength of the
electromagnetic and chromomagnetic interactions and $i,j$ are
flavor indices.  For $i>j$, as
$\bar{b}_R\sigma^{\mu\nu}F_{\mu\nu}s$, the SM contribution to the
Wilson coefficients of the opposite chirality operators
$O_{7\gamma,8g}'$ is suppressed by the corresponding quark mass
ratio, and thus negligible. This might turn out to be a unique
feature of the SM not shared by NP contributions. It is therefore
instructive to study new physics contributions of any flavor model
to the operators $O_{7\gamma,8g}$ and to the opposite chirality
operators, $O_{7\gamma,8g}'$, and compare with experimental
results. In the following we show that, differently from flavor
anarchic models, the RS-A${}_4$ model shares the SM features, with
no enhancement of the opposite chirality operators.

\subsection{Flavor structure of Dipole operators}\label{subsec:EDM}
\begin{figure}
\begin{minipage}[t]{0.5\linewidth}
\begin{center}
\includegraphics[scale=0.60]{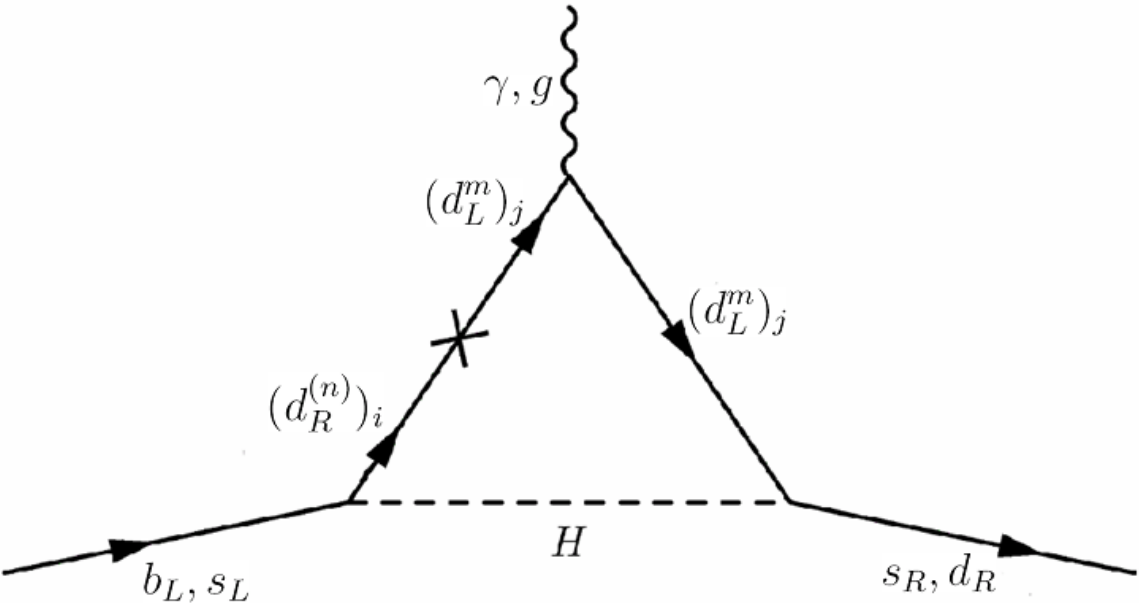}
\caption{One-loop down-type (neutral Higgs) contribution
to $b\rightarrow s\gamma$, $\epsilon'/\epsilon_K$ and the neutron EDM (for
external $d$ quarks). The analogous one-loop up-type contribution (charged Higgs)
contains internal up-type KK modes. }
\label{fcncLoop}
\end{center}
\end{minipage}
\hspace{0.2cm}
\begin{minipage}[t]{0.5\linewidth}
\begin{center}
\includegraphics[scale=0.26]{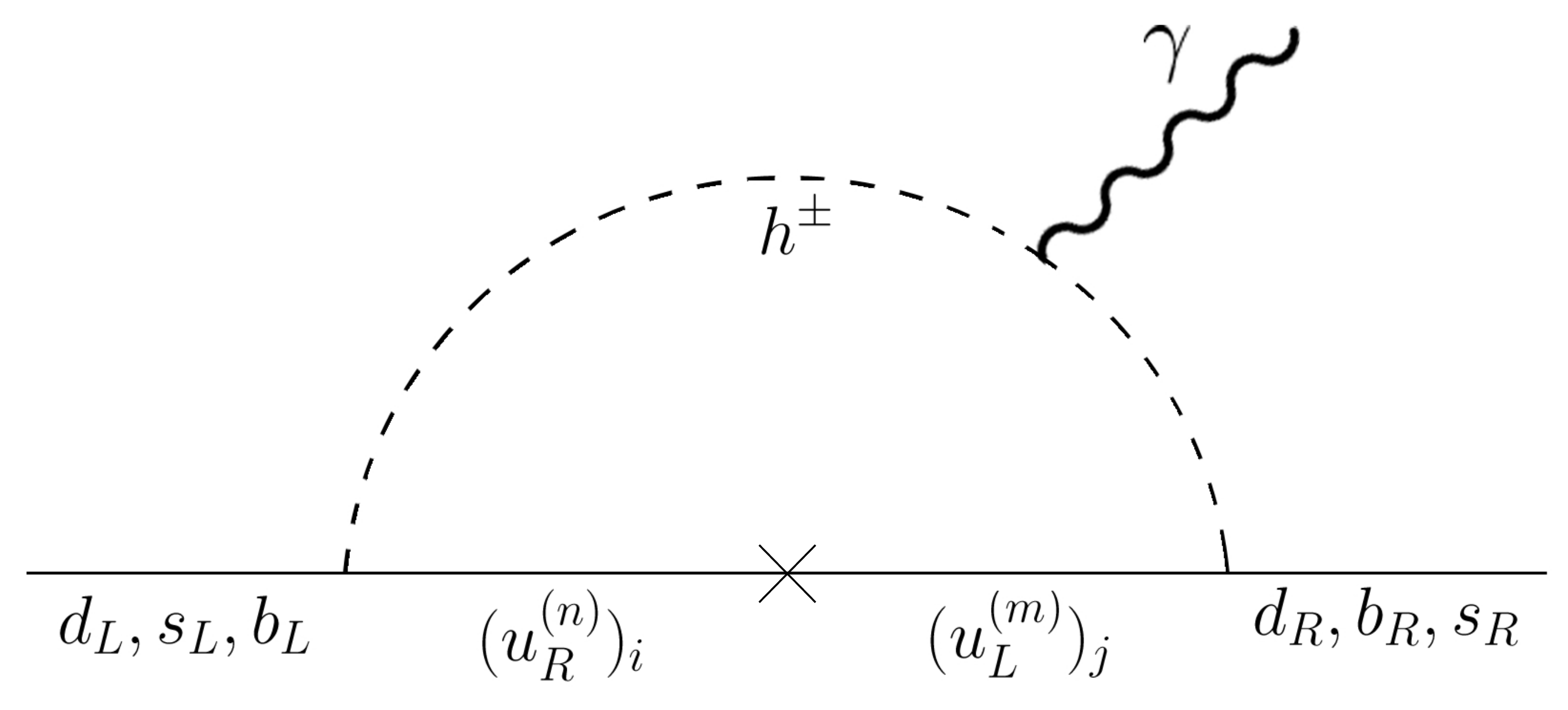}
\caption{Charged Higgs one-loop contribution to $b\rightarrow
s\gamma$ and  the neutron EDM. The latter has external $d$
quarks. }
\label{fcncLoopcharged}
\end{center}
\end{minipage}
\end{figure}

The new physics contributions to the  FCNC processes we are
interested in are generated at one-loop by the Yukawa interactions
between SM fermions  and their KK excitations, leading to the
diagrams shown in Fig.~\ref{fcncLoop}  and
Fig.~\ref{fcncLoopcharged}.
 To obtain the flavor structure for the Wilson
coefficients of the corresponding dipole operators we first recall the spurion analysis in the mass insertion approximation of
\cite{Agashe:2004cp}, corresponding to the IR localized Higgs
case. The contributions associated with internal KK down quarks
in the special interaction basis\footnote{The basis in which $F_{Q,u,d}$
are real and diagonal is referred to as the special interaction basis in
\cite{Agashe:2004cp} and the rest of this paper.}  can  be written as
\begin{equation}
(C_{7\gamma(8g)}^{d-type})_{ij}=A^{1L}\frac{v}{M_{KK}}\,\left(F_Q\,
\hat{Y}_d\hat{Y}_d^\dagger
\hat{Y}_d\,F_d\right)_{ij}\,,\label{C7DY}\end{equation}
 where
$v\equiv v_H^{4D}=174$ GeV denotes the Higgs VEV, $\hat{Y}_d$ the
5D Yukawa matrices and the fermion profile matrices are
$F_{Q,u,d}=diag(f_{Q_i,u_j,d_j}^{-1})$ -- see also appendix
\ref{app:A}. Finally, the factor $A^{1L}= 1/(64\pi^2M_{KK})$ comes
from the one-loop integral for the diagram in
Fig.~\ref{fcncLoop}\footnote{Notice that we assumed degenerate KK
masses with common mass $M_{KK}$ and the result is valid in the
limit $m_H\ll M_{KK}$. We also disregard subdominant W/Z mediated
diagrams.} and the factor $v/M_{KK}$ comes from the mass insertion
approximation. The contributions associated with internal up-type
KK quarks (and a charged Higgs) in Fig.~\ref{fcncLoop} will
analogously be given by:
\begin{equation}
(C_{7\gamma(8g)}^{u-type})_{ij}=A^{1L}\frac{v}{M_{KK}}\,\left(F_Q\,\hat{Y}_u\hat{Y}_u^\dagger
\hat{Y}_d\,F_d\right)_{ij}\,,\label{C7UY}\end{equation}
written again in the special interaction basis.
The neutron EDM and  $b\rightarrow s\gamma$ receive an
additional up-type contribution from the diagram in
Fig.~\ref{fcncLoopcharged}, which carries the same flavor (and overlap)
structure of the up-type diagram in Fig.~\ref{fcncLoop} and a one-loop amplitude
that differs by a sign to a very good approximation\footnote{Neglected contributions
are suppressed by mass ratios $m_{d_i}/M_{KK}$, see \cite{Azatov} for a derivation of
those terms.}.  Hence, the total up-type contribution is obtained by replacing $A^{1L}$ with
\begin{equation}
(\tilde{A}^{1L}_u)=A^{1L}Q_u+A^{1L}_{H\gamma}Q_{h_{-}}=A^{1L}(\frac{2}{3}+1)
=\frac{5}{3}\frac{1}{64\pi^2M_{KK}}
\label{A1Ladd},\end{equation}
where $Q_u$ and $Q_{h-}$ are the
electric charges of an up-type quark and the negatively charged
Higgs, respectively.

\noindent Thus far we have not considered the modifications of the above
spurion structures due to the overlap of internal KK quarks,
 external fermion zero modes, and bulk scalar fields $\Phi$, $\chi$ and Higgs field, encoded in
the various $r_{nm}$ factors in Eq.~(\ref{lagrangian}). Since the
bulk nature of all 5D fields is an essential feature of  our model,  the effect
of all overlaps should be taken into account.
In the following section we derive the analogous of Eqs.~(\ref{C7DY}) and (\ref{C7UY}), corrected
 by the overlap factors in our model.
Subsequently, we show that conservatively reducing the overlap corrections to
an overall multiplicative factor will
suffice for a conservative estimate of most of the flavor
violation bounds on the KK scale in our model, and it will be instructive for the comparison with other
 flavor scenarios and in particular warped flavor anarchic models.
\subsection{The spurion-overlap approximation}\label{Sec:overspurion}
Observing the Yukawa Lagrangian of Eq.~(\ref{lagrangian}), we
realize that the spurion analysis in the mass insertion approximation can only directly account
 for the interactions (and related overlaps) associated with $(++)$ KK modes or a
combination of $(-+)$ and $(++)$ KK modes as internal
states in the Feynman diagrams of Figs.~\ref{fcncLoop} and \ref{fcncLoopcharged}.
However, since the first KK masses of each fermion are nearly
degenerate (see appendix \ref{app:B}), they almost maximally mix.
Therefore, we expect that the contributions of the three distinct KK
modes of each given fermion, can be estimated to a good
approximation by only considering the modes directly entering the spurion analysis.
The corresponding overlap correction factors now enter in the spurion structures of
Eqs.~(\ref{C7DY}) and (\ref{C7UY})  to
yield in the special interaction basis
\begin{eqnarray}
(C_7^{d,u})_{(++)}&\propto&
F_Q\hat{Y}_{d,u}\,r_{01}(c_{Q_i},c_{d_{\ell_1},u_{\ell_1}},\beta)\,\hat{Y}_{d,u}^\dagger\,r_{11}
(c_{d_{\ell_1},u_{\ell_1}},c_{Q_{\ell_2}},\beta)
\,\hat{Y}_{d,d}\,r_{10}(c_{Q_{\ell_2}},c_{d_j,d_j})F_{d}\nonumber\,,\\
(C_7^{d,u})_{(-+)}  &\propto&
F_Q\hat{Y}_{u,d}\,r_{01^{-+}}(c_{Q_i},c_{u_{\ell_1},d_{\ell_1}},\beta)\,\hat{Y}_{u,d}^\dagger\,
r_{1^{-+}1}(c_{u_{\ell_1},d_{\ell_1}},c_{Q_{\ell_2}},\beta)\,
\hat{Y}_{d,d}\,r_{10}(c_{Q_{\ell_2}},c_{d_j,d_j},\beta)F_{d}\,,\nonumber\\
\label{Spurion+Overlap}\end{eqnarray}
where  $\beta=\beta_H +\beta_\Phi$ and $\ell_i$, $i$ and $j$ are flavor indices. Notice
that we have omitted the flavor independent prefactor
$\,vA^{1L}/M_{KK}$ to ease the notation. From
Eq.~(\ref{lagrangian}) and Fig.~\ref{fcncLoop}, it is clear that the
4D Yukawa matrices $\hat{Y}$ carry the same flavor indices
as the adjacent overlap correction factors. Notice also that
the $c_{u_j,d_j}$ dependence of the $(-+)$ overlaps is opposite to
the one of the $(++)$ ones. This is not surprising since they
arise from the Yukawa interactions with $\tilde{u}_i$ and
$\tilde{d}_i$, the first (and higher) level KK excitations of the
$SU(2)_R$ partners of $d_{R_i}$ and $u_{R_i}$, respectively.

\noindent  In this context it is important to
mention the work of \cite{IsidoriPLB}, based on the method
developed in \cite{Azatov}, which involves the direct
diagonalization of the zero modes and first KK modes mass matrix; the latter is a $4\times 4$ matrix for a single generation in our case, and it reduces to a $3\times 3$ matrix when no custodial symmetry is imposed.
The interesting
result in flavor anarchic models for the one generation case, and to a good approximation
for the three generation case, is that the most dominant contributions, in terms of the
perturbative parameter $x=vY/M_{KK}$ with generic Yukawa $Y$, solely arise
 from the $(-\,-)$ modes. Namely, the dominant contribution turns out to be
proportional to the overlap structure $r_{01}r_{1^-1^-}r_{10}$, and it is not accounted for in the naive spurion analysis. This shows the limits of
the spurion analysis in the mass insertion approximation and the need to {\em a priori}
account for the mixing between all KK modes of the same generation, in addition to
their intergenerational mixing. This is especially important in the RS-A${}_4$ setup,
where custodial symmetry also induces  an extra degree of freedom for each ``RH" 5D fermion.

\noindent Nevertheless, because of the relative smallness of $r_{1^-1^-}$
compared to  $r_{11}$ and $r_{11^{-+}}$ in our setup (see
appendix \ref{app:A1}), the accuracy of the spurion-overlap approximation is
still satisfactory for the purpose of imposing constraints on the KK
mass scale and the physical Higgs mass, as long as the
corresponding contribution turns out to be non vanishing in this approximation. If vanishing -- as it is true for the down type contribution
to the neutron EDM \cite{A4Warped} -- one has to fully account for the flavor structure
 and mixing of all zero modes and first KK modes in order to provide an estimate of
 the dominant contributions. This is done in section \ref{Sec:MassDiag}.
\subsection{Explicit structure of dipole contributions in the spurion-overlap
approximation}\label{Sec:3spurion}
In this section we analyze in more detail  the most general flavor structure of up-
and down-type  contributions to dipole operators and study the simplifications induced
by the RS-A${}_4$ setup. We limit the analysis to the first level $n=1$ KK states,
since $n=2$ states will give rise to
$\mathcal{O}(25\%)$ effects and for $n=3$ and higher the
theory is strongly coupled and
cannot be treated perturbatively \cite{Agashe:2004cp}.
Inverting the relation
$(\hat{m}_{u,d})_{ij}=v\,f^{-1}_{Q_i}\hat{Y}^{u,d}_{ij}f^{-1}_{u_j,d_j}r_{00}(\beta
c_{Q_i},c_{u_j,d_j})$  and rotating the mass matrix (see also Appendix~\ref{app:A}),  we express the Yukawa couplings in terms of the diagonal physical mass matrices
 \begin{equation}
\hat{Y}^{u(d)}_{ij}=\frac{1}{v}r_{00}^{-1}(c_{Q_i},c_{u_j(d_j)},\beta)\left(F^{-1}_{Q}V_L^{u(d)}
\mbox{diag}(m_{u,c,t(d,s,b)})V_R^{u(d)\dagger}F^{-1}_{u(d)}\right)_{ij}.\label{massrelation}\end{equation}
Promoting the overlap corrections to matrices $\hat{r}$ in flavor space,  the
down-type contributions to the dipole amplitude rotated to the ZMA mass
basis can be written as
\begin{eqnarray}
(C_{7\gamma(8g)}^{d-type})_{ij}&=&\frac{A_{1L}}{v^2M_{KK}}\left[(V_L^d)^\dagger_{i\ell}
(\hat{r}_{0n}^d)_{\ell\ell_1}(\hat{r}_{00}^d)^{-1}_{\ell\ell_1}\left(V_L^d{\mbox
diag}(m_{d,s,b})(V_R^d)^\dagger\mbox{diag}(f_{d,s,b}^2)\right)_{\ell\ell_1}\right.\nonumber\\
&&\left. \times(\hat{r}_{nm}^d)_{\ell_2\ell_1}
(\hat{r}_{00}^d)^{-1}_{\ell_2\ell_1}\left(V_R^d{\mbox
diag}(m_{d,s,b})V_L^{d\dagger}{\mbox
diag}(f_{Q1,Q2,Q3}^2)\right)_{\ell_1\ell_2}(\hat{r}_{m0}^d)_{\ell_2\ell_3}\right.\nonumber\\
&&\left.\times(\hat{r}_{00}^d)^{-1}_{\ell_2\ell_3}
\left(V_L^d{\mbox
diag}(m_{d,s,b})(V_R^d)^\dagger\right)_{\ell_2\ell_3}(V_R^d)_{\ell_3j}\right]\label{Dspur},
\end{eqnarray}
with an implicit sum over all allowed first level KK modes, and where
$(\hat{r}_{nm}^{u,d})_{ij}=r_{nm}(c_{Q_i},c_{u_j,d_j},\beta)$.
In the above equation flavor indices are written explicitly, in
order to clarify the exact flavor structure of the overlap matrices. Analogously,
 we obtain the up-type contributions to dipole operators
\begin{eqnarray}
(C_{7\gamma(8g)}^{u-type})_{ij}&=&\frac{A_{1L}}{v^2M_{KK}}\left[(V_L^d)^\dagger_{i\ell}
(\hat{r}_{0n}^u)_{\ell\ell_1}(\hat{r}_{00}^u)^{-1}_{\ell\ell_1}\left(V_L^u{\mbox
diag}(m_{u,c,t})(V_R^u)^\dagger\mbox{diag}(f_{u,c,t}^2)\right)_{\ell\ell_1}\right.\nonumber\\
&&\left. \times(\hat{r}_{nm}^u)_{\ell_2\ell_1}
(\hat{r}_{00}^u)^{-1}_{\ell_2\ell_1}\left(V_R^u{\mbox
diag}(m_{u,c,t})V_L^{u\dagger}{\mbox
diag}(f_{Q1,Q2,Q3}^2)\right)_{\ell_1\ell_2}(\hat{r}_{m0}^d)_{\ell_2\ell_3}\right.\nonumber\\
&&\left.\times(\hat{r}_{00}^d)^{-1}_{\ell_2\ell_3}
\left(V_L^d{\mbox
diag}(m_{d,s,b})(V_R^d)^\dagger\right)_{\ell_2\ell_3}(V_R^d)_{\ell_3j}\right]\label{Uspur}.
\end{eqnarray}
Notice that the IR localized Higgs case can be obtained by simply setting all
overlap matrices in Eqs.~(\ref{Dspur}) and~(\ref{Uspur}) to be proportional to the identity matrix.
The above equations are valid for generic textures of the Yukawa couplings and patterns
of the bulk profiles $c_{Q_i,u_i,d_i}$, rendering the spurion-overlap formulae in
 Eqs.~(\ref{Dspur}) and~(\ref{Uspur})  directly applicable to generic flavor scenarios.

\noindent In the RS-A${}_4$ framework,  a simplification comes from the degeneracy of
left-handed bulk mass parameters, thus $F_Q=diag(f^{-1}_{Q_i})=f^{-1}_Q\cdot\mathbbm{1}$.
 For the same reason the overlap
correction matrices of Eqs.~(\ref{Dspur}) and (\ref{Uspur}) simplify
\begin{eqnarray}
     \hat{r}_{00,10,01}^{u,d}={\mbox
diag}({\large r}_{00,10,01}(c_q^L,c_{u_i,d_i},\beta)) &
\hat{r}_{11}^{u,d}={\mbox
diag}(r_{11}(c_{u_i,d_i},c_q^L,\beta))\nonumber\\
\hat{r}_{01^{-+}}^{u,d}={\mbox
diag}(r_{01^{-+}}(c_q^L,c_{d_j,u_j},\beta)) &
\hat{r}_{1^{-+}1}^{u,d}={\mbox
diag}(r_{1^{-+}1}(c_{d_i,u_i},c_q^L,\beta))\label{overlapMatrix}.
\end{eqnarray}
The resulting structure of the down-type
contributions in the mass basis follows straightforwardly
\begin{eqnarray} (C_{7\gamma(8g)}^{d-type})_{ij}&=&
\frac{m_{d_i}A^{1L}f_Q^2}{v^2M_{KK}} \left [V_R^{d\dagger}{\mbox
diag}(f^2_{d,s,b})(\hat{r}_{00}^d)^{-1}
\tilde{r}_{01}^d\,\tilde{r}_{11}^d(\hat{r}_{00}^d)^{-1}
V_R^d\,{\mbox diag} (m^2_{d,s,b})\right.\nonumber\\&&\left. \times
V_R^{d\dagger}\,(\hat{r}_{00}^d)^{-1}\hat{r}_{10}^d\,
 V_R^{d\phantom{\dagger}}\right]_{ij}
 ,\label{C7Doverlap}
\end{eqnarray}
where
$\tilde{r}_{01}^{u,d}\tilde{r}_{11}^{u,d}=\hat{r}_{01}^{u,d}\hat{r}_{11}^{u,d}+
\hat{r}_{01^{-+}}^{u,d}\hat{r}_{1^{-+}1}^{u,d}$ and all overlap matrices are diagonal.
 Similarly, we obtain the  up-type contributions to dipole
operators in the mass basis
\begin{eqnarray}
(C_{7\gamma(8g)}^{u-type})_{ij}&=&\frac{A^{1L}f_Q^2}{v^2M_{KK}}\left
[V_{CKM}^\dagger\, {\mbox diag}(m_{u,c,t}) V_R^{u\dagger}{\mbox
diag}(f^2_{u,c,t})(\hat{r}_{00}^u)^{-1}\,\tilde{r}_{01}^{u}\tilde{r}_{11}^u
(\hat{r}_{00}^u)^{-1}\right.\nonumber\\&&\left.\times
V_R^u\,{\mbox diag} (m_{u,c,t})V_{CKM}\, {\mbox diag}(m_{d,s,b})
V_R^{d\dagger}(\hat{r}_{00}^d)^{-1}\hat{r}_{10}^d\,V_R^d \right
]_{ij} \, .\label{C7Uoverlap}
\end{eqnarray}
Since all overlap corrections are real and enter through diagonal
matrices, the resulting modifications to the IR localized Higgs
case are limited, in particular their effect on the imaginary
parts relevant for CP violating processes. Qualitatively, this
result can be understood from the fact that the new (real valued)
overlap correction matrices appear always together with the
diagonal $f$'s, with patterns
$V_{L,R}^{u,d\dagger}r_1f_1r_2f_2r_3V_{L,R}^{u,d}$. Given the
structure of $V_{L,R}^{u,d}$ (see Eqs.~(\ref{VLQ}) and
(\ref{VRQ})), it can be shown that the presence of the $r$'s will
have no effect on the cancellation of imaginary parts of diagonal
dipole operators to
$\mathcal{O}(f_\chi^{u_i,d_i}\tilde{x}_i^{u,d},f_\chi^{u_i,d_i}\tilde{y}_i^{u,d})$.
At the following order, $\mathcal{O}((f_\chi^{u_i,d_i})^2)$,  the cancellation pattern
of imaginary parts of
the diagonal elements of $C_7^{u,d}$ in the IR localized Higgs
case will be modified by terms that are suppressed by linear or
quadratic quark mass ratios, coming from $V_R^{u,d}$ and
proportional to differences between  overlap correction factors.
We will provide an explicit example in the case of the neutron EDM.

\noindent As shown in appendix \ref{app:A1},  the generational
flavor dependence of the overlap factors is anyway very small
and the largest difference of $\mathcal{O}(5\%)$ is associated
with  $t_R$ and its $SU(2)_R$ partner $\tilde{b}$. In addition, as
will be shown explicitly below, most of the dominant contributions
will be proportional to the first generation ``inverted" zero mode
profiles $f_{u,d}^2\propto (\hat{\chi}_0^{u,d})^{-2} $, due to the
large hierarchy of quark masses.
In general, the modifications induced by the slight
generational dependence of overlap effects are thus expected  to be
less significant (numerically) than the ones arising from second
order corrections, $\mathcal{O}((f_\chi^{u_i,d_i})^2)$, to the NLO
Yukawa matrices, with $f_\chi\approx0.05$. For this reason, and barring zeros of the amplitudes, one should
expect to obtain a fairly conservative estimate of the
contributions to the neutron EDM, $\epsilon'/\epsilon$ and
$b\rightarrow s\gamma$, by parametrizing the effect of the
overlap corrections by an overall multiplicative factor for the
up- and down-type contributions to the dipole operators. Defining
the overall factor $ B_P^{u,d}$ as the maximum for each element of
the overlap correction matrices 
\begin{equation}B_P^{u,d}={\mbox
max}\left((\hat{r}_{00}^{u,d})^{-3}(\hat{r}_{01}^{u,d}\hat{r}_{11}^{u,d}+\hat{r}_{01^{-+}}^{u,d}
\hat{r}_{1^{-+}1}^{u,d})\,\hat{r}_{10}^{u,d}\right),\label{BPdef}\end{equation}
the down-type contributions reduce to
\begin{eqnarray}(C_{7\gamma(8g)}^{d-type})_{ij}
&=&\frac{A^{1L}m_{d_i}m_{d_j}B_P^d}{v^2M_{KK}}\left [
V_R^{d\dagger}{\mbox diag}(f^{2}_{d_1,d_2,d_3}) V_R^d{\mbox diag}
(m_{d,s,b})
 V_L^{d\dagger}{\mbox diag}(f^{2}_{Q_1,Q_2,Q_3}) V_L^d \right ]_{ij}\nonumber
\\ &=& \frac{A^{1L}f_Q^2
m_{d_i}m^2_{d_j}B_P^d}{v^2M_{KK}}\sum_{n=1}^{3}(V_R^d)^*_{ni}(V_R^d)_{nj}f_{d_n}^2\,
,\label{C7D}
\end{eqnarray}
while the contributions associated with internal up-type KK quarks
have a slightly more complicated structure
\begin{eqnarray}
(C_{7\gamma(8g)}^{u-type})_{ij}&=&\frac{A^{1L}m_{d_j}B_P^u}{v^2M_{KK}}\left
[V_{CKM}^\dagger {\mbox diag}(m_{u,c,t}) V_R^{u\dagger}{\mbox
diag}(f^{2}_{u_i}) V_R^u{\mbox diag} (m_{u,c,t})
V_L^{u\dagger}{\mbox diag}(f^{2}_{Q_i}) V_L^d \right ]_{ij}\nonumber\\
&=&\frac{A^{1L}f_Q^2 m_{d_j}B_P^u}{v^2M_{KK}}  \left
[V_{CKM}^\dagger {\mbox diag}(m_{u,c,t}) V_R^{u\dagger}{\mbox
diag}(f^{2}_{u_i}) V_R^u {\mbox diag}(m_{u,c,t})V_{CKM}\right
]_{ij} \, .\label{C7U}
\end{eqnarray}
It is evident from Eqs.~(\ref{C7D}) and (\ref{C7U}) that, if we
restrict ourselves to the LO Yukawa interactions in Eq.~(\ref{LOYukawa}),
we  have $V_L^{u,d}=U(\omega)$ and $V_R^{u,d}=\mathbbm{1}$.
Hence, both up- and down-type NP contributions to $C_{7ij}$ reduce to real diagonal matrices and  generate no  corrections to
the processes we are interested in, as already anticipated in \cite{A4Warped}.
 This situation typically changes when we also consider  the NLO Yukawa interactions
in Eq.~(\ref{NLOYukawa}) and the corresponding diagonalization
matrices in Eqs.~(\ref{VLQ}) and~(\ref{VRQ}). As we said, small additional corrections can also be induced at leading order by the slight generational non degeneracy of overlap factors. In principle, both sources have to be taken into account when estimating deviations from zero of  the NP contributions. In practice, the latter source is typically suppressed by roughly an order of magnitude in comparison with the corrections generatd by NLO Yukawa interactions. This can also be inferred from Eqs.(\ref{C7Doverlap}) and
(\ref{C7Uoverlap}), where the terms generated by LO Yukawa
interactions in RS-A${}_4$ carry through a systematic
cancellation pattern of the form $1+\omega+\omega^2$ between
nearly degenerate quadratic functions of the overlap correction
factors, which originates from $U(\omega)$.

\noindent In the flavor anarchic case, a direct diagonalization of the one generation
KK mass matrix, augmented with generational mixing factors
derived in the mass insertion approximation, yields reliable
predictions due to the lack of structure of the Yukawa couplings
and the bulk mass parameters, as shown in \cite{Azatov}.
On the other hand, when considering  flavor symmetries we particularly care  for the
three-generation structure. To go beyond the
mass insertion approximation and the one-generation case requires diagonalizing a
$9\times9$ mass matrix in the non custodial setup and a
$12\times12$ mass matrix in the custodial one, leaving limited space for a fully analytical
 description. For this reason the spurion-overlap analysis remains an appealing tool for understanding the cancellation mechanisms induced by a particular flavor pattern, as is the case when a discrete flavor symmetry such as A${}_4$ is imposed.
\section{Beyond the mass insertion approximation}\label{Sec:MassDiag}
To go beyond the mass insertion approximation  and account
for the complete generational mixing requires the direct diagonalization of the full
 $12\times 12$ KK mass matrix in the custodial case. The mass matrix can be perturbatively
 diagonalized to
first order in the parameter $x=v Y/M_{KK}$, which
measures the relative strength of Yukawa interactions with the
Higgs compared to the masses of the first level KK modes.
A lower level of approximation is obtained by disregarding the mixing among generations
and work with one-generation mass matrices.
This was done in \cite{Azatov} and \cite{IsidoriPLB} for the flavor anarchic non-custodial case.
Already at this level, the diagonalization of the one-generation mass matrix enables one to
account for the
contribution of the $(-\,-)$ KK modes to the dipole operators, not realized in
the spurion-overlap analysis within the mass insertion approximation.
In addition, it was numerically verified \cite{Azatov} that  within the flavor anarchic
non-custodial framework
of \cite{Agashe:2004cp} the difference between the results in the one-generation  and the
three-generation case is rather mild.
This is expected, and stems from the fact that all
Yukawa couplings are $\mathcal{O}(1)$ and no pattern is
present in the phases of these couplings. Consequently, the
structure of each diagonal and off-diagonal block in the full
$9\times9$ mass matrix is identical up to the profiles $f_{q,u,d}$'s and
the slight variation of the overlap corrections over the three generations.
The texture of Yukawa couplings and bulk profiles in RS-A${}_4$  induces different
 patterns of the results and gives more significance to the comparison between the
one-generation and three-generation analysis.

\subsection{Direct diagonalization of the one-generation mass matrix}
\label{Subsec:KKdiag1}
The study described in this section  is also instructive for flavor
anarchic models with custodial symmetry, which contain a separate
$SU(2)_R$ doublet for each 5D fermion with a RH zero mode. This case was not considered
in \cite{ Agashe:2004cp, Azatov,IsidoriPLB}.
The LO  mass matrix for the first generation in the down-type sector, including the zero
modes and first level KK modes, is of the form
\begin{equation}\frac{\hat{\mathbf{M}}_d^{KK}}{(M_{KK})}=
\left(\begin{array}{c}
\bar{Q}_L^{d(0)}\\\bar{d}_L^{(1^{--})} \\ \bar{Q}^{d(1)}_L\\
\bar{\tilde{d}}_L^{(1^{+-})} \end{array} \right)^T \left(
\begin{array}{cccc}
\breve{y}_df^{-1}_Q f^{-1}_d r_{00} x & 0 & \breve{y}_df^{-1}_Q r_{01} x &
\breve{y}_uf^{-1}_Q r_{101} x \\
 0 &  \breve{y}_d^*r_{22} x & 1 & 0 \\
 \breve{y}_df^{-1}_d r_{10} x & 1 &  \breve{y}_dr_{11} x &  \breve{y}_ur_{111} x \\
 0 &  \breve{y}_u^*r_{222} x & 0 & 1
\end{array}
\right)\left(\begin{array}{c}
d_R^{(0)}\\Q_R^{d(1^{--})} \\ d^{(1)}_R\\
\tilde{d}_R^{(1^{-+})} \end{array} \right),\label{M4KK}
\end{equation}
where we factorized a common KK mass scale $M_{KK}$,
$\breve{y}_{u,d}\equiv (\hat{Y}_{LO}^{u,d})_{11}=
2y_{u,d}v_\Phi^{4D}e^{k\pi R}/k$ and the perturbative expansion
parameter is defined as $x\equiv v/M_{KK}$. In
the above equation $r_{111}\equiv r_{11^{-+}}$, $r_{101}\equiv r_{01^{-+}}$,
$r_{22}\equiv r_{1^-1^-}$, $r_{222}\equiv r_{1^{-}1^{+-}}$ and the notation
for the rest of the overlaps is the same as in
Eq.~(\ref{lagrangian}).
The $c$ dependence of
the overlap corrections was suppressed to ease the notation and
can be inferred from the labelling of the rows and columns. Since
the overlaps vary little among generations, the structure of
the mass matrix will be almost identical for all three generations
of the up and down sectors, up to the zero mode profiles denoted
by the $f_{q,u,d}$ and the Yukawa couplings, $\breve{y}_{u,d}$.  Notice that the NLO Yukawa interactions
are suppressed by $f_\chi^{u_i,d_i}$ compared to the LO
contributions, rendering them to be approximately
$\mathcal{O}(x^2)$ numerically and thus in principle safe to
neglect when working to $\mathcal{O}(x)$. In order to include NLO
Yukawa interactions in the above matrix one should simply replace
$y_{u,d}\rightarrow
y_{u,d}+f_\chi^{u,d}(\tilde{x}_1^{u,d}+\tilde{y}_1^{u,d})$ in
$\breve{y}_{u,d}$, following Eqs.~(\ref{LOYukawa}) and~(\ref{NLOYukawa}), and analogously for the other matrices.
Despite their relative smallness, it is still
important to study the generational modifications associated with
NLO Yukawa interactions, which are essential for matching the quark mixing data in the ZMA.

\noindent Notice that the anarchic case is simply obtained from Eq.~(\ref{M4KK})
by setting $\breve{y}_{u,d}=Y$ for all generations, where $Y$ is a $\mathcal{O}(1)$
Yukawa coupling which can be absorbed in $x$.
In RS-A$_4$, when considering the mass matrices for the
second and third generation, we encounter additional
$\omega$ factors coming from the LO Yukawa matrix of
Eq.~(\ref{LOYukawa}).
In addition, the approximation of degenerate KK masses will turn out to be fair only
for two out of the three KK masses in Eq.~(\ref{M4KK}), given the bulk mass
assignments of the RS-A$_4$ setup.
 In appendix~\ref{app:B1} we perform the
diagonalization of each of the one-generation mass matrices for
the up and down sectors to first order in $x$, before proceeding
to the approximate analytical diagonalization of the full $12\times12$ up and down
mass matrices in appendix~\ref{app:B3}.

\noindent The $4\times4$ one-generation diagonalization matrices,
$O_{L}^{(u,c,t,d,s,b)_{KK}}$ and $O_{R}^{(u,c,t,d,s,b)_{KK}}$, are
defined as follows
\begin{equation}(O_L^{(u_i,d_i)_{KK}})^\dagger\,\hat{\bf{M}}_{u_i,d_i}^{KK}
\,(O_R^{(u_i,d_i)_{KK}})
=\hat{\bf{M}}_{u_i,d_i}^{KK_{diag}}\, . \label{OKKdef}\end{equation}
Once the above diagonalization matrices are obtained, the ground is
set for the estimation of physical couplings between light and
heavy modes in the flavor anarchic custodial case. This is done by simply
transforming the charged and neutral Higgs Yukawa interaction
matrices to the mass basis using $O_{L,R}^{(u_i,d_i)_{KK}}$,
while the generational mixing factors can be
estimated in the mass insertion approximation, as also done in
\cite{Azatov,IsidoriPLB}.

\noindent  In the RS-A$_4$ setup
we can extract  the overlap dependence of the coupling between the
zero mode and the three first level KK modes of each generation in the same way.
Then, to account for generational mixing and the underlying flavor pattern, this
information is combined with the spurion-overlap analysis in the mass insertion approximation
of Eqs.~(\ref{C7D}) and~(\ref{C7U}), where it provides a
redefinition of the overall overlap factors $B_P^{u,d}$. As already noticed, the
latter approximation of reducing the overlap structure to a common overall factor,
is justified since
the almost degenerate KK modes mix almost maximally and are hence
well approximated by one representative for each type of BC. The
new down-type  $B_P^{d}$ factors for the dipole operators, corresponding to the
process in Fig.~\ref{fcncLoop} with a neutral Higgs, will thus be extracted from
\begin{equation}
(\mathcal{A}_{ij})^{overlap}_D=\left. \frac{ \sum_n( (O_L^{(d_i)_{KK}})^\dagger\hat{Y}^{d_i}_{KK}
 O_R^{(d_i)_{KK}})_{1n}
( (O_L^{(d_i)_{KK}})^\dagger\hat{Y}^{d_j}_{KK} O_R^{(d_i)_{KK}}  )_{n1}   }
{\left (  M_{KK}^{d_i\, (n)}/M_{KK}   \right )}   \right|_{overlap}\, ,\label{KK1genAppD}
\end{equation}
where $|_{overlap}$ denotes taking the overlap part of the
corresponding expression by assigning all Yukawas to one. The indices $i,j$ denote the flavor
of the external SM physical zero modes, while $n$ runs over the three KK states for the given
generation. The components (1n), (n1) of the Yukawa matrices rotated to the mass basis indicate
the coupling between a zero mode and the n-th KK mode of the same generation.
The new $B_P^u$ factors, corresponding to the amplitudes in Fig.~\ref{fcncLoop} and
Fig.~\ref{fcncLoopcharged} with a
charged Higgs, will analogously be extracted from
\begin{equation}
(\mathcal{A}_{ij})^{overlap}_U=\left. \frac{\sum_n(  (O_L^{(d_i)_{KK}})^\dagger
 \hat{Y}^{(h_-)d_i}_{KK}   O_R^{(u_i)_{KK}})_{1n}
(  (O_L^{(u_i)_{KK}})^\dagger  \hat{Y}^{(h_+)d_j}_{KK}  O_R^{(d_i)_{KK}})_{n1} }
{\left (  M_{KK}^{u_i\, (n)}/M_{KK}   \right )}
\right|_{overlap}\, .
\label{KK1genAppU}
\end{equation}
Notice that the one-loop factor $A^{1L}$ in Eqs.~(\ref{C7D}) and~(\ref{C7U}) is
calculated at the reference
KK mass $M_{KK}\simeq 2.55\, (R')^{-1}$, while the non-degeneracy of KK states is taken into
account by the rescaling  $(M_{KK}^{d_i,u_i\,(n)}/M_{KK})$.
It is useful to mention the explicit structure of the
down-type Yukawa couplings with a neutral Higgs in the interaction basis
\begin{equation}\hat{Y}^d_{KK}=
\left(\begin{array}{c}
\bar{Q}_L^{d(0)}\\\bar{d}_L^{(1^{--})} \\ \bar{Q}^{d(1)}_L\\
\bar{\tilde{d}}_L^{(1^{+-})} \end{array} \right)^T \left(
\begin{array}{cccc}
\breve{y}_df^{-1}_Q f^{-1}_d r_{00}  & 0 & \breve{y}_df^{-1}_Q r_{01}&\breve{y}_uf^{-1}_Q r_{101}  \\
 0 & \breve{y}_d^* r_{22}  & 0 & 0 \\
 \breve{y}_df^{-1}_d r_{10}  & 0 & \breve{y}_d r_{11}  & \breve{y}_u r_{111}  \\
 0 &\breve{y}_u^* r_{222}  & 0 & 0
\end{array}
\right)\left(\begin{array}{c}
d_R^{(0)}\\Q_R^{d(1^{--})} \\ d^{(1)}_R\\
\tilde{d}_R^{(1^{-+})} \end{array} \right).\label{Y4KK}
\end{equation}
Similarly, the Yukawa interactions with the charged Higgs $h^{-}$  are
\begin{equation}\hat{Y}^{(h_-)d}_{KK}=\left(\begin{array}{c}
\bar{Q}_L^{d(0)}\\\bar{d}_L^{(1^{--})} \\ \bar{Q}^{d(1)}_L\\
\bar{\tilde{d}}_L^{(1^{+-})} \end{array} \right)^T \left(
\begin{array}{cccc}
 -\breve{y}_uf^{-1}_Q f^{-1}_u r_{00}  & 0 & - \breve{y}_uf^{-1}_Q r_{01} & -\breve{y}_df^{-1}_Q
 r_{101}
   \\
 0 &  \breve{y}_d^*r_{22}  & 0 & 0 \\
 -\breve{y}_uf^{-1}_u r_{10}  & 0 & - \breve{y}_u r_{11} & - \breve{y}_dr_{111}  \\
 0 &  \breve{y}_u^*r_{222} & 0 & 0
\end{array}\right) \left(\begin{array}{c}
u_R^{(0)}\\Q_R^{u(1^{--})} \\ u^{(1)}_R\\
\tilde{u}_R^{(1^{-+})} \end{array}
\right)\, ,\label{HUDmatrix}\end{equation}
and for $h^{+}$ they are given by the replacement $ \breve{y}_{u,d}\to -\breve{y}_{d,u}$
and $f_{u,d}\to f_{d,u} $
\begin{equation}\hat{Y}^{(h_+)d}_{KK}=\left(\begin{array}{c}
\bar{Q}_L^{u(0)}\\\bar{u}_L^{(1^{--})} \\ \bar{Q}^{u(1)}_L\\
\bar{\tilde{u}}_L^{(1^{+-})} \end{array} \right)^T \left(
\begin{array}{cccc}
 \breve{y}_df^{-1}_Q f^{-1}_d r_{00}  & 0 & \breve{y}_df^{-1}_Q r_{01} & \breve{y}_uf^{-1}_Q
 r_{101}
   \\
 0 & - \breve{y}_u^* r_{22} & 0 & 0 \\
 \breve{y}_d f^{-1}_d r_{10} & 0 &  \breve{y}_dr_{11}  &  \breve{y}_u r_{111} \\
 0 & -\breve{y}_d^*r_{222}  & 0 & 0
\end{array}\right) \left(\begin{array}{c}
d_R^{(0)}\\Q_R^{d(1^{--})} \\ d^{(1)}_R\\
\tilde{d}_R^{(1^{-+})} \end{array}
\right)\, .\label{HDUmatrix}\end{equation}
The physical Yukawa couplings between zero modes and
KK modes are then obtained by  the $O_{L,R}^{u,d}$ rotations. Once inserted in
Eqs.~(\ref{KK1genAppD}) and (\ref{KK1genAppU}), they provide the new
overlap factors $B_{P}^{u,d}$ to be inserted in the spurion-overlap formulae Eqs.~(\ref{C7D})
and~(\ref{C7U}), for each  dipole operator.
The results of this analysis -- diagonalization of the one-generation mass matrices
combined with the spurion-overlap procedure -- will be considered separately for each
process, while the details of the derivation and the $B_{P}^{u,d}$ factors can be found
in appendices \ref{app:B1} and \ref{app:B2}.

\noindent It is however important to recall that  the complete A${}_4$ flavor structure in the
full $12\times12$ up and down mass matrices may still induce deviations from the
approximations considered till now.  Differences may arise from inter- and intra-generational
mixing, non-degeneracy of KK states and overlaps. All these effects are numerically more
significant when involving the third generation. On the other hand, the drawback of a fully
numerical treatment of the 12x12 mass matrix is that it does not allow to easily discriminate
 among different orders in the $x$-parameter expansion, and it fails to provide insightful
information on  the flavor patterns and cancellation mechanisms of the numerical results.
Eventually, such a numerical treatment will turn out to induce more
sizable contributions to the neutron EDM and less significantly so for
 other processes.  This situation illustrates the
importance of a full three-generation diagonalization and its comparison with
approximate analytical estimates when a
flavor texture is present in the Yukawa matrices.

\subsection{Approximate analytical diagonalization of the $12\times 12$ mass matrix}\label{Subsec:KKFulldiag}
It is clear that a complete description of the contributions to physical processes in the
three-generation RS-A${}_4$ can only be achieved by a direct diagonalization
of the full $12\times12$ up and down
mass matrices, including  first level, $n=1$, KK modes. Using the
$12\times 12$  rotation matrices we can obtain all the couplings
between each zero mode and KK modes of all generations, thus
establishing an a priori more reliable way to describe the flavor patterns of A${}_4$.
However, the size of the matrices, the large number of
parameters even in the minimal case and the near degeneracy
of most of the KK masses, render the diagonalization hard to
perform analytically. For this reason the three-generation case was
considered only numerically in \cite{Azatov}, for flavor anarchic models. A fully numerical
diagonalization of the $12\times 12$ mass matrices in  RS-A${}_4$
may provide an estimate of contributions possibly missed by other approximations, but
fails to give us insight on the flavor pattern of the three-generation A${}_4$ case.
 In addition, since the one-generation 4D mass matrices have
been themselves derived and diagonalized linearly in the A${}_4$
parameters $\tilde{x}_i^{u,d},\, \tilde{y}_i^{u,d}$, the most
appropriate diagonalization should always be performed to the same
order. Instead, a numerical treatment will inevitably include
higher order contributions in an a priori uncontrolled way.

\noindent Given all the above reasons, one should still attempt an
approximate analytical diagonalization, as described below. We
first decompose  the  $12\times12$ mass matrix of the RS-A${}_4$
down sector in terms of the one-generation matrices in
Eq.~(\ref{M4KK})
\begin{equation}
\hat{\bf{M}}^{D}_{Full}=M_{KK}\left(\begin{array}{ccc}
\hat{\bf{M}}_d^{KK}/M_{KK} &
x\hat{Y}^{s}_{KK}(\hat{y}^{LO}_{12},f_s)
&x\hat{Y}^{b}_{KK}(\hat{y}^{LO}_{13},f_b)\\
x\hat{Y}^{d}_{KK}(\hat{y}^{LO}_{21},f_d) &
\hat{\bf{M}}_s^{KK}/M_{KK} &
x\hat{Y}^{b}_{KK}(\hat{y}^{LO}_{23},f_b)\\x\hat{Y}^{d}_{KK}(\hat{y}^{LO}_{31},f_d)
&x\hat{Y}^{s}_{KK}(\hat{y}^{LO}_{32},f_s)&\hat{\bf{M}}_b^{KK}/M_{KK}
\end{array}\right)\,,\label{MFullD}\end{equation}
where the expression in brackets of each off-diagonal element
denotes the replacements to be made in
Eq.~(\ref{Y4KK}). The $s$ and $b$ one-generation mass matrices $\hat{\bf{M}}_{s,b}^{KK}$
are obtained by obvious replacements in Eq.~(\ref{M4KK}).
 To account for NLO Yukawa interactions, the replacement $\hat{y}^{LO}_{ij}\rightarrow
\hat{y}^{LO}_{ij}+\hat{y}^{NLO}_{ij}$ applies. In the above equation, $M_{KK}$ is the KK mass
corresponding to the degenerate left-handed bulk mass parameter,
$c_q^L$, and we normalize all matrices accordingly while keeping non-degenerate KK modes. The only
significant deviations from degeneracy lie in the third generation mass matrix
due to $\tilde{b}$, the $SU(2)_R$ partner of $t_R$ .

\noindent Using the
$4\times 4$ diagonalization matrices for each generation, we construct the matrices
$\hat{\bf{O}}^{D_{KK}}_{L,R}={\mbox
diag}(O_{L,R}^{d_{KK}},O_{L,R}^{s_{KK}},O_{L,R}^{b_{KK}})$ to
first diagonalize the diagonal entries of $\hat{\bf{M}}^{D}_{
Full}$.
The main difficulty in achieving the diagonalization of the full mass matrix in
Eq.~(\ref{MFullD}) is the near degeneracy of 6 out of 9 KK masses
which also survives the $(\hat{\bf{O}}^{D_{KK}}_{L})^\dagger
\hat{\bf{M}}^{D}_{Full}\hat{\bf{O}}^{D_{KK}}_{R}$ rotation,
rendering non degenerate perturbation theory useless in the
corresponding subspace. Therefore, the nearly
degenerate subspace is first diagonalized non perturbatively to
find a new basis, in which  non-degenerate perturbation theory can
be used. Off-diagonal elements in the non-degenerate subspaces can
obviously be treated in the conventional way. Since this task is
hard to perform analytically when all parameters are
unassigned, we look for some symmetry property of the A${}_4$
 structure in $\hat{\bf{M}}_{Full}^{D}$ that might
supplement us with a shortcut.

\noindent
Given the structure of Eq.~(\ref{MFullD}),  we then construct new
rotation matrices using $V_{L,R}^{u,d}$ from Eq.~(\ref{VRQ}). The
new A${}_4$ rotation matrices are thus defined as the direct
product
\begin{equation}
\hat{\bf{O}}^{(U,D)_{A_4}}_{L,R} = V_{L,R}^{u,d}\otimes \tilde{\mathbbm{1}}_{4\times 4}\, ,
\end{equation}
where $\tilde{\mathbbm{1}}_{4\times 4}={\mbox diag}(1,1()^*,1,1()^*)$ and $()^*$ denotes complex conjugation of
the coefficient that multiplies the corresponding element, namely
$(V_{L,R}^{u,d})_{ij}$. In appendix~\ref{app:B3} we show that using
$\hat{\bf{O}}^{(U,D)_{A_4}}_{L,R}$ and
$\hat{\bf{O}}^{D_{KK}}_{L,R}$ to rotate $\hat{\bf{M}}_{Full}^D$,
one obtains an approximately diagonalized degenerate subspace,
which in turn enables to generate the remnant rotation by acting
with non-degenerate perturbation theory on
$(\hat{\bf{O}}^{D_{A_4}}_L)^\dagger(\hat{\bf{O}}^{D_{KK}}_{L})^\dagger\hat{\bf{M}}_{Full}^D
\hat{\bf{O}}^{D_{KK}}_{R}\hat{\bf{O}}^{D_{A_4}}_{R}$. The
analogous procedure is followed in the up sector. More details are
collected in  appendix~\ref{app:B3}. Once the diagonalization
matrices are obtained, the contribution to the Wilson coefficient
of a given dipole operator will be a generalization of the
one-generation case and can generically be written as follows
\begin{equation}
\left (C_{7\gamma (8g)}^{d-type}\right )_{ij} = A^{1L}(M_{KK}) \frac{\sum_n(
(O_L^{(d_i)})^\dagger\hat{Y}^{d_i}_{KK}  O_R^{(d_i)})_{(4i-3)\,n}
( (O_L^{(d_i)})^\dagger\hat{Y}^{d_j}_{KK} O_R^{(d_i)}  )_{n\,(4j-3)}   }
{ (M_{KK}^{(n)}/M_{KK}) }
\label{KK3gdown},\end{equation}
for down-type contributions, and
\begin{equation}
\left (C_{7\gamma (8g)}^{u-type}\right )_{ij} = \tilde{A}^{1L}_u(M_{KK})
 \frac{\sum_n( (O_L^{(d_i)})^\dagger\hat{Y}^{d_i}_{KK}  O_R^{(d_i)})_{(4i-3)\,n}
((O_L^{(d_i)})^\dagger\hat{Y}^{d_j}_{KK} O_R^{(d_i)}  )_{n\,(4j-3)}   }
{ (M_{KK}^{(n)}/M_{KK}) }
\label{KK3gup},\end{equation}
for up-type contributions. The matrices $O_{L,R}^{(u_i,d_i)}$
diagonalize the $12\times 12$ mass matrices. The index $n$ runs
over the KK modes of the three generations, thus $n\neq 1,5,9$,
and the indices $(4i-3)\,n$ and $n\,(4j-3)$ select the couplings
of the external zero mode to the internal KK states. The one-loop
factors are calculated at the reference KK mass $M_{KK}\simeq 2.55\,(R')^{-1}$,
 while the non-degeneracy of KK states is taken into
account by the rescaling  $(M_{KK}^{(n)}/M_{KK})$. However, as
expected, the resulting expressions for the physical couplings
between zero modes and KK modes are long functions of all overlap
correction factors and are therefore not stated explicitly.
Instead, we explore the couplings and the resulting predictions
for assigned values of the parameters, and compare them with the
results of a fully numerical diagonalization for varying values of
the KK scale, $M_{KK}$.  This will be done separately for each process
in section \ref{Sec:NumericalResults}.

\section{Numerical Results and Experimental Bounds for Dipole
Operators}\label{Sec:NumericalResults} In this section we analyze
FCNC processes in the RS-A${}_4$ model, using the approximations
described in sections \ref{sec:Dipole} and \ref{Sec:MassDiag}, and
compare them with a fully numerical analysis based on the
diagonalization of the $12\times 12$ mass matrices for the zero
modes and first level KK modes. We focus on those processes
mediated by dipole operators and known to provide the most
stringent constraints on new physics contributions, and thus the
KK scale, in the context of flavor models in warped geometry:
these are the neutron EDM in section \ref{sec:EDM}, $\epsilon'/
\epsilon_K$ in section \ref{subsec:Epsilon} and the radiative
decay $b\to s\gamma$ in section \ref{subsec:BSGamma}. Finally,
tree level Higgs mediated FCNC contributions are considered in
section \ref{sec:higgs}.

\subsection{New physics contributions to the neutron EDM}
\label{sec:EDM}
 The new physics contributions to the
neutron EDM are mediated by the dipole operator
$e\bar{d}_L\sigma^{\mu\nu}F_{\mu\nu}d_R$. In particular, we need
to compute  the imaginary part of the $i=j=1$ component of the
Wilson coefficients defined in Eqs. (\ref{C7D}) and (\ref{C7U}).
They are of dimension $[{\mbox mass}]^{-1}$ and can thus be
directly compared to the experimental bound $|d_n|<3\times
10^{-26}e\cdot cm$ \cite{EDMBound}. We have already anticipated in
\cite{A4Warped} that the down-type contribution to the EDM is
vanishing in RS-A${}_4$, due to the fact that $V_L^d$ disappears
from the down-type contributions. This conclusion can also be
reached by inspecting  Eq.~(\ref{C7Doverlap}). This leaves us with
the up-type contributions encoded in ${\mbox
Im}[(C_7^{u-type})_{11}]$. In order to isolate the dominant terms
in Eq.~(\ref{C7U}), where we approximate the overlap factors with
an overall constant, we recall that the hierarchy of quark masses
is translated into the  inverse hierarchy of the right-handed
profiles $f_{q_i}$, since there is only one left-handed bulk mass
parameter. More specifically, $f_{u}\simeq 4.48\times10^4$,
$f_{d}\simeq 2.25\times10^4$, $f_{s}\simeq 1.36\times10^3$,
$f_{c}\simeq 1.22\times10^2$, $f_{b}\simeq 28.8$, $f_{t}\simeq
1.22$ and $f_{Q}\simeq 3.14$. In addition, we recall that the
off-diagonal elements of $V_R^u$ in
Eqs.~(\ref{DeltaR1})-(\ref{DeltaR3}) are suppressed by up-type
quark mass ratios.  Hence, the most dominant contribution to the
neutron EDM from Eq.~(\ref{C7U}) turns out to be 
\begin{equation}
{\mbox Im}[(C_7^{u-type})_{11}]\simeq{\mbox
Im}\left[\frac{5B_P^uA^{1L}m_df_Q^2F_{EDM}^{u}
}{3v^2M_{KK}}\right], \label{EDM}\end{equation}
where the factor $5/3$ comes from the electric charge of an up-type quark
and a charged Higgs from Eq.~(\ref{A1Ladd}).  The factor $F_{EDM}^u$
is obtained from Eq.~(\ref{C7U}) and given by
\begin{eqnarray}
F_{EDM}^u&=&\left[V_{CKM}^{\dagger}diag(m_{u,c,t})V_R^{u\dagger}diag(f_{u_i}^2)V_R^udiag(m_{u,c,t})V_{CKM}
\right]_{11}\nonumber\\&=&
\sum_{k,l=1}^{3}(V_{CKM}^\dagger)_{1k}(\tilde{R}_u)_{k\ell}(V_{CKM})_{\ell1}\,\,\Rightarrow\,\,
{\mbox Im}(F_{EDM}^u)=0,\label{EDMstruc.}\end{eqnarray}
with
\begin{equation}
(\tilde{R}_u)_{ij}=\left(diag(m_{u,c,t})V_R^{u\dagger}diag(f_{u_i}^2)V_R^udiag(m_{u,c,t})\right)_{ij}=
\sum_{n=1}^{3}m_im_j(V_R^u)^*_{ni}(V_R^u)_{nj}f_{u_n}^2\, .\end{equation}
Given that  the matrix $\tilde{R}_u$ is hermitian and that in RS-A${}_4$ to
$O(\tilde{x}^{u,d}_i,\tilde{y}^{u,d}_i)$ one has $(V_{CKM})_{ii}=1$,
and $(V_{CKM})_{ij}=-(V_{CKM})^*_{ji}$ for $i\neq j$ (see Eqs.~(\ref{VLQ})\,--\,(\ref{Vub})),
 we conclude that $F_{EDM}^u$ has no imaginary part if we disregard the tiny non-degeneracy
of overlap factors by replacing them with the overall coefficient $B_P^u$.
For later convenience we anyway look at what terms are dominant in the cancellation pattern;
 they are proportional to $f_u^2$ or  $f_c^2$
\begin{eqnarray}F_{EDM}^u&=&f_u^2[m_u^2-m_um_cV_{us}(\Delta_1^u)^*-m_um_cV_{us}^*
\Delta_1^u \nonumber\\
&-&(f_c^2/f_u^2)m_c^2V_{us}V_{us}^*-m_um_tV_{ub}^*\Delta_2^u-m_um_tV_{ub}(\Delta_2^u)^*]\, .
\label{FEDM}\end{eqnarray}
The first and fourth terms are real, while the second and third
terms and the fifth and sixth terms cancel each other's imaginary
parts. All other contributions to $F_{EDM}^u$ are suppressed by at
least two orders of magnitude, and exhibit the same cancellation
pattern of imaginary parts. Thus, in order to obtain a
conservative estimate of the contribution to the neutron EDM in
our setup, we must fully account for non degeneracies and possibly
go beyond the mass insertion approximation.

\noindent It is also important to recall that $V_{L,R}^{u,d}$ have
been determined \cite{A4Warped} to linear order in the Yukawa
parameters $\tilde{x}^{u,d}_if_\chi^{u,d},\,\tilde{y}^{u,d}_i
f_\chi^{u,d}$ and that the neglected
$\mathcal{O}((x_i^{u,d}f_\chi^{u,d})^2,
(y_i^{u,d}f_\chi^{u,d})^2)$ corrections can  also a priori modify
the cancellation pattern of imaginary parts for up- and down-type
contributions to the neutron EDM. However, a realistic estimate of
these corrections would require to perform a new matching with the
experimentally determined CKM matrix. We can still provide a
conservative order of magnitude estimate of these effects by
evaluating the size of a generic term in Eq.~(\ref{FEDM}). In
particular, labelling the second (or third) term, proportional to
the largest CKM element $V_{us}$, as $\Delta{F}_{EDM}^{u}$ one
obtains
\begin{eqnarray}
\Delta {F}_{EDM}^{u}=f_u^2m_u m_c
V_{us}(\Delta_1^u)^{*}\Rightarrow d_n&\approx& {\mbox Im}
\left[\frac{5B_P^uA^{1L}m_df_Q^2\Delta F_{EDM}^{u}} {3v^2M_{KK}}
\right]\nonumber\\ &\approx&  4.1\times10^{-27} \left (
\frac{3\,{\mbox TeV} }{M_{KK}} \right )^2Y^2\, e\cdot cm
,\label{EDMparam2nd}\end{eqnarray} 
for  $B_P^u\simeq1.5$ and assuming $|\Delta_1^u|\approx
|V_{us}|m_u/m_c$, with maximal phase for $\Delta {F}_{EDM}^u$. The symbol
$Y$ denotes the overall scale of the 5D Yukawa couplings, defined as
$y_{u,d,c,s,t,b}\to Y\,y_{u,d,c,s,t,b}$
with reference values $y_{u,d,c,s,b}=1$ and $y_t=2.8$.
The predicted contribution is suppressed by one order of
magnitude compared to the experimental bound, however, an
enhancement induced by a coherent sum of many higher order terms
in $\tilde{x}^{u,d}_i,\tilde{y}^{u,d}_i$ cannot be excluded at
this level. Translating the above result into a conservative
constraint on the KK mass scale yields
\begin{equation}
(M_{KK}^{EDM})^{spur.}_{cons.}\gtrsim1.1\,Y \, {\mbox TeV}\, .
\label{EDMspurionbound}
\end{equation}
\noindent When taking into account the non-degeneracy of the
overlap factors in the spurion-overlap approximation of
Eqs.~(\ref{C7Doverlap}) and (\ref{C7Uoverlap}), the contributions
to the EDM are still vanishing to $O(C_\chi f_\chi )$ --
corresponding to $O(\lambda)$ of the Wolfenstein parametrization
of the CKM matrix. Negligible non vanishing contributions appear
in the  up sector at $O(\lambda^2)$. The most dominant non
vanishing contribution in the up sector, together with its exact
generational dependence of the overlap correction factors is as follows
\begin{equation}
F_{EDM}^u=((r_{00}^b)^{-1}r_{10}^b-(r_{00}^d)^{-1}r_{10}^d)(r_{00}^t)^{-2}(r_{01}^tr_{11}^t+
r_{101}^tr_{111}^t)f_t^2m_t^2V_{ub}(\Delta_2^u)^*\, .
\label{FEDMnodeg}\end{equation} 
 It provides the estimate 
\begin{equation}
(d_n)_{RS-A_4}^{{\mbox spurion}}\approx 2\times10^{-29}  \left (
\frac{3\,{\mbox TeV} }{M_{KK} }\right )^2Y^2 \,  e\cdot cm \,
,\nonumber\end{equation} 
and a lower bound on the KK mass scale $M_{KK}\gtrsim 0.07\,Y$ TeV,
when comparing with the experimental result.

\noindent We can improve upon the previous estimate by directly
diagonalizing the mass matrices and work with KK mass eigenstates.
In the estimates below, the effects of generational mixing,
non-degeneracy of overlaps and KK states are described to various
degrees of approximation. We first state the results obtained by
the procedure of section \ref{Subsec:KKdiag1}, where we use the
diagonalization of the one-generation mass matrices combined with
the spurion-overlap analysis to account for generational mixing.
The contribution from $\Delta F^u_{EDM}$ in
Eq.~(\ref{EDMparam2nd}) to the neutron EDM will be obtained by the
replacement $B_P^{u}\rightarrow(B_P^u)_{EDM}^{KK(1gen)}$, see appendix~\ref{app:B2}). Using
$(B_P^u)_{EDM}^{KK(1gen)}\simeq5.2$, we obtain:
\begin{equation}
  (d_n)_{RS-A4}^{1-gen(cons.)}\simeq
1.42*10^{-26}\left(\frac{3TeV}{M_{KK}}\right)^2Y^2\, e\cdot cm \, ,
\label{EDM1gen}\end{equation}
which implies the bound $M_{KK}\gtrsim 2.1 \, Y$ TeV.
 The modification of the contribution in Eq.~(\ref{FEDMnodeg})
 will be obtained by the same replacement and leads to
\begin{equation}
(d_n)_{RS-A_4}^{1-gen}\simeq 8\times10^{-29}  \left (
\frac{3\,{\mbox TeV} }{M_{KK}} \right )^2 Y^2 \, e\cdot cm \,
, \label{EDM1gensp}\end{equation} 
implying a weak lower bound on the KK mass scale $M_{KK} \gtrsim
0.15 \,Y$ TeV, when comparing with the experimental result.

\noindent Secondly, we consider the three-generation case, where
the $12\times 12$ mass matrices can be approximately diagonalized
analytically as described in section \ref{Subsec:KKFulldiag}, or
they can be diagonalized numerically. In both cases we assign the
bulk masses and Yukawa couplings according to
Eq.~(\ref{CKMAssignment}) and the assignments in Appendix
\ref{app:A} and use the corresponding values of the overlap
correction factors obtained in Appendix \ref{app:A1}. In the
 numerical case, we perform a scan in the KK mass scale
$M_{KK}$ in the range $1-10$ TeV and $Y$ in the range $[0.3,5]$.
We have also verified the stability of the results against
modifications of the phases of the NLO Yukawa couplings and  in
particular for the assignments  of Eqs.~(\ref{DeltaBounds}),
(\ref{EpsilonAssignmentWPS}) and~(\ref{BSassignmentU}). Notice
that differences with the previous estimates can be attributed to
the presence of higher order corrections in the perturbative
parameter $x$ (beyond the mass insertion approximation) and to the
partial, and a priori uncontrolled, contamination of higher
order terms in $\tilde{x}^{u,d}_i,\,\tilde{y}^{u,d}_i$. The first
source can be estimated by performing a scan over the values of
$x$ and match the dominant linear behavior in the vicinity of
$x=0.037$. We recall \cite{A4Warped} that the latter value
corresponds to $(R')^{-1} \simeq 1.8$ TeV and $M_{KK}\simeq
2.55\, (R')^{-1} \simeq 4.6$ TeV, the value for which RS-A${}_4$
predicts a NP correction to the $Zb_L\bar{b}_L$ coupling within $67\%$
CL for the bulk parameters in Eq.~(\ref{BulkAssignments}).

\noindent To obtain the
explicit contributions for each of the processes of interest, we use the
$12\times12$ analogues of Eqs.~(\ref{KK1genAppD})
and~(\ref{KK1genAppU}). The fully numerical diagonalization
procedure, and without truncation in the $x$-parameter expansion, leads to the prediction for the neutron EDM
\begin{equation}
{\mbox Im}\left[(C_7^d)_{EDM}^{Num}\right]\simeq
3.1\times10^{-29}e\cdot cm \qquad  {\mbox
Im}\left[(C_7^u)_{EDM}^{Num}\right]\simeq -1.66\times
10^{-28}e\cdot cm \label{EDMNum},\end{equation}
\begin{equation}(d_n)_{RS-A_4}^{Num}\simeq1.65\times10^{-28}  e\cdot
cm\, ,\nonumber\end{equation} 
while a scan in $x$ and matching to the linear behavior leads to
\begin{equation}
{\mbox Im}\left[(C_7^d)_{EDM}^{Num}\right]\simeq
3.3\times10^{-29}e\cdot cm \qquad  {\mbox
Im}\left[(C_7^u)_{EDM}^{Num}\right]\simeq -1.75\times
10^{-28}e\cdot cm\label{EDMNumlin},\end{equation}
\begin{equation}(d_n)_{RS-A_4}^{Num}\simeq 1.7\times10^{-28}  e\cdot
cm\, ,\nonumber\end{equation}
 where the up- and down-type contributions
were summed in quadrature.
Both results saturate the experimental bound for $M_{KK} \approx
0.3$ TeV. For $M_{KK} \simeq 4.6$ TeV the neutron EDM is smaller  than
the experimental bound by two orders of magnitude. What is also
relevant is that the resultant constraint on $M_{KK}$  deviates by
a $O(1)$ factor from the constraint implied by
Eq.~(\ref{FEDMnodeg}). The characteristic strength of the
numerical results is rather stable against modifications of
$\tilde{y}_i^{u,d}$, $\tilde{x}_i^{u,d}$, $f_{Q,u_i,d_i}$ and
$y_{u_i,d_i}$ that still yield physical quark masses and CKM
elements. In addition, when varying the parameters up to $O(3)$ in magnitude away from their CKM values, the variation of the predicted neutron EDM and all other observables stays within a factor two.

\noindent The contributions predicted by the approximate analytical
diagonalization procedure described in
Sec.~\ref{Subsec:KKFulldiag} consist of extremely long expressions, which we do not state
explicitly. Instead, we study them as a function of $x=v/M_{KK}$
only, with all other parameters assigned to yield the physical
quark masses and CKM matrix elements.  Finally, we can  compare the resulting
predictions with the results of the fully numerical analysis. The approximate
analytical diagonalization for the neutron EDM provides
\begin{equation} {\mbox
Im}\left[(C_7^d)_{EDM}^{12\times12th})\right]\simeq-2.5\times10^{-9}A_{1L}
\qquad {\mbox
Im}\left[(C_7^u)_{EDM}^{12\times12th})\right]\simeq-2\times10^{-9}\tilde{A}_{1L}^u
\label{EDM12X12Th}\end{equation}
and
\begin{equation}(d_n)_{RS-A_4}^{12\times12th}\simeq 1.6\times10^{-28}
\left ( \frac{4.3\,{\mbox TeV} }
{M_{KK}} \right )^2Y^2\,  e\cdot cm  \, ,\nonumber\end{equation} 
in good agreement with the  estimate of Eq.~(\ref{FEDMnodeg})
and the numerical results in Eqs.~(\ref{EDMNum}) and (\ref{EDMNumlin}).
The discrepancy between the numerical and the semianalytical approach for
the three-generation case
will turn out to be  larger for other observables. This is due to the fact
that the semianalytical diagonalization described in Sec.~\ref{Subsec:KKFulldiag}
works better within the first generation, while more significant off-diagonal terms
still appear in the second and third generation. In the case of
the EDM, a  cancellation mechanism at leading order is indeed in place.

\subsection{New physics and $(\epsilon'/\epsilon )$}
\label{subsec:Epsilon}
In this section we derive the new physics contributions to
Re$(\epsilon '/\epsilon )$ in RS-A${}_4$, generalizing the flavor
anarchic analysis of \cite{IsidoriPLB} to our setup. We show that
the bound induced on the KK mass scale by this quantity is relaxed
even below the bounds obtained from EWPM, differently from what
happens in the flavor anarchic case. The current experimental
average, measured by KTeV and the NA48 collaborations, is  ${\mbox
Re}(\epsilon'/\epsilon)_{exp}=(1.65\pm 0.26)\times 10^{-3}$
\cite{PDG}. Given the uncertainties still affecting the standard
model prediction ${\mbox Re}(\epsilon'/\epsilon)_{SM}$
\cite{EpsilonUncertainty}, we adopt the most conservative approach
as also done in \cite{IsidoriPLB} and assume $0<{\mbox
Re}(\epsilon'/\epsilon)_{SM}<3.3\times 10^{-3}$.

\noindent The potentially large new physics contributions to
 ${\mbox Re}(\epsilon'/\epsilon)$ in the RS setup are induced by the two
effective chromomagnetic operators with opposite chirality
\begin{equation}
O_g=g_sH^\dagger\bar{s}_R\sigma^{\mu\nu}T^aG_{\mu\nu}^ad_L\,,
\qquad
O_g'=g_sH\bar{s}_L\sigma^{\mu\nu}T^aG_{\mu\nu}^ad_R\,,\label{Og}\end{equation}
 generated by the one-loop amplitude in Fig.~\ref{fcncLoop}\footnote{
The presence of $H$ in the definition of the 4D effective operators in Eq.~(\ref{Og}),
 tells us that the Wilson coefficients $C_g$ and $C'_g$ for these operators should be obtained by
dividing the spurion analysis result by the Higgs VEV $v$ , thus obtaining
Wilson coefficients of mass dimension -2.}.
 The imaginary part of the corresponding Wilson coefficients $C_g$ and $C'_g$ contributes to
${\mbox Re}(\epsilon'/\epsilon)\propto Im (C_g-C'_g)$.

\noindent In the spurion-overlap analysis and neglecting the generational
dependence of the overlap functions, we need to compute the  (12) and (21)
elements in Eqs.~(\ref{C7D}) and~(\ref{C7U}), respectively.
Again, given the dominance
of terms proportional to $f_{u,d}^2$ and the
suppressions by mass ratios in $V_{R}^{u,d}$, there
are only a few dominant contributions for each of the above
elements in the up and down sectors, while all other
contributions are suppressed by at least an order of magnitude.
However, since the dominant contributions to
$(C_{8g}^{up-type})_{12(21)}$ are proportional to
$V_{us}(V_{us}^*)$ and they are real if $\tilde{x}_2^{u,d}$ and
$\tilde{y}_2^{u,d}$ are assigned according to
Eq.~(\ref{CKMAssignment}),  we also consider the leading
 sub-dominant contributions in the up and down sectors. We obtain
\begin{eqnarray}\label{CEpsilon} \!\!\!\!\!\!\!\!\!\!{\mbox
Im}(C_g-C_g')={\mbox
Im}\left[(C_{8g}^{u-type}+C_{8g}^{d-type})_{12}-(C_{8g}^{u-type}+C_{8g}^{d-type})_{21}\right]\\
\simeq\frac{A^{1L}f_Q^2}{v^3M_{KK}}{\mbox
Im}\left[B_P^u\left(m_sF_{\epsilon_{12}}^u-m_dF_{\epsilon_{21}}^u\right)
+B_P^dm_dm_s\left((F_{\epsilon_{12}}^d-F_{\epsilon_{21}}^d)
\right)\right]\, ,\nonumber\\
\simeq\frac{1}{4.3\pi^2v^3M_{KK}^2}{\mbox
Im}\left[\left(m_sF_{\epsilon_{12}}^u-m_dF_{\epsilon_{21}}^u\right)
+\,m_dm_s\left(F_{\epsilon_{12}}^d-F_{\epsilon_{21}}^d
\right)\right]\, ,\nonumber
\end{eqnarray}
where in the last line we used  $A_{1L}=1/(64\pi^2M_{KK})$,
$f_Q^2\simeq 9.9$ and $B_P^{u,d}\simeq 1.5$. The dominant
contributions to the functions  $F_{\epsilon_{12,21}}^{u,d}$ are
\begin{equation}
F_{\epsilon_{12}}^{d}=m_s(f_d^2-f_s^2)\Delta_1^d\, \qquad
F_{\epsilon_{21}}^{d}=m_d(f_d^2-f_s^2)(\Delta_1^d)^*
\label{FepsilonD},\end{equation}
in the down sector and
\begin{equation}
F_{\epsilon_{12}}^u\simeq
(f_u^2m_u^2-f_c^2m_c^2)V_{us}+(f_u^2-f_c^2)m_um_c\Delta_1^u+f_c^2m_c^2V_{us}
\left((\tilde{x}_3^u+\omega\tilde{y}_3^u)f_\chi^t+((\tilde{x}_2^u)^*+
\omega(\tilde{y}_2^u)^*)f_\chi^c\right)
\label{FepsilonU1},
\end{equation}
\begin{equation}
F_{\epsilon_{21}}^u\simeq(f_u^2m_u^2-f_c^2m_c^2)V_{us}^*+(f_u^2-f_c^2)
m_um_c(\Delta_1^u)^*+f_c^2m_c^2V_{us}^*
\left(((\tilde{x}_3^u)^*+\omega^2(\tilde{y}_3^u)^*)
f_\chi^t+(\tilde{x}_2^u+\omega^2\tilde{y}_2^u)f_\chi^c\right)\,
\label{Fepsilon2}\end{equation}
in the up sector.
It is evident from Eqs.~(\ref{CEpsilon})\,--\,(\ref{Fepsilon2})
that  the contributions associated with
$F_{\epsilon_{21}}^u$ are suppressed by $m_d/m_s$ compared to
those arising from $F_{\epsilon_{12}}^u$, and the contributions
from $F_{\epsilon_{21}}^d$ are similarly suppressed compared to
those from $F_{\epsilon_{12}}^d$. This is analogous to the standard model
 pattern and opposite to what happens in flavor anarchic models. Therefore,
 to a good approximation, we only need to
estimate the imaginary parts of
$F_{\epsilon_{12}}^{u,d}$ as functions of the input parameters, in
order to obtain bounds on the KK mass scale in RS-A${}_4$.
Since
$f_c^2m_c^2\simeq f_u^2m_u^2$, and more precisely
$f_c^2m_c^2/(f_u^2m_u^2)-1\simeq O(10^{-4})$, the first term of
$F_{\epsilon_{12}}^{u}$ in
Eq.~(\ref{FepsilonU1}) approximately vanishes, and  the third term is roughly
suppressed by $|V_{us}|=0.2257$ compared to the second term.
Consequently, the imaginary part of  $F_{\epsilon_{12}}^{u,d}$ is largest
for $\Delta_{1}^d$ and $\Delta_{1}^u$  pure imaginary.
We first find  the maximal possible
contributions by spanning the $(\tilde{x}^{u,d}_i$,\,$\tilde{y}^{u,d}_i)$
parameter space, and only later we impose the constraints arising from
matching the CKM matrix. Denoting the magnitude of all $\tilde{x}_i^{u,d}$ and
$\tilde{y}_i^{u,d}$ parameters collectively by $\tilde{y}_{U,D}$ we obtain
 \begin{equation}
 max\left[|{\mbox
Im}(\Delta_1^d)|\right]=2\frac{m_d}{m_s}(f_\chi^d+f_\chi^s)\tilde{y}_D
\qquad max\left[|{\mbox
 Im}(\Delta_1^u)|\right]=2\frac{m_u}{m_c}(f_\chi^u+f_\chi^c)\tilde{y}_U
\label{ImFepsilon}\,,\end{equation}
with the assignment
\begin{equation}
(\tilde{x}_1^{u,d})^*=\omega^2(\tilde{y}_1^{u,d})^*=\pm
i\tilde{y}_{U,D}\qquad\tilde{x}_2^{u,d}=\tilde{y}_2^{u,d}=\pm
i\tilde{y}_{U,D} \, .\label{EpsilonAssignmentWPS}\end{equation}
From
Eq.~(\ref{CKMAssignment}),   in order to get
$|V_{us}|=|V_{cd}|=0.2257$, we must require
$\tilde{x}_2^d=\tilde{y}_2^d=-\tilde{x}_2^u=-\tilde{y}_2^u$ and
$\tilde{y}_{U,D}=1$. The resulting maximal imaginary parts in
Eq.~(\ref{ImFepsilon}) will then be reduced by a factor 2 for realistic
CKM assignments, due to
the exact cancellation of terms proportional to
$\tilde{x}_2^{u,d}$ and $\tilde{y}_2^{u,d}$. For
$\tilde{y}_{U,D}\neq 1$, we should correspondingly rescale the
$\chi$ VEV to maintain $|V_{us}|=0.2257$. However, this will have
significant implications on the neutrino mass spectrum even for
$\mathcal{O}(1)$ rescaling, as was shown in \cite{A4Warped}.
The two (dominant) terms in Eq.~(\ref{ImFepsilon})
will add up maximally for
$\tilde{x}_2^d=\tilde{y}_2^d=\tilde{x}_2^u=\tilde{y}_2^u$, which
corresponds to a vanishing $V_{us}$ according to Eq.~(\ref{Vus}).

\noindent  Focusing now  on the
third term in Eq.~(\ref{FepsilonU1}), and considering the
assignment of Eq.~(\ref{EpsilonAssignmentWPS}), we realize that
for $V_{us}$ pure imaginary we should maximize the real part
of the expression adjacent to it. The only parameters left to be
assigned are $\tilde{x}_3^{u,d}$ and $\tilde{y}_3^{u,d}$ leading to
\begin{equation}
{\mbox max}[|{\mbox
Re}\left((\tilde{x}_3^u+\omega\tilde{y}_3^u)f_\chi^t+((\tilde{x}_2^u)^*
+\omega(\tilde{y}_2^u)^*)f_\chi^c\right)\!|]=
(2f_\chi^t-(\sqrt{3}/2)f_\chi^c)\tilde{y}_U,
\end{equation}
for the parameter choice
$\tilde{x}_3^u=\omega\tilde{y}_3^u=\tilde{y}_U$.  Notice that this
contribution,  proportional to $V_{us}$, will vanish for the
choice $V_{us}=0$ that maximizes the sum of the  terms in
Eq.~(\ref{ImFepsilon}). Instead, for the assignments that lead to
a realistic CKM matrix such as the one in
Eq.~(\ref{CKMAssignment}), the contribution of the above terms
will be suppressed by roughly an order of magnitude compared to
the ($\Delta_1^{u,d}$)  terms, and can be safely neglected.

\noindent From Eq.~(\ref{FepsilonU1}), using $f_c\ll f_u$ and
$B_P^{u,d}\simeq1.5$, we obtain the following upper bound on the
up-type contribution to $\epsilon'/\epsilon$ 
\begin{equation}
{\mbox
Im}(C_g-C_g')^{u}\simeq\frac{Y^2f_u^2m_u^2m_s}{11.5\pi^2v^3M_{KK}^2}
2(f_\chi^u+f_\chi^c)\tilde{y}_U\label{FinalEUP}.\end{equation} 
It scales with $Y^2$, where $Y$
generically denotes the overall scale of the 5D LO Yukawa
couplings associated with the $H\Phi$ interactions.
Using again the assignment in Eq.~(\ref{EpsilonAssignmentWPS}) the down type contribution is
given by
\begin{equation}
{\mbox
Im}(C_g-C_g')^{d}\simeq\frac{Y^2f_d^2m_d^2m_s}{11.5\pi^2v^3M_{KK}^2}2
(f_\chi^d+f_\chi^s)\tilde{y}_D\,.\label{FinalEDown}
\end{equation}
The NP contributions to ${\mbox Re}(\epsilon'/\epsilon)$ are
directly constrained by the experiment. In order to extract such a
constraint we construct the difference $\delta_\epsilon' = (
{\mbox Re}(\epsilon'/\epsilon)_{NP} - {\mbox
Re}(\epsilon'/\epsilon)_{SM})/ {\mbox
Re}(\epsilon'/\epsilon)_{exp}$, as done in \cite{IsidoriPLB}.
Assuming  ${\mbox Re}(\epsilon'/\epsilon)_{SM} =0$ one can write
\begin{equation}
\delta_{\epsilon}'=\frac{\omega_\epsilon\langle(2\pi)_{I=0}|
\lambda_sO_g|K^0\rangle)}{\sqrt{2}{\mbox Re}A_0 {\mbox
Re}(\epsilon'/\epsilon)_{exp}|\epsilon|_{exp}}\left[\frac{{\mbox
Im}(C_g-C_g')}{\lambda_s}\right]\simeq(58{\mbox
TeV})^2B_G\left[\frac{{\mbox Im}(C_g-C_g')}{\lambda_s}\right]
\label{EpsilonExp},\end{equation}
where $B_G$, the hadronic bag parameter \cite{HadronBag}, is given by
\begin{equation}
\langle(2\pi)_{I=0}|\lambda_sO_g|K^0\rangle=\sqrt{\frac{3}{2}}
\frac{11}{4}\frac{m_\pi^2m_K^2}
{F_\pi}B_G,\label{BagDef}\end{equation}
and we set $B_G=1$ as in \cite{IsidoriPLB}. The
parameter $\lambda_s$ is the SM Yukawa coupling of the $s$ quark,
namely $\lambda_s\times 174\, {\mbox GeV}=m_s\simeq50\,{\mbox MeV}$. The quantities $A_0$
and $A_2$ denote the amplitudes for the $(2\pi)_{I=0}$ and
$(2\pi)_{I=2}$ decay channels of the $K^0$ meson, respectively.
We take $F_\pi=131\,{\mbox MeV}$, ${\mbox
Re}(A_0)=3.3\times 10^{-4}$MeV, $\omega_\epsilon=|A_2/A_0|=0.045$
and $|\epsilon_{exp}|=2.23\times 10^{-3}$.
Imposing
$|\delta_\epsilon'|<1$ and using Eqs.~(\ref{FinalEUP}) and (\ref{FinalEDown})
 directly leads to a bound for the KK scale
\begin{equation}
(M_{KK})_{\epsilon'/\epsilon}^{Spurion-overlap}\gtrsim
1.4\,Y~{\mbox TeV}. \label{EpsilonBound}\end{equation} 
Using the results for the overlap dependence of the up- and down-type
contributions to $\epsilon'/\epsilon$ from the one-generation
KK diagonalization scheme in Eq.~(\ref{KK1genFactros}), the up-
and down-type contributions get enhanced and suppressed
respectively and the resulting bound on $M_{KK}$ is
\begin{equation}
(M_{KK})_{\epsilon'/\epsilon}^{KK(1gen)} \gtrsim 1.25\,Y~{\mbox
TeV}\label{EpsilonBoundKK}.\end{equation} 
This bound  is the most stringent in the
$(\tilde{x}_i^{u,d},\tilde{y}_i^{u,d})$ parameter space and
corresponds to an highly unnatural set of parameter assignments, for which
$V_{us}=0$, and maximize the sum of up and down imaginary
parts. Nevertheless, this bound is still less stringent than the
one in the flavor anarchic case considered in \cite{IsidoriPLB},
and it allows for $O(1\,{\mbox TeV})$ KK masses for $O(1)$ Yukawa
couplings $Y\approx 1$, $\tilde{y}_{U,D}\simeq1$. The latter
values correspond to $\chi_0=0.155M_{Pl}^{3/2}$, a value allowed
by the neutrino oscillation data \cite{A4Warped}. Notice also that
in the flavor anarchic case \cite{IsidoriPLB} the most stringent
bound on the KK scale arises from the combined Yukawa coupling
dependence of the new physics contributions to
$\epsilon'/\epsilon_K$ and $\epsilon_K$. The latter contribution
comes from the tree level KK gluon exchange, it is inversely
proportional to the Yukawa coupling, and provides the most
stringent bound on the KK scale for $O(1)$ Yukawas. The combined
bound on $M_{KK}$ in \cite{IsidoriPLB} has been obtained for
Yukawa couplings of $\mathcal{O}(6)$, implied by the constraint
from the tree level KK gluon exchange contribution to
$\epsilon_K$. This contribution vanishes identically  in
RS-A${}_4$ \cite{A4Warped}, thus relaxing one of the most
stringent constraints of flavor anarchic models.

\noindent A more natural bound on the KK mass scale is obtained
for  the parameters assignment of Eq.~(\ref{CKMAssignment}) that
provides an almost realistic CKM matrix, while still choosing
$\tilde{x}_1^{u,d}$ and $\tilde{y}_1^{u,d}$ to maximize the
imaginary parts of up- and down-type contributions according to
Eq.~(\ref{EpsilonAssignmentWPS}). The resulting bound on $M_{KK}$
will be further suppressed by at most a factor of $\sqrt{2}$
compared to the one in Eq.~(\ref{EpsilonBoundKK}). This bound (for
$Y=1$) is again significantly lower than the one implied by
constraints arising from EWPM\,\cite{A4Warped}, in particular the
$Zb_L\bar{b}_L$ coupling.
This is a pleasing result in RS-A${}_4$, indicating that constraints arising from new physics
contributions to FCNC processes tend to be weaker than in
flavor anarchic models, see for example
\cite{Agashe:2004cp,Csaki:2009wc,IsidoriPLB} and references
therein. Another important difference between RS-A${}_4$ and
anarchic frameworks stems from the dominance of SM-like dipole
operators, and the lack of enhancement of the opposite chirality
operators.

\noindent To conclude the analysis of the constraints arising
from $\epsilon'/\epsilon$ we state the results from the numerical
and semianalytical diagonalization of the three-generation $12\times12$ mass matrices. For
$x=0.037$ the numerical result
is
\begin{equation}(\delta_{\epsilon'})_{RS-A4}^{Num.}\simeq0.1\quad
(M_{KK})_{RS-A4}^{Num.}\gtrsim0.98
\, {\mbox TeV}\, ,\label{EpsilonNumResult}\end{equation}
 close to the spurion-overlap and  the one-generation mass-matrix diagonalization
approximations. The results of the semianalytical
diagonalization for the RS-A$_4$ contributions to
$\epsilon'/\epsilon$ are given by
\begin{equation}
(\delta_{\epsilon'})_{RS-A4}^{12\times12th}\simeq0.3\qquad(M_{KK})_{RS-A4}^{12\times12
th}\simeq2.5\, Y\,{\mbox TeV}\, .\end{equation}
As anticipated, the deviation of the semianalytical $12\times12$ estimate from
the numerical one can be attributed to the fact that the mass matrices in the
first case are only approximately
diagonalized, and a residual contamination of $\mathcal{O}(x)$ is noticed to be
still present
 in the second- and third-generation off-diagonal blocks of the approximately
diagonalized mass matrix. Nonetheless,  the importance of the semianalytical method
 stays in providing some insight into the way the A$_4$ flavor
structure induces cancellation patterns, due to the explicit phase
structure of the LO Yukawa interactions in Eq.~(\ref{LOYukawa}).
\subsection{How stringent is the constraint from $b \rightarrow
s\gamma$?}\label{subsec:BSGamma}
Measurements of the branching ratio BR$(B \rightarrow X_s\gamma)$
are already accurate enough to provide stringent constraints on
new physics contributions to $b \rightarrow s\gamma$. It is thus
instructive to obtain the NP physics contributions to  $b
\rightarrow s\gamma$ in RS-A${}_4$, derive the corresponding bound
on the KK mass scale and compare with the flavor anarchic results
of \cite{IsidoriPLB} and \cite{Azatov}.

\noindent As it is also true for $\epsilon'/\epsilon$, and in contrast to flavor
anarchic models, the largest contribution to $b
\rightarrow s \gamma$ in RS-A${}_4$ is generated by the single-chirality effective
dipole operator
\begin{equation}
O_7=\frac{em_b}{8\pi^2}\bar{b}_R\sigma^{\mu\nu}F_{\mu\nu}s_L\, ,
\label{O7}\end{equation}
while the contribution arising from the opposite chirality operator
$O_7'$ is suppressed by $m_s/m_b$ as in the SM. This can easily be inferred by looking at the
$i=2,j=3$ (for $O_7$) and $i=3,j=2$ (for $O_7'$) components of the Wilson
coefficients defined in Eqs.~(\ref{C7D}) and~(\ref{C7U}).

\noindent The Wilson coefficient $C_7$ for $O_7$ is generated by the
loop amplitudes in Figs.~\ref{fcncLoop}  and~\ref{fcncLoopcharged} and we follow the same procedure as for $\epsilon'/\epsilon$
 to  estimate the dominant contributions to $C_7$\footnote{
Considering the definition of $O_7$, we should correspondingly
rescale the  contributions coming from Eqs.~(\ref{C7D})
and~(\ref{C7U}) by $(8\pi^2/m_b)$, supplementing us again with a
coefficient of dimension $[mass]^{-2}$.}.
For the down-type contributions we obtain the following dominant
terms \begin{equation}
(C_7)^{d-type}\simeq -\frac{1}{3}
\frac{B_P^df_Q^2m_sm_b}{8v^2M_{KK}^2}\left(f_d^2(\Delta_1^d)^*
\Delta_2^d+f_s^2(\Delta_3^d)\right)
\label{BSgammaD},\end{equation}
 where
the factor $-1/3$ comes from the charge of a down-type quark.
Considering Eqs.~(\ref{DeltaR1})--(\ref{DeltaR3}), it is clear that the second term
in Eq.~(\ref {BSgammaD}) is dominant, despite the presence of $f_d^2$ in the first one.
 As before, we assign a collective magnitude to the
$\tilde{x}_i^{u,d}$, $\tilde{y}_i^{u,d}$ parameters in terms of
$\tilde{y}_{U,D}$ and look for the phase assignments that
maximize the total contribution. We thus first rewrite  the maximal
magnitudes of $\Delta_{1,2,3}^d$, defined in
Eqs.~(\ref{DeltaR1})--(\ref{DeltaR3}), in terms of $\tilde{y}_D$.
It is straightforward to obtain
\begin{eqnarray}
&&{\mbox
max}(|\Delta_{1}^d|)=\frac{m_d}{m_s}\left(2f_\chi^d+2f_\chi^s\right)\tilde{y}_D\,\,\,\,\, {\mbox
max}(|\Delta_{2}^d|)=\frac{m_d}{m_b}\left(2f_\chi^d+2f_\chi^b\right)
\tilde{y}_D\nonumber\\
&&{\mbox
max}(|\Delta_{3}^d|) =\frac{m_s}{m_b}\left(2f_\chi^s+2f_\chi^b\right)
\tilde{y}_D,\label{DeltaBounds}\end{eqnarray}
and similar bounds for the up-type right-handed diagonalization
matrices are obtained via the replacement
$(d,s,b,\tilde{y}_D)\rightarrow(u,c,t,\tilde{y}_U)$ in the above
equation. In particular, the bound in Eq.~(\ref{DeltaBounds}) on
$\Delta_3^d$, to which the dominant term in Eq.~(\ref{BSgammaD})
is proportional, is obtained with the assignment
\begin{equation}
\tilde{x}_2^d=\omega\tilde{y}_2^d=\tilde{x}_3^d=\omega\tilde{y}_3^d=
\tilde{y}_De^{i\delta_{bsD}}.\label{BSassignmentD}
\end{equation}
The above bounds can now be used to estimate the down-type
contribution to $O_7$ leading to
\begin{equation}
(C_7)^{d-type}\simeq-\frac{Y^2\tilde{y}_D}{4v^2M_{KK}^2}\left(m_s^2f_s^2(f_\chi^s+f_\chi^b)
+2m_d^2f_d^2(f_\chi^d+f_\chi^b)(f_\chi^d+f_\chi^s))\right),\label{BsgammaD1}
\end{equation}
where we used $f_Q^2\simeq9.9$, and $B_P^d\simeq 1.5$ and made
explicit the $Y^2$ scaling behavior. In the up sector, we learn
from Eq.~(\ref{C7U}) that there are two dominant contributions to
the effective coupling of $O_7$, leading to
 \begin{equation}
 (C_7)^{u-type}\simeq\frac{5}{3}\frac{B_P^uf_Q^2}{8v^2M_{KK}^2}
\left((f_c^2m_c^2-f_t^2m_t^2)V_{cb}+
f_c^2m_c^2(f_\chi^c((\tilde{x}_2^u)^*+\omega(\tilde{y}_2^u)^*)+
f_\chi^t(\tilde{x}_3^{u}+\omega\tilde{y}_3^{u})\right)
,\label{BSgammaU}\end{equation}
where the factor $5/3$ comes from the electric charge of an
internal up-type quark in Fig.~\ref{fcncLoop} and a negatively charged Higgs
in Fig.~\ref{fcncLoopcharged}.
All the other up-type contributions to $C_7$ are
suppressed by at least another order of magnitude. To obtain a
conservative bound on NP contributions to $b\to s\gamma$, we can
again find the maximal values of the combined up and down
contributions in the model parameter space. We first use
Eq.~(\ref{Vcb}) to rewrite the up-type contribution as
\begin{eqnarray}
 (C_7)^{u-type}&\simeq&\frac{5}{3}\frac{B_P^uf_Q^2}{8v^2M_{KK}^2}\left[\phantom{\frac{1}{2}}
 \!\!\!-f_t^2m_t^2\left(f_\chi^b(\tilde{x}_3^d+\omega\tilde{y}_3^d)-f_\chi^t
(\tilde{x}_3^u+\omega\tilde{y}_3^u)\right)\right.\nonumber\\&&\left.
+f_c^2m_c^2\left(f_\chi^c((\tilde{x}_2^u)^*+\omega(\tilde{y}_2^u)^*)+f_\chi^b(\tilde{x}_3^{d}+
\omega\tilde{y}_3^{d})\right)\phantom{\frac{1}{2}}\!\!\!\!\right]\, ,
\label{BSgammaUU}\end{eqnarray}
where we use $f_c^2m_c^2\ll f_t^2m_t^2$, and in particular
$f_t^2m_t^2=(y_t^2/y_c^2)(r_{00}^t/r_{00}^c)^2f_c^2m_c^2\simeq
6.35\, f_c^2m_c^2$, so that the first term in
Eq.~(\ref{BSgammaUU}) is dominant
 for $\mathcal{O}(1)$ parameter assignments.

\noindent The maximal combined up and down contributions to $C_7$
would be realized when both are real and negative.
 This corresponds to $\delta_{bsD}=0$ in Eq.~(\ref{BSassignmentD}) and  for the up
 sector:
\begin{equation}
\tilde{x}_3^u=\omega\tilde{y}_3^u=\tilde{x}_2^u=\omega\tilde{y}_2^u=-\tilde{y}_U\, .
\label{BSassignmentU}\end{equation}
With this assignment the second (subdominant) term in
Eq.~(\ref{BSgammaUU}) vanishes. Using $B_P^u\simeq 1.5$ and
$f_Q^2\simeq 9.9$  we obtain
\begin{equation}
(C_7)^{u-type}\simeq-\frac{5Y^2\tilde{y}_U}{4v^2M_{KK}^2}\left(f_t^2m_t^2(f_\chi^t+f_\chi^b)
\right)\, .\label{BSgammaU1}
\end{equation}
It thus turns out that the  up-type contribution to $b\rightarrow
s\gamma$ dominates over the down-type by roughly an order of
magnitude.

\noindent A bound on the KK mass scale can be extracted by comparing with the
 SM contribution to $b\to s\gamma$ and the corresponding experimental bound.
Contrary to flavor anarchic models, the dominant contributions
from NP and SM both come from the single chirality Wilson
coefficient $C_7$. The SM contribution evaluated at the $W$ scale
can be written as follows \cite{Buras}
\begin{equation}
C_{7}^{SM}(\mu_W )\simeq\frac{g^2|V_{tb}V_{ts}^*|D_0'(m_t)}{M_W^2}\simeq
1.06\times 10^{-7}({\mbox GeV})^{-2}\label{BSgammaSM},
\end{equation}
where $D_0'(m_t)\sim0.4$, $M_W=80.4$ GeV and $g\simeq0.65$. Following
the analysis in \cite{Azatov}  we conveniently define the
ratio between the NP and SM contributions as
\begin{equation}
\delta_7\equiv  \frac{C_{7}^{RS-A_4}(M_{KK})}{C_7^{SM}(\mu_W)}\simeq \frac{1.3\times
10^{-7}(GeV)^{-2}}{1.06\times
10^{-7}(GeV)^{-2}}\times\frac{(1\,{\mbox
TeV})^2}{M_{KK}^2}Y^2\tilde{y}_U\, , \label{BSgammaCompare}
\end{equation}
which  is thus a function of $\tilde{y}_U$ and $M_{KK}$. An
analogous definition holds for $C'_7$. We realize that, even for
KK masses as low as 3 TeV and for the largest Yukawa allowed by
perturbativity bounds $Y=4\pi/\sqrt{N_{KK}}\sim 9$, the RS-A${}_4$
new physics contribution predicted in the spurion-overlap
approximation is at most comparable to the SM one, and it is
suppressed by roughly an order of magnitude for the parameter
assignments that yield a realistic CKM matrix. To impose a
conservative bound on the KK mass scale we proceed as in the
analysis of flavor anarchic models \cite{Azatov, IsidoriPLB}. To
compare with the experiment, we use the model independent ratio
\cite{Azatov} $\Gamma^{total}(b\to s\gamma )/\Gamma^{SM}(b\to
s\gamma ) \approx 1+0.68 Re (\delta_7 ) + 0.11 |\delta'_7|$ which
takes into account the running from $M_{KK}$ down to $\mu_b$, with
$\Gamma^{total} \propto |C_7(\mu_b)|^2 + |C'_7(\mu_b)|^2$. Given
that the running of the Wilson coefficients from the KK scale down
to $\mu_b$ remains an $\mathcal{O}(1)$ suppression effect,  and
using the experimental value for $BR(b\rightarrow s\gamma)$
affected by $\sim 10\%$ uncertainty, a $\mathcal{O}(20\%)$ departure
of NP contributions from the SM prediction is still allowed.
Considering separately the contributions from $C_7$ and $C'_7$,
the allowed window translates into \cite{Azatov}
$Re(\delta_7)\lesssim 0.3$ and $|\delta_7'|\lesssim 1.4$. Since
the contribution to $C_7'$ in our setup is further suppressed by
$m_s/m_b$ compared to the one of $C_7$ and the bound on
$\delta_7'$ is far less stringent, the constraint on the KK mass
scale will come from $\delta_7$. Substituting in
Eq.~(\ref{BSgammaCompare}), we obtain a conservative bound, which
 do not correspond to a realistic CKM matrix
\begin{equation}
(M_{KK})_{b\rightarrow s\gamma}^{cons.}\gtrsim2.1\,
Y\sqrt{\tilde{y}_U}\,{\mbox TeV}\, .\label{BSgammaBound1}\end{equation} 
Using instead the parameter assignment of Eq.~(\ref{CKMAssignment}) to obtain
a realistic CKM matrix, leads to the more realistic constraint
\begin{equation}
(M_{KK})_{b\rightarrow s\gamma}^{{CKM}}\gtrsim 1.4\,
Y\sqrt{\tilde{y}_U}\, {\mbox TeV}\, . \label{BSgammaCKMbound}\end{equation}
 One important difference between RS-A${}_4$ and flavor anarchic
models resides in the dominance of $C_7$ in the new physics contributions. This can
obviously lead to different patterns of interference between NP
and SM contributions in direct CP asymmetries. Hence, a study of
the latter might discriminate among NP models more efficiently
than the measure of branching ratios. In RS-A${}_4$, a non trivial
pattern of interference between $C_7^{RS-A_4}$ and $C_7^{SM}$
might be in place.

\begin{figure}[t]
\centering
\includegraphics[width=10 truecm]{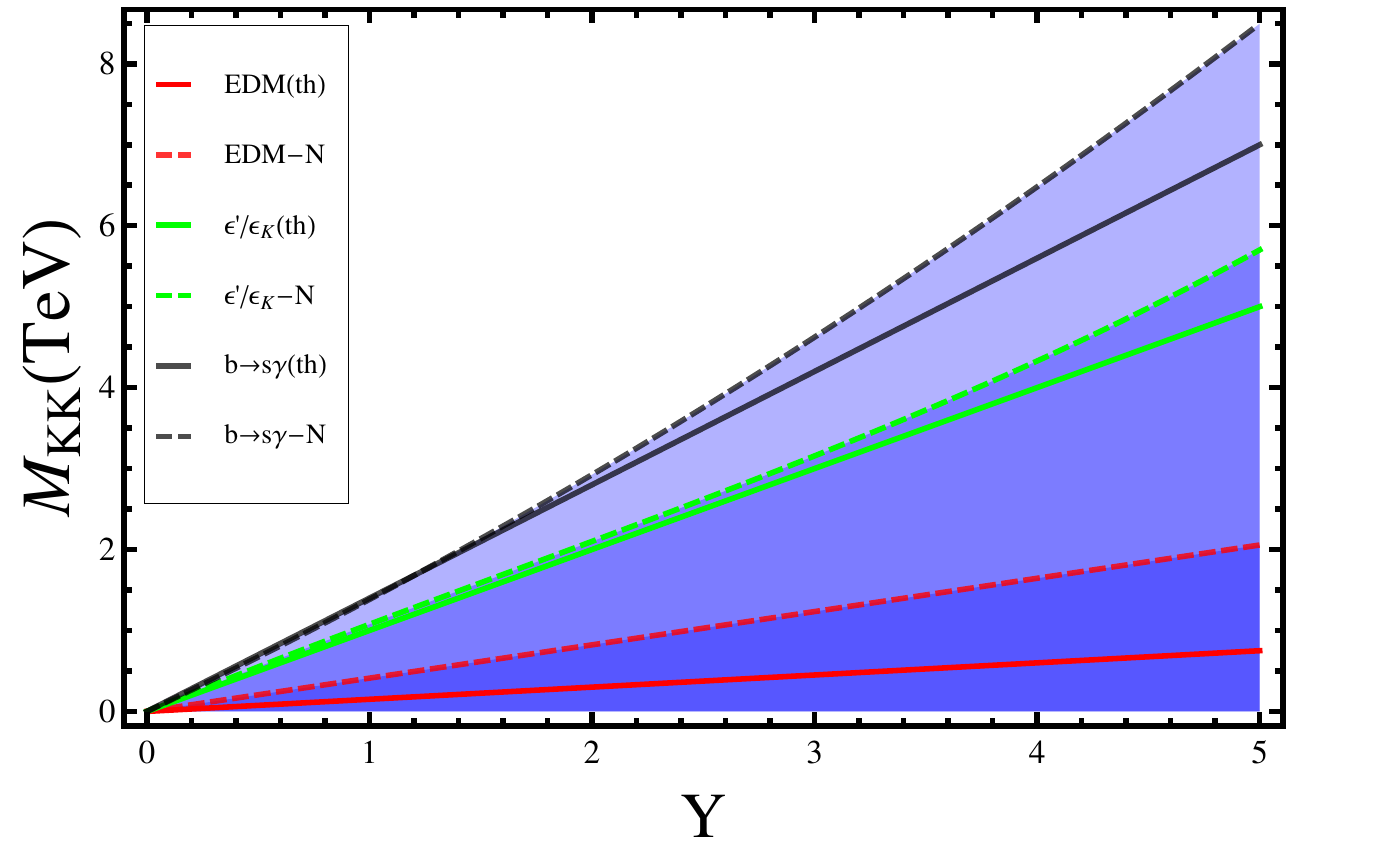}
\caption{ Bounds on the KK mass scale $M_{KK}$ as a function of the overall
Yukawa scale $Y$ for  the neutron EDM (bottom, red) $\epsilon'/\epsilon$
 (centre, green) and $b \rightarrow s\gamma$ (top, black). The analytical
(th) results (solid lines) are obtained within the one-generation
approximation combined with the spurion-overlap analysis and compared
 with the numerical (N) results (dashed lines) of the three-generation case.
In both cases, predictions are obtained for the model parameters that lead
to a realistic CKM matrix.}\label{PlotNumerical}
\end{figure}

\noindent Since the above constraint is the most significant we have
encountered so far, we go beyond the spurion-overlap approximation
and use the results of the analytical diagonalization of the
one-generation  mass matrices. In this way we are
able to obtain a better description of the overlap dependence of the above
process. The modifications to the overlap factor $B_P^u$ for
$b\rightarrow s\gamma$  are obtained in appendix~\ref{app:B2} and
the resulting new overall correction factor is
$(B_P^u)_{b\rightarrow s\gamma}^{KK(1 gen)}\simeq-1.54$, which has
a very moderate effect on the  KK mass scale bound. The dependence
of the resulting constraint on the size of Yukawa couplings is
illustrated together with the constraints from
$\epsilon'/\epsilon$ and the neutron EDM, in
Fig.~\ref{PlotNumerical}.
\noindent Finally, we state the results of the  numerical and semianalytical
approach to the three-generation case.
Numerically, and by adding up- and down-type
contributions in quadrature, we obtain
\begin{equation}
(\delta_7)_{RS-A4}^{Num.}\simeq0.03 \qquad
(M_{KK})_{RS-A4}^{b\rightarrow s\gamma(Num.)}\gtrsim 1.27\, {\mbox  TeV}\, ,
\label{bsgammaNum}\end{equation}
while the semianalytical diagonalization scheme leads to
\begin{equation}
(\delta_7)_{RS-A4}^{12\times12th}\simeq0.06 \qquad
(M_{KK})_{RS-A4}^{12\times12th}\gtrsim 1.8\, Y\, {\mbox TeV}\, .
\end{equation}
 Fig.~\ref{PlotNumerical} provides a summary of the results obtained in
this section. We compare the bounds on the KK mass scale $M_{KK}$ as a
function of the Yukawa coupling for  the neutron EDM, $\epsilon'/\epsilon$
and $b \rightarrow s\gamma$. Differently from flavor anarchic models, the
 most stringent constraint eventually comes from $b\to s\gamma$. For an
overall Yukawa scale $Y\simeq 1$\footnote{We recall that the overall Yukawa
 scale is defined as $y_{u,d,c,s,t,b}\to Y\,y_{u,d,c,s,t,b}$,
with reference values $y_{u,d,c,s,b}=1$ and $y_t=2.8$.} all constraints are
weaker than the one implied by EWPM, in particular $Zb_L\bar{b}_L$.
In Fig.~\ref{PlotNumerical} we also compare the
analytical prediction $(th)$, obtained in the combined spurion-overlap
and one-generation diagonalization scheme described in Sec.~\ref{Subsec:KKdiag1},
 with the {\em exact} numerical analysis $(N)$ of all three generations.
Both predictions are obtained for the model parameters that lead to a
realistic CKM matrix. It is worth to recall that the analytical
prediction for the neutron EDM reported in Fig.~\ref{PlotNumerical}
and Eq.~(\ref{EDM1gensp}), represents a very conservative estimate coming
 from the up sector contribution in Eq.~(\ref{FEDMnodeg}) and entirely
due to the non degeneracy of overlap factors. Differently from other quantities,
 the neutron EDM identically vanishes in the spurion-overlap approximation with
degenerate overlap factors.

\noindent We conclude that the constraints on new physics contributions
from the neutron EDM,
$(\epsilon'/\epsilon)$ and $b \rightarrow s\gamma$ are relaxed in
our setup compared to generic warped flavor  anarchic models.
 In order to impose significant bounds on the KK mass scale in RS-A${}_4$ using
the above processes we must wait for more precise
measurements of these observables. This might not be the situation
with Higgs mediated FCNCs, considered in the next
section.
\section{New physics from Higgs mediated FCNCs}
\label{sec:higgs}
It has been pointed out, both  in the context of a composite Higgs
sector of strong dynamics \cite{HiggsContino} and warped extra
dimensions \cite{HiggsFCNC}, that higher dimensional operators in
the low energy 4D effective theory with extra insertions of a
Higgs field generally leads to a misalignment between mass and
Yukawa matrices and consequently to tree level Higgs mediated
FCNCs. The presence of a misalignment is a quite general and model
independent result.

\noindent In the RS-A${}_4$ framework, once A${}_4$ is completely
broken by ``cross-talk" interactions, the 4D effective Yukawa
couplings originate from  5D Yukawa operators that involve the
Higgs, and one or both flavons $\Phi$ and $\chi$. This is relevant
to determine the typical strength of the effective 4D
interactions. The operators that generate the misalignment between
the Higgs Yukawa couplings and SM fermion mass matrices in the 4D
effective theory, are of dimension six and can be written in terms
of the 4D fields as follows \cite{HiggsFCNC}:
\begin{equation}
A_{ij}^{u,d}\frac{H^2}{\Lambda^2}H\bar{Q}_{L_i}(U_{R_j},D_{R_j}),\qquad
B_{ij}\frac{H^2}{\Lambda^2}\bar{D}_{R_i}\slashed{\partial}D_{R_j},\qquad
C_{ij}\frac{H^2}{\Lambda^2}\bar{U}_{R_i}\slashed{\partial}U_{R_j},\qquad
K_{ij}\frac{H^2}{\Lambda^2}\bar{Q}_{L_i}\slashed{\partial}Q_{L_j},
\label{HiggsOperators}\end{equation}
where $Q_{L_i}$ and $D_{R_i}$($U_{R_i}$) are the ${\mbox
SU}(2)_L$ SM fermion doublets and singlets, respectively, and the
Higgs field $H=v+{h}$ is a 4D field containing the
physical Higgs $h$. The scale $\Lambda$ is the 4D cutoff and the
coefficients $A_{ij}$, $B_{ij}$, $C_{ij}$ and $K_{ij}$ are in
general complex. The indices $i,j$ denote flavors of the SM
quarks. Once the electroweak symmetry is broken at the Higgs VEV
scale, $v=174$\,GeV, the above operators will induce corrections
to the fermion masses, Yukawa couplings and kinetic terms. The
corrected mass and kinetic terms can be generically parametrized
as \cite{HiggsFCNC} 
\begin{equation}
v\left(\hat{y}_{ij_{d}}^{\tiny{SM}}+A_{ij}^d\frac{v^2}{\Lambda^2}\right)
\bar{Q}_{L_i}D_{R_j}\,,\qquad
\left(\frac{\delta_{ij}}{2}+K_{ij}\frac{v^2}{\Lambda^2}\right)\bar{Q}_{L_i}
\slashed{\partial}Q_{L_j}\,,\qquad
\left(\frac{\delta_{ij}}{2}+B_{ij}\frac{v^2}{\Lambda^2}\right)\bar{D}_{R_i}
\slashed{\partial}D_{R_j}\,,
\label{HiggsCorrectionGeneral}
\end{equation}
while the corrected Yukawa interactions with the physical 4D Higgs $h$ are generally given by
\begin{equation}
\left(\hat{y}_{ij_{d}}^{\tiny{SM}}+3A_{ij}^d\frac{v^2}{\Lambda^2}\right){h}\bar{Q}_{L_i}D_{R_j}\,,
\label{HiggsYukawaGeneral}
\end{equation}
and analogously for the up contributions $A_{ij}^u$ and $C_{ij}$, where
$\hat{y}_{ij_{u,d}}^{SM}\simeq\hat{Y}_{ij_{u,d}}^{LO}f_{Q_i}^{-1}f_{u_j,d_j}^{-1}
r_{00}^{H\Phi}(c_{Q_i},c_{u_j,d_j},\beta)$
are the SM leading order Yukawa couplings.
\begin{figure}
\begin{center}
\includegraphics[width=10 truecm]{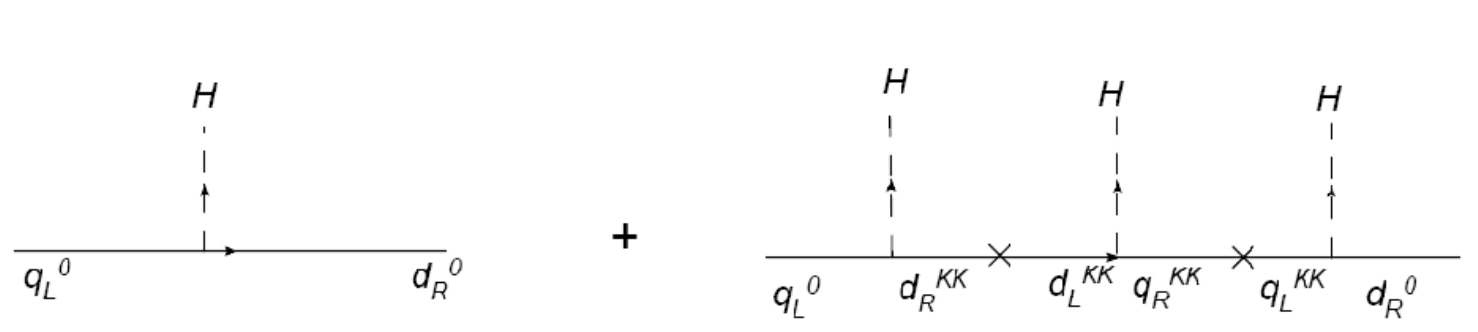}
\caption{Higgs mediated corrections to masses and Yukawa couplings of
SM fermions in
the mass insertion approximation. The 4D effective Higgs field is defined
here as $H=v + h$ and  contains the physical Higgs field $h$.}\label{insertion}
\end{center}
\end{figure}
\begin{figure}
\begin{center}
\includegraphics[width=10 truecm]{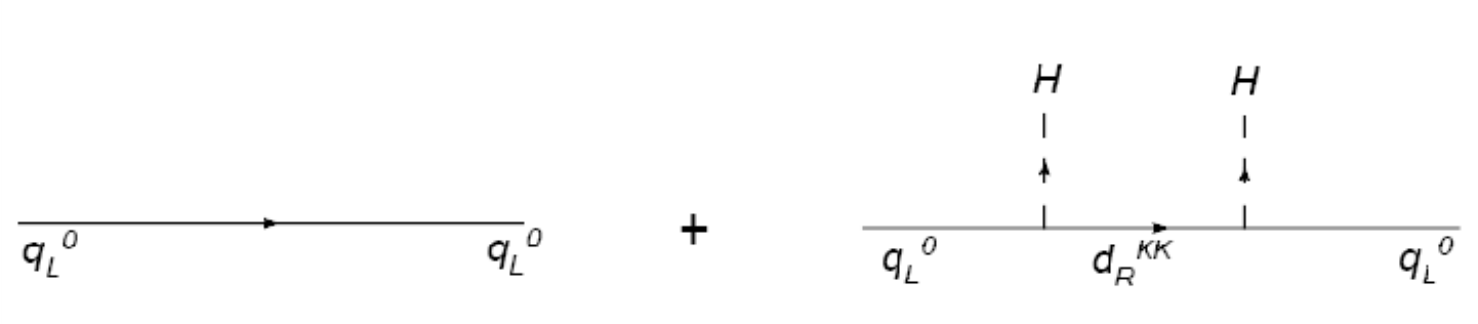}
\caption{Higgs mediated corrections to kinetic terms of SM fermions in the mass insertion
approximation.}\label{insertion1}
\end{center}
\end{figure}
The origin of the misalignment between Yukawa couplings and SM
masses \cite{HiggsFCNC} resides in the simple fact that an
additional multiplicity factor 3 is associated with the corrected
Yukawa couplings to the physical Higgs $h$. In the mass insertion
approximation, the leading corrections to Yukawa couplings and
fermion masses are generated by the second diagram in
Fig.~\ref{insertion}, while the corrections to the fermion kinetic
terms are generated by the second diagram in Fig.~\ref{insertion1}
via the exchange of KK modes.

\noindent After redefining the fermion fields to canonically
normalize the corrected fermion action, the total misalignment
between SM masses and Yukawa couplings in the mass insertion
approximation is given by \cite{HiggsFCNC}
\begin{equation}
\hat{\Delta}^{u,d}_{H}=\hat{m}^{SM}_{u,d}-\hat{y}^{SM}_{{u,d}}\,v=
\hat{\Delta}_{H_1}^{u,d}+\hat{\Delta}_{H_2}^{u,d}
\label{GeneralShift1},\end{equation}
where
\begin{equation}
\hat{\Delta}_{H_1}^{u,d}=-2A_{ij}^{u,d}\frac{\,v^2}{\Lambda^2}
\qquad
\hat{\Delta}_{H_2}^{u(d)}=- \left(K_{ij}\frac{v^2}
{\Lambda^2}+C_{ij}(B_{ij})\frac{v^2}{\Lambda^2}\right)\, .
\label{GeneralShift2}
\end{equation}
It is clear that, after the shift in Eq.~(\ref{GeneralShift1}),
the SM mass matrix and Yukawa couplings are in general not
diagonalized by the same biunitary transformation. Thus, in the
diagonal mass basis, non-diagonal Yukawa interactions are in
general present and induce FCNC processes by tree level Higgs
exchange.

\noindent In \cite{Azatov} the contributions of Higgs mediated
FCNC to $\Delta F =2$ processes were estimated in the framework of
flavor anarchy. It was found that the dominant contribution to the
misalignment is in this case due to $(--)$ KK modes, and does not
vanish for an IR localized Higgs, contrary to the conclusions of
previous analyses \cite{BurasHiggs}.  Also, the overall
misalignment was calculated by mass diagonalization in the
one-generation approximation, and generalized to three generations
using a spurion analysis in the mass insertion approximation
\cite{Agashe:2004cp}. Here, we analyze the same misalignment in
the context of RS-A${}_4$ to establish whether a significant
suppression of Higgs mediated FCNC contributions to $\Delta F=2$
processes can be induced by the particular structure of up and
down diagonalization matrices of A${}_4$.

\noindent We start with writing the explicit flavor structure of
the corrections in Eq.~(\ref{GeneralShift2}), in the IR localized
case and using the spurion analysis in the mass insertion
approximation. In the special interaction basis one would in this
case obtain
\begin{equation}
\left(\hat{\Delta}_{H_1}^{u,d}\right)_{ij}= -\frac{2v^3R'^2}{3}
\left[{F}_Q\hat{Y}_{u,d}\hat{Y}_{u,d}^\dagger
\hat{Y}_{u,d}{F}_{u,d}\right]_{ij}\,,\label{HiggsSpurion}
\end{equation}
\begin{equation}
\left(\hat{\Delta}_{H_2}^{u,d}\right)_{ij}= - R'^2\left[
\hat{m}_{u,d}\left(\hat{m}_{u,d}^\dagger
\hat{K}(c_q^L)+\hat{K}(c_{u_i,d_i})
\hat{m}_{u,d}^\dagger\right)\hat{m}_{u,d}\right]_{ij}
\label{HiggsSpurion1},\end{equation}
where $\hat{K}(c_i)={\mbox diag}(K(c_i))$ and
 \begin{equation}
K(c)=\frac{1}{1-2c}\left[-\frac{1}{\varepsilon^{2c-1}-1}+
\frac{\varepsilon^{2c-1}-\varepsilon^2}
{(\varepsilon^{2c-1}-1)(3-2c)}+\frac{\varepsilon^{1-2c}-
\varepsilon^2}{(1+2c)(\varepsilon^{2c-1}-1)}\right]\, ,
\label{KineticFactor}\end{equation}
with $(R')^{-1}=k\, e^{-k\pi R}\approx 1.8$ TeV the UV cutoff of
the 4D theory and $\varepsilon = R/R' = e^{-k\pi R}$. The function
$K(c)$  was obtained in \cite{HiggsFCNC} by taking the brane
localized Higgs limit $(\beta_H\rightarrow\infty)$ of the original
bulk function for one generation of down-type quarks. The $c$ dependence of $K(c)$
renders the shift from kinetic terms subdominant w.r.t. the one
arising from Yukawa interactions, in the case of  first and second
generation quarks that interest us.

\noindent To account for bulk effects in the three-generation case
we use the spurion-overlap approximation of
Sec.~\ref{Sec:3spurion}.  One can easily realize that the
general flavor structure of the shift in Eq.~(\ref{HiggsSpurion})
for the up and down sectors is identical to the one of the
down-type  contributions to dipole operators. This means that we
can rewrite Eq.~(\ref{HiggsSpurion}) in the form of
Eq.~(\ref{C7D}) with a new factor $B_{P_{u,d}}^H$ that quantifies
to a good approximation the overall effect of overlap corrections
present  in the second diagram of Fig.~\ref{insertion}. The
difference with the dipole operator case in
Eq.~(\ref{Spurion+Overlap}) stems from the different type of KK
modes: $d_L^{KK}$ and $q_R^{KK}$ in Fig.~\ref{insertion} denote
$(--)$ and  $(+-)$ KK states in custodial RS-A${}_4$. The flavor
structure of the dominant contribution to the misalignment
$\hat{\Delta_{H_1}^{u,d}}$ thus contains two terms 
\begin{eqnarray}
(\hat{\Delta_{H_1}^{d,u}})_{(++)}&\propto&
F_Q\hat{Y}_{d,u}\,r_{01}(c_{Q_i},c_{d_{\ell_1},u_{\ell_1}},\beta)\,
\hat{Y}_{d,u}^\dagger\,r_{1^-1^-}
(c_{d_{\ell_1},u_{\ell_1}},c_{Q_{\ell_2}},\beta)
\,\hat{Y}_{d,u}\,r_{10}(c_{Q_{\ell_2}},c_{d_j,u_j})F_{d,u}\nonumber\,,\\
(\hat{\Delta}_{H_1}^{d,u})_{(-+)}  &\propto&
F_Q\hat{Y}_{u,d}\,r_{01^{-+}}(c_{Q_i},c_{u_{\ell_1},d_{\ell_1}},\beta)\,
\hat{Y}_{u,d}^\dagger\,
r_{1^{+-}1}(c_{u_{\ell_1},d_{\ell_1}},c_{Q_{\ell_2}},\beta)\,
\hat{Y}_{d,u}\,r_{10}(c_{Q_{\ell_2}},c_{d_j,u_j},\beta)F_{d,u}\,.\nonumber\\
\label{Spurion+OverlapHiggs}\end{eqnarray}
Given the almost degeneracy of overlap factors as shown in appendix
\ref{app:A1}, we can again work in the approximation analogous to
Eq.~(\ref{C7D}), and define $B_{P_{u,d}}^H$
\begin{equation}B_{P_{u,d}}^H={\mbox
max}\left((\hat{r}_{00}^{u,d})^{-3}(\hat{r}_{01}^{u,d}
\hat{r}_{1^-1^-}^{u,d}+\hat{r}_{01^{-+}}^{u,d}
\hat{r}_{1^{+-}1^-}^{u,d})\,\hat{r}_{10}^{u,d}\right)
 \label{BPdefH}\end{equation}
as an overall multiplicative overlap factor. It is now important
to notice that the IR peaked profile of both $\Phi$ and the Higgs
and the vanishing of the $(--)$ and $(+-)$ profiles at the IR
brane, provide a suppression by almost an order of magnitude of
the overlap factors $r_{1^-1^-}$ and $r_{1^{+-}1^{-}}$ compared to
$r_{11}$ and $r_{1^{-+}1}$, rendering $\hat{\Delta}_{H_1}^{u,d}$
smaller, but still dominant over $\hat{\Delta}_{H_2}^{u,d}$. In
the same approximation of Eq.~(\ref{C7D}) and in the mass basis
$\hat{\Delta}_{H_1}^{u,d}$ reduces to
\begin{eqnarray}(\hat{\Delta}_{H_1}^{u,d})_{ij}
 &=& -\frac{2R'^{2}f_Q^2
m_{u_i,d_i}m^2_{u_j,d_j}B_{P_{u,d}}^H}{3}\sum_{n=1}^{3}
(V_R^{u,d})^*_{ni}(V_R^{u,d})_{nj}f_{u_n,d_n}^2\, .\label{HMDelta}
\end{eqnarray}
Disregarding $\hat{\Delta}_{H_2}^{u,d}$, the off-diagonal Yukawa
couplings in the mass basis are then obtained by dividing the
contribution in Eq.~(\ref{HMDelta}) by the Higgs VEV, $v$.
Recalling the structure of the right-handed diagonalization
matrices in Eq.~(\ref{VRQ}) and the hierarchy of $f_{u_i,d_i}$, it
is straightforward to identify  the dominant corrections. Defining
$(\Delta\hat{y}_{SM}^{u,d})_{ij}\equiv a_{ij}^{u,d}\equiv
(\hat{\Delta}_{H_1}^{u,d})_{ij}/v$ we obtain
\begin{equation}
a_{ij}^{u,d}\simeq\frac{-2R'^2f_Q^2
m_{u_i,d_i}m^2_{u_j,d_j}B_{P_{u,d}}^{H}}{3v}\left(\begin{array}{ccc}
f_{u,d}^2 &
 f_{u,d}^2\Delta_1^{u,d} & f_{u,d}^2\Delta_2^{u,d}\\
-f_{u,d}^2(\Delta_1^{u,d})^* &
f_{c,s}^2 & f_{c,s}^2\Delta_3^{u,d}\\
-f_{u,d}^2(\Delta_2^{u,d})^* & -f_{c,s}^2(\Delta_3^{u,d})^* &
f_{t,b}^2
\end{array}\right)_{ij}\, .\label{HMShift}
\end{equation}
Hence, new 4D effective operators of the form
$a_{ij}^{d}h\bar{d}^i_{L}\bar{d}^j_R + ({d\to u}) + {\mbox h.c.}$
will induce tree level Higgs mediated FCNC. One can already notice
that the suppression of third-generation couplings, as for
$\bar{t}t$, is much milder than in the flavor anarchic case
\cite{HiggsFCNC}. For $R'\simeq 1.8$ TeV, the suppression amounts
to $\Delta y_t/y_t\sim 4\times 10^{-3}$. This is due to the
degeneracy of the left-handed fermion profiles $f_Q$ and the
consequent factorization of the left-handed matrices $V_L$.
\subsection{Low energy physics bounds from $\Delta F=2$ processes}
\begin{figure}[t]
\centering
\includegraphics[width=10 truecm]{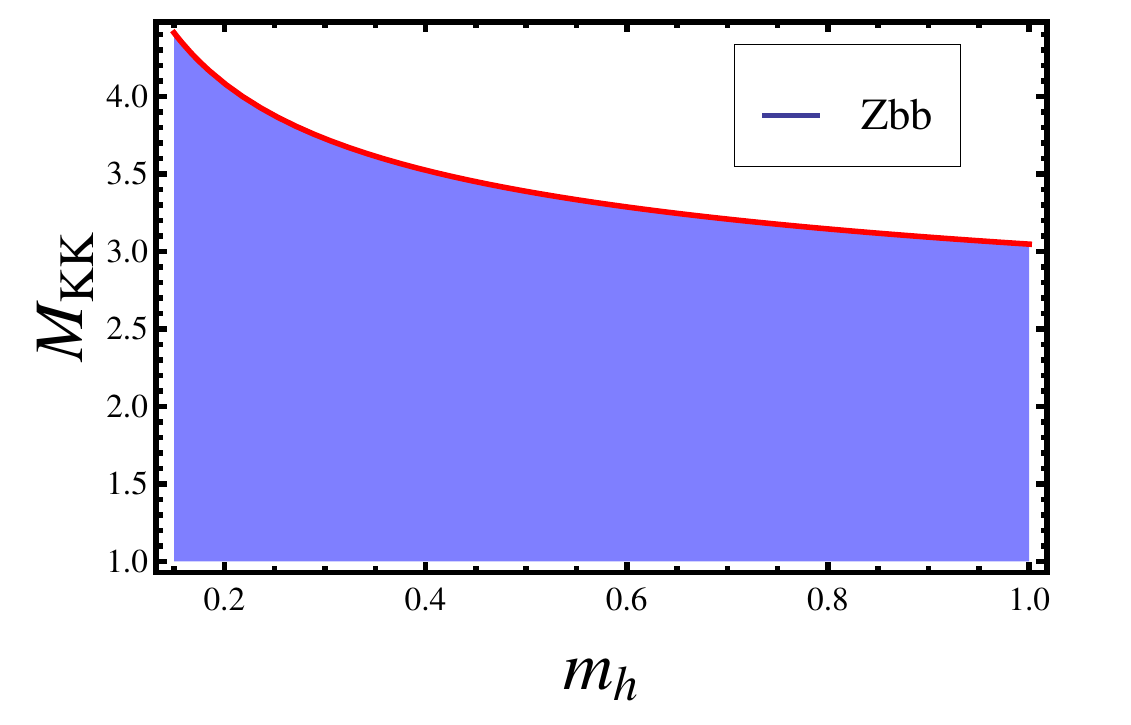}
\caption{Bound on the KK mass scale $M_{KK}$(TeV) from $Zb_L\bar{b}_L$
as a function of the Higgs mass, $m_h$(TeV). }\label{Plot:Bounds}
\end{figure}

The Higgs flavor violating couplings can induce tree level FCNC
contributions to various observables. The most stringent
constraints on their size may come from experimental bounds on
$\Delta F=2$ processes, such as $K-\bar{K}$,
$B_{d,s}-\bar{B}_{d,s}$ and $D-\bar{D}$ mixing. $\Delta F=2$
processes are described by the general effective Hamiltonian
\cite{Buras,UTfit}:
\begin{equation}\mathcal{H}^{\Delta F=2}_{eff}=\sum_{a=1}^{5}C_aO_a^{q_iq_j}+
\sum_{a=1}^{3}C'_aO_a^{'q_iq_j}
\label{DF2Ham},\end{equation}
where
\begin{eqnarray}
O_1^{q_iq_j}=\bar{q}_{jL}^{\alpha}\gamma_\mu
q_{iL}^\alpha\bar{q}_{jL}^\beta\gamma^\mu q_{iL}^\beta, &
O_2^{q_iq_j}=\bar{q}_{jR}^\alpha q_{iL}^\alpha\bar{q}_{jR}^\beta
q_{iL}^\beta, & O_3^{q_iq_j}=\bar{q}_{jR}^\alpha
q_{iL}^\beta\bar{q}_{jR}^\beta q_{iL}^\alpha \nonumber\\&
O_4^{q_iq_j}=\bar{q}_{jR}^\alpha q_{iL}^\alpha\bar{q}_{jL}^\beta
q_{iR}^\beta ,& O_5^{q_iq_j}=\bar{q}_{jR}^\alpha
q_{iL}^\beta\bar{q}_{jL}^\beta q_{iR}^\alpha\, , \label{DF2operators}
\end{eqnarray}
\begin{figure}[h]
\center
\includegraphics[width=9 truecm]{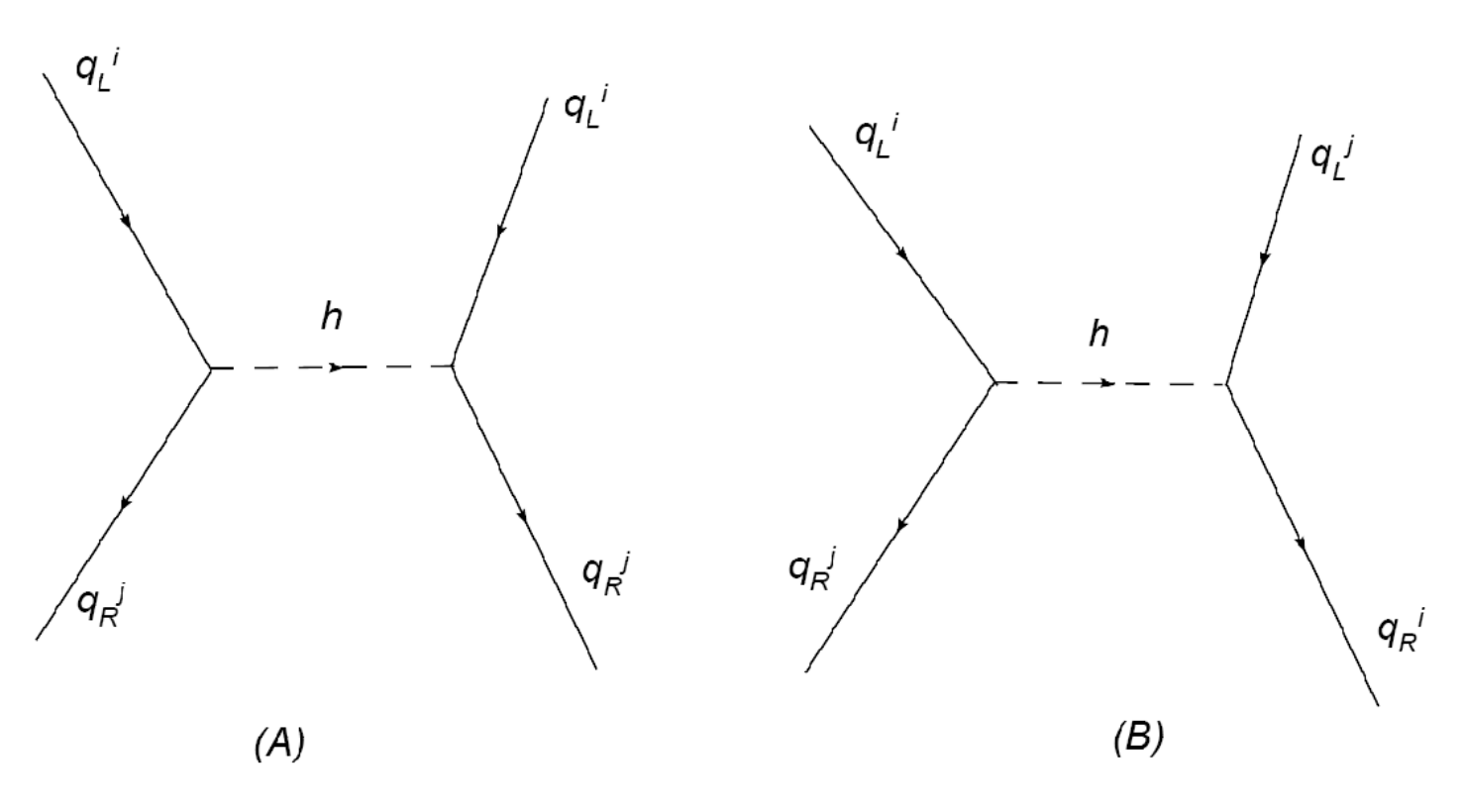}
\vspace{-.2cm} \caption{Contributions to $\Delta F = 2$ processes
from Higgs exchange at tree level.}\label{higgsexchange}
\end{figure}
with color indices $\alpha$, $\beta$. The opposite chirality
operators are denoted with a $'$ and obtained from the operators
above by the replacement $L\leftrightarrow R$. For $K-\bar{K}$,
$B_d-\bar{B}_d$, $B_s-\bar{B}_s$ and $D-\bar{D}$ mixing we have
$q_iq_j=sd, bd, bs$ and $uc$, respectively. In particular, Higgs
mediated tree-level processes as in Fig.~\ref{higgsexchange}
generate new contributions to $C_2$, $C_2'$
(Fig.~\ref{higgsexchange}(A)) and $C_4$
(Fig.~\ref{higgsexchange}(B)) \cite{HiggsFCNC}. They read as follows
\begin{equation}
C_2^h=\frac{a_{ij}^2}{m_h^2} \qquad
C_2^{\,'h}=\frac{a_{ji}^2}{m_h^2} \qquad
C_4^h=\frac{a_{ij}a_{ji}}{m_h^2}
\label{Hcoeeficients},\end{equation}
 where $m_h$ denotes the mass
of the physical Higgs. We  adopt the model independent
bounds of \cite{UTfit},  renormalized at the scale $\mu_h=200$
GeV as in \cite{HiggsFCNC}, to make the comparison with the
flavor anarchic results in \cite{HiggsFCNC} transparent. The bounds
\begin{eqnarray}
{\mbox Im}C_2^K\leq 2.04\times 10^{-16}GeV^{-2}, & {\mbox
Im}C_4^K\leq 5.9\times 10^{-17}GeV^{-2}, & |C_2^D|\leq 2.77\times
10^{-13}GeV^{-2}, \nonumber\\ |C_4^D|\leq 1.18\times
10^{-13}GeV^{-2}, & |C_2^{B_d}|\leq1.23\times10^{-12}GeV^{-2}, &
|C_4^{B_d}|\leq5.1\times10^{-13}GeV^{-2},\nonumber\\
|C_2^{B_s}|\leq 1\times10^{-10}GeV^{-2}, & |C_4^{B_s}|\leq
3.46\times 10^{-11}GeV^{-2}\label{HMbounds}
\end{eqnarray}
directly  constrain both $m_h$ and $a_{ij}$, and provide lower
bounds for the KK mass scale. The most stringent  bound in
RS-A${}_4$ should be  provided by ${\mbox Im}C_4^{K}$ and comes
from $\epsilon_K$. Using Eqs.~(\ref{Hcoeeficients}) and
(\ref{HMShift}) in the spurion-overlap approximation with an
overall overlap factor,  one obtains
\begin{equation}
{\mbox Im}(C_4^K)_{RS-A_4}\simeq{\mbox
Im}\left(\frac{(2B_P^{H_d}f_Q^2R'^2)^2}{m_h^2(3v)^2}f_d^4m_d^3m_s^3
\Delta_1^d\Delta_1^{d*}\right)=0
\end{equation}
 for any parameter assignment to first order in
$\tilde{x}_i^d$ and $\tilde{y}_i^d$. The next strongest
constraint should come from $C_2^K$. Using again
Eqs.~(\ref{Hcoeeficients}) and (\ref{HMShift}), assuming ${\mbox
Im} ((\Delta_1^d)^2)=({\mbox
max}(|\Delta_1^d|))^2=4\tilde{y}_D^2(f_\chi^d+f_\chi^s)^2$,
choosing $m_h=300$ GeV and $R'=(1.8\, {\mbox TeV})^{-1}$ as reference values,
the largest  contribution to $C_2^K$ in the same approximation is
\begin{equation}
{\mbox Im}(C_2^K)_{RS-{A_4}}\simeq{\mbox
Im}\left(\frac{(2B_P^{H_d}f_Q^2R'^2)^2}{m_h^2(3v)^2}f_d^4m_d^2m_s^4
\Delta_1^d\Delta_1^{d}\right)\simeq
1.4\times10^{-20}\,{\mbox GeV}^{-2}\left(\frac{Y^2\tilde{y}_D(1.8\,{\mbox TeV}R')^4}{(m_h/ 300\,{\mbox GeV})^2}
\right)\label{C2Kbound}\end{equation}
where we used $f_d=2.24\times10^4$, $f_Q=3.13$ and
$B_P^{H_d}\simeq0.18$. Thus, both constraints from $C_4$ and $C_2$
are strongly  suppressed  in our setup. This is again due to the A${}_4$ pattern of the Yukawa matrices. Suppression factors come from the mass ratios in
$V_R^{d}$,  due to the mass hierarchy
and the consequent hierarchy of right-handed fermion profiles in A${}_4$, the presence of $f_\chi^{d,s,b}$ suppression factors also in $V_L^{d}$, and the suppressed overlap of $(--)$ and $(+-)$
fermion KK modes with the IR peaked VEV of $\Phi$ and the bulk
Higgs. The same sources of suppression are at work  in the up
sector.
For completeness, we obtain the largest possible
estimation of the NP contribution to $C_2^D$, the most
constraining bound in the up sector. Assuming ${\mbox Im}
((\Delta_1^u)^2)=({\mbox
max}(|\Delta_1^u|))^2=4\tilde{y}_U^2(f_\chi^u+f_\chi^c)^2$, and
using the same reference values for $m_h$ and $R'$ as before, we obtain
\begin{equation}
|(C_2^D)|_{RS-{A_4}}=
\left|\frac{(2B_P^{H_u}f_Q^2R'^2)^2}{m_h^2(3v)^2}f_u^4m_u^2m_c^4\Delta_1^u
\Delta_1^{u}\right|\simeq
2.4\times10^{-18}{\mbox GeV}^{-2}\left(\frac{Y^2\tilde{y}_U(1.8\,{\mbox TeV}R')^4}{(m_h/300\,{\mbox GeV})^2}
\right),\label{C2Dbound}\end{equation}
where we  used $B_P^{H_{u,d}}=0.18$ from Eq.~(90). The above
contribution is almost six orders of magnitude suppressed compared
to the model independent bound in Eq.~(\ref{HMbounds}). Higher
order corrections in $\tilde{x}_i^{u,d}$ and $\tilde{y}_i^{u,d}$
will induce terms which are suppressed by at least
$O(f_\chi^{u_i,d_i})$ compared to the first order terms and they
will generally combine incoherently. Hence, it seems safe to
expect that  Higgs mediated FCNC contributions to $\Delta F=2$
processes do not provide the most stringent bounds on the KK mass
scale in the RS-A${}_4$ model, even going beyond the
spurion-overlap approximation and taking full account of
generational mixing. Far more stringent  constraint thus remains the
one coming from the $Zb_L\bar{b}_L$ coupling,
which is fairly satisfied for the choice $c_q^L=0.4$,
$R'^{-1}=1.8$ TeV, $m_h = 150$ GeV and order one Yukawa couplings.
Fig.~\ref{Plot:Bounds} also shows how the bound on the KK
mass $M_{KK}\simeq 2.55\,R'^{-1}$ becomes weaker upon increasing the Higgs mass. 
\section{Conclusions}
\label{sec:conclusion}

We have illustrated how the presence of an additional A${}_4$ flavor symmetry
 in the bulk of a warped two-brane scenario allows to relax the most stringent
lower bounds on the KK mass scale typical of flavor anarchic models. The most
 relevant difference between the RS-A${}_4$ model proposed in \cite{A4Warped}
and flavor anarchy stems from the degeneracy of the left-handed fermion bulk
 profiles, and the consequent factorization of the left-handed rotation matrices
 in many contributions to dipole operators.
The flavor hierarchy of the Standard Model is induced by the A${}_4$ texture of
 the 5D Yukawa couplings and the bulk mass parameters
of the right-handed fermions.

\noindent At leading order in an expansion in powers of the UV cutoff of
RS-A${}_4$, i.e. in the absence of cross-talk interactions \cite{A4Warped},
the CKM matrix is the unit matrix and no quark mixing is generated in the
effective 4D theory.
At next-to-leading order, an almost realistic CKM matrix is obtained in a
rather economical way, due to the presence of cross-talk higher-dimensional
operators and cross-brane interactions.
We have also shown in \cite{A4Warped} that the structure of the leading order
Yukawa couplings may induce exact cancellations in the contributions to dipole operators.
It is hence natural
to expect the suppression of many contributions to the same operators, once the next-to-leading order corrections to the Yukawa couplings are taken into account, as
compared with flavor anarchic descriptions.

\noindent It should also be noticed that whenever a flavor symmetry is
present in the 5D theory, it is important -- and more relevant than in
flavor anarchic models -- to fully account for non-degeneracies and KK
 mixing patterns within the same generation and among generations.
For this reason, we have considered various analytical approximations
and compared their prediction with an {\em exact} analysis based on a
 fully numerical diagonalization of the complete KK mass matrix, a
$12\times 12$ matrix in the custodial case for three generations.

\noindent Concerning flavor violating processes, the first relevant
difference with flavor anarchic models is the fact that  new physics contributions  are dominated by the same chirality operators as in the Standard Model and no enhancement of the opposite chirality operators is in place. Another relevant feature is the vanishing of the dominant
 new-physics contribution to $\epsilon_K$, mediated by a KK gluon exchange
 at tree level. This has striking consequences, due to the fact that this
contribution to $\epsilon_K$ is inversely proportional to the Yukawa scale,
 while other relevant observables such as $\epsilon'/\epsilon_K$, $b\to s\gamma$
and the neutron EDM are directly proportional to the Yukawa scale.
The consequence in flavor anarchic models \cite{IsidoriPLB} is that
the combined constraints from $\epsilon_K$, $\epsilon'/\epsilon_K$
 and $b\to s\gamma$
force large Yukawa couplings, closer to the perturbativity bound,
and an overall bound $M_{KK}\gtrsim 7.5$ TeV. In addition, the little
 CP problem \cite{Agashe:2004cp}, related to the generation of a far
 too large neutron EDM in flavor anarchy, remains to be solved.
In contrast, given the vanishing of the leading new physics contributions
 to $\epsilon_K$ in RS-A${}_4$, the most relevant constraints on the new
 physics scale should come from the remaining FCNC processes, while
relaxing the constraints on the size of the Yukawa couplings.

\noindent Fig.~\ref{PlotNumerical} is a summary of the most relevant
results for FCNC processes in RS-A${}_4$, expressed in terms of the
lower bounds on the KK mass scale $M_{KK}\simeq 2.55\, R'^{-1}$ and
by varying the typical size of Yukawa couplings. Given the absence of
the constraint from $\epsilon_K$, an $O(1)$ Yukawa coupling is allowed,
 providing the overall bound $M_{KK} \gtrsim 1.3$ TeV, induced by
$b\to s\gamma$. The latter bound is weaker than any flavor anarchic
bound and less stringent than the bound $M_{KK}\gtrsim 4.6$ TeV, from
$Zb_L\bar{b}_L$ in RS-A${}_4$ \cite{A4Warped}.
Another salient feature in Fig.~\ref{PlotNumerical} is the substantial
suppression of new physics contributions to the neutron EDM. This stems
from the A${}_4$-induced degeneracies in the left-handed fermion sector,
 which determine the vanishing of these contributions to the EDM also at
next-leading-order in the Yukawa couplings, in the
 spurion-overlap analysis within the mass insertion approximation.

\noindent The pattern of HMFCNC in RS-A${}_4$ shows a much milder suppression
 of the top Yukawa coupling if compared with flavor anarchic models, and more
in general the A${}_4$ flavor structure guarantees weak bounds on the KK mass
scale induced by Higgs mediated FCNC processes. We defer to future work the
study of potentially interesting features of an extended $P_{LR}$ custodial
symmetry \cite{Casagrande_ECust,Agashe:2006at} within a A${}_4$ warped flavor model. Such an additional symmetry is known \cite{Agashe:2006at} to relax the constraints from  $Zb_L\bar{b}_L$.

\noindent We conclude that the little CP problem related to the neutron EDM
in flavor anarchic models is avoided in the custodial RS-A${}_4$, while the
most stringent bounds on the KK mass scale come from EWPM, in particular
 the $Zb_L\bar{b}_L$ coupling. The dominance of the constraint induced by
$b\to s\gamma$ over the constraints from $\epsilon'/\epsilon_K$ and the
neutron EDM mainly stems from the amount of non-degeneracy of the
third-generation Yukawa coupling, in turn induced by the degeneracy of
the left-handed 5D profiles of all quarks in A${}_4$.

\vspace*{1.0truecm}
{\bf \noindent Acknowledgments}
\vspace*{0.2truecm}

\noindent We thank Aleksander Azatov, Yuval Grossman and Gilad Perez for useful discussions.
The work of A.K is supported in part by the Ubbo Emmius scholarship program at the University of Groningen.

\appendix

\section{Explicit Calculation of Overlap Corrections}
\label{app:A}
In this appendix we define and obtain explicitly the various
overlap correction factors introduced in Eq.~(\ref{lagrangian})
and discussed through the text. We start by some definitions. The
bulk geometry is a slice of $AdS_5$ compactified on an orbifold
$S_1/Z_2$ and can be described by the proper distance metric
\begin{equation}
ds^2=dy^2+e^{-2k|y|}\eta_{\mu\nu}dx^{\mu}dx^{\nu}\label{metric},
\end{equation}
where $k\simeq M_{Pl}$ is the $AdS_5$ curvature scale and
$-\pi R\leq y\leq\pi R$. The UV and IR branes are located at the
orbifold fixed points $y=0$ and $y=\pi R$, respectively.
The same problem is also studied by many authors in an interval setup with
 conformal coordinates. The corresponding metric is in this case
\begin{equation}
ds^2=\left(\frac{R}{z}\right)^2(-dz^2+\eta_{\mu\nu}dx^{\mu}dx^{\nu})\label{metricCFT}\,,
\end{equation}
where $z=e^{k y}/k$, defined on the interval $(R,R')$ with $R=z_h=1/k$ and $R'=z_v=e^{k\pi
R}/k$. One feature of the interval setup is that it
naturally allows for more general boundary conditions (BC) for the bulk fields, as
compared to the orbifold case. On the other hand, only the orbifold fixed points can be
 naturally interpreted as the location of physical branes due to source terms originating from the
``jump" of derivatives at the fixed points; in the
interval picture branes can only be assumed to be located at
the edges of the interval, namely $z_h$ and $z_v$.
Since in the orbifold case the behavior of all bulk fields in the
interval $[-\pi R,0]$ is determined by their transformation law
under the orbifold $Z_2$ symmetry, we normalize all wave functions and perform all integrals
on the interval $[0,\pi R]$ without loss of
generality.

\noindent The  normalized wave function for a fermion left-handed zero mode
as a function of its bulk mass is
\begin{equation}
\chi_0(c_{f_i},y)=\sqrt{\frac{ 2 k(1/2-c_{f_i})}{e^{\pi
kR(1-2c_{f_i})}-1}}e^{(2-c_{f_i})k|y|},\label{ZeroMode}\end{equation}
where $f_i=Q_i,\ell_i,u_i,d_i,e_i $ and $c_{f_i}$ is the
corresponding bulk mass given in units of $k$. A right-handed
zero mode is obtained with the replacement $c\rightarrow -c$.
The canonically normalized wave function on the IR brane,
$\hat{\chi}_0$, is defined as $\hat{\chi}_{0}(c_{f_i},\pi
R)=e^{(-3/2)k\pi R}\chi_{0}(c_{f_i},\pi R)$.

\noindent The induced 4D VEV's of the IR peaked bulk scalars $\Phi$ and
$H$,  required for the definition of the IR localized
Higgs limit, are related to the 5D VEV's as follows
\begin{equation}
H_0(\Phi_0)=\sqrt{\frac{2k(1+\beta_{H(\Phi)})}{1-e^{-k\pi
R(2+2\beta_{H(\Phi)})}}}v_{H(\Phi)}^{4D}e^{k\pi R}\simeq
\sqrt{2k(1+\beta_{H(\Phi)})}v_{H(\Phi)}^{4D}e^{k\pi R}\, ,
\end{equation}
 where $\beta_{H,\Phi}=\sqrt{4+\mu_{H,\Phi}^2}$ tunes the amount of localization
 from the UV to the IR and $\mu_{H,\Phi}$ is the corresponding bulk mass in units of $k$.
The above relation is obtained by integrating the solution (for $H$ or $\Phi$) of
 the bulk equation of motion along the extra dimension \cite{WiseScalar}.
In the IR localized Higgs and $\Phi$ case, the charged fermion masses arising
from the Yukawa coupling to the Higgs, and before diagonalization, are obtained
via~\cite{bulkfermion1}:
\begin{equation}
(\hat{m}_{f})_{IR}^{ij}=(\hat{y}_{ij}^{f})_{LO}\frac{ v v_{\Phi}^{4D}e^{k\pi R}}{k^2}
\hat{\chi}_0(c_{\ell_i}(c_{Q_i}),\pi R)
\hat{\chi_0}(c_{e_{R_j}}(c_{q_{R_j}}),\pi R)=
\frac{2(\hat{y}_{ij}^{f})_{LO}\,vv_\Phi^{4D}e^{k\pi
R}}{kf_{\ell_i,Q_i}f_{e_j,u_j(d_j)}} \label{IRMass},\end{equation}
where $q=u,d$, the matrix $(\hat{y}_{ij}^{f})_{LO}$ of dimensionless 5D couplings is defined in
Eq.~(\ref{LOYukawa}) and $v\equiv v_H^{4D}=174$ GeV is the Higgs VEV. In
the second equality we write the fermion masses in similar
notations to \cite{Agashe:2004cp}, where
$f_{f_i}=\sqrt{2k}/\hat{\chi}_{0_{f_i}}$ to make the comparison
with their results more transparent.
 In the setup we use, where
charged fermion masses are generated by the Yukawa interactions
with bulk $H$ and $\Phi$ and where LH fermion bulk masses are degenerate, we
have to consider the overlap of scalar VEV profiles and zero mode fermion
profiles, leading to the following masses before diagonalization:
\begin{equation}
(\hat{m}_f)_{Bulk}^{ij}=(\hat{y}_{ij}^{f})_{LO}\frac{H_0\Phi_0}{k^2}
\int_{0}^{\pi R} dy e^{-4k|y|} e^{(4+\beta_H+\beta_{\Phi})(k|y|-\pi
R)}\chi_0(c_{\ell_i}(c_{Q_i}),y)\chi_0(c_{e_{R_j}}(c_{q_{R_j}}),y)\, .
\label{BulkMass}\end{equation}
 As a natural choice for the bulk scalar profile, we
assume \cite{WiseScalar} $\beta_{H,\Phi}\simeq 2+\epsilon$, with
$\epsilon$ a small parameter for stabilisation purposes. To obtain the physical quark masses
at the scale $ke^{-k\pi R}\simeq1.8$ TeV  we
used the following assignment \cite{A4Warped}:
\begin{equation}
c_q^L=0.4,\quad c_u=0.79,\quad c_d=0.77,\quad c_s=0.683,\quad
c_c=0.602 ,\quad c_b=0.557 ,\quad c_t=-0.17\,,
\label{BulkAssignments}\end{equation}
 with $y_{u,c,d,s,b}=1$ and
$y_t\simeq2.8$, which is still required to match
$m_t(\mu=1.8 {\mbox TeV})\simeq 140$ GeV.
 The integration in Eq.~(\ref{BulkMass}) is straightforward, given that all functions
are simple exponentials and only depend on
$\beta=\beta_H+\beta_\Phi$. Dividing Eq.~(\ref{BulkMass}) by
Eq.~(\ref{IRMass}), we obtain the definition of the (LO) RS-A$_4$ zero-zero overlap
correction factors
\begin{equation}
(r_{00}^{H\Phi})_{ij}=
\frac{(\hat{m}_{f})_{Bulk}^{ij}}{(\hat{m}_f)_{IR}^{ij}}\simeq
   \frac{(2\sqrt{(1+\beta_H)(1+\beta_\Phi)} v_\Phi^{4D}ve^{2k\pi
R})}{(4+\beta_H+\beta_\Phi-c_q^L-c_{u_i,d_i})v_{\Phi}^{4D}
ve^{2k\pi R}}\simeq\frac{6}{8-c_q^L-c_{u_i,d_i}}
\label{r00},\end{equation}
where we used
$\beta_H=\beta_\Phi\simeq2$, $H_0=0.396M_{Pl}^{3/2}$,
$\Phi_0=0.577M_{Pl}^{3/2}$, $k\simeq M_{Pl}=2.44\times10^{18}TeV$ and
$k\pi R\simeq 34.8$. The numerical values of the bulk parameters in
 Eq.~(\ref{BulkAssignments})
lead to
\begin{equation}
r_{00}^{u,d}\simeq0.88\quad
r_{00}^{s}\simeq0.87\quad r_{00}^{c}\simeq0.86 \quad
r_{00}^b\simeq0.85\quad r_{00}^t\simeq0.77\, ,
\label{r00result}\end{equation}
with a rather mild flavor dependence.
In an analogous way we obtain $r_{00}^{H\Phi\chi}$, the overlaps associated
with the NLO Yukawa interactions of Eq.(\ref{NLOYukawa}), for
which the corresponding integration for $\beta_{\chi}=2$ was
already performed in \cite{A4Warped}:
\begin{eqnarray}
r_{00}^{H\Phi\chi}(\beta_{H,\Phi,\chi},c_q^L,c_{u_i,d_i})&\simeq&
\frac{4\sqrt{2}\beta_\chi\sqrt{(1+\beta_H)(1+\beta_{\Phi})(1+\beta_{\chi})}}
{(6+\beta_H+\beta_\Phi+\beta_\chi-c_q^L-c_{u_i,d_i})(6+\beta_H+
\beta_\Phi-\beta_\chi-c_q^L-c_{u_i,d_i})}
\nonumber\\&\simeq&
\frac{24\sqrt{6}}{(12-c_q^L-c_{u_i,d_i})(8-c_q^L-c_{u_i,d_i})}
\label{r00chi},\end{eqnarray}
 where in the second equality we used
$\beta_{H,\Phi ,\chi}=2+\epsilon_{H,\Phi,\chi}$ and
$\epsilon_{H,\Phi,\chi}<<1$.
Notice that the interactions with $\chi$ are vanishing
identically  on the IR brane, for $\beta_\chi=2$, due to the VEV
profile of $\chi$; thus the IR localized limit of Yukawa
interactions involving $\chi$ is naturally suppressed. The
correction $r_{00}^{H\Phi\chi}$, as defined in Eq.~(\ref{r00chi}), is just a way for us to
parametrize the bulk NLO Yukawa interactions in a  way similar
to the LO Yukawa interactions. The same goes for the
definition of the 4D VEV for the $\chi$ field, $\chi_0=v_{\chi}^{4D}e^{k\pi
R}\sqrt{(1+\beta_\chi)}$. The overlap correction factors from Eq.~(\ref{r00chi})
range from $(r_{00}^{H\Phi\chi})_t\simeq0.64$ to
$(r_{00}^{H\Phi\chi})_u\simeq 0.8$. Finally, the function
$f_\chi^{u_i,d_i}$, defined after Eq.~(\ref{NLOYukawa}) and
measuring the relative strength of NLO and LO Yukawa
interactions in the bulk case for generic $\beta_{H,\Phi,\chi}$,
is given by: \begin{equation}
f_\chi^{u_i,d_i}\equiv\frac{(v_\chi^{4D}e^{k\pi
R}/k)\,r_{00}^{H\Phi\chi}(c_q^L,c_{u_i,d_i},\beta_{H\Phi\chi})}{r_{00}^{H\Phi}
(c_q^L,c_{u_i,d_i},\beta_{H \Phi})}\simeq\frac{2\beta_\chi
C_\chi}{6+\beta_H+\beta_\Phi+\beta_\chi-c_q^L-c_{u_i,d_i}}.
\label{Fchi}\end{equation}
We then consider the overlap correction factors associated with the interaction of KK
and zero mode fermions. They enter at each Higgs vertex (and mass insertion)
 in the one-loop diagrams of Figs.~\ref{fcncLoop}  and \ref{fcncLoopcharged}.
The wave functions for the KK fermion modes are  \cite{bulkfermion1}:
\begin{equation}
\chi_n(c,y)=\frac{e^{5k|y|/2}}{N_n\sqrt{\pi R}}\left[J_\alpha
\left(m_n\frac{e^{k|y|}}{k}\right)+b_\alpha(m_n) Y_\alpha
\left(m_n\frac{e^{k|y|}}{k}\right)\right],\label{KKWaves}\end{equation}
where $\alpha=|c+1/2|$ and $\chi_n$ denotes the normalized wave
function of the level $n$ KK mode.
The coefficients $b_\alpha(m_n)$ and the mass spectrum $m_n$ are determined
 by the BC  imposed on the corresponding fermion. For $(++)$ BC, one obtains
 \cite{bulkfermion1,Agashe:2004cp}
\begin{equation}
-b_n=\frac{J_{\alpha -1}(m_n /k)}{Y_{\alpha -1}(m_n /k)}
=\frac{J_{\alpha -1}(m_ne^{k\pi R}/k)}{Y_{\alpha -1}(m_ne^{k\pi R}
/k)}\label{KKCoeff1}.\end{equation}
The coefficient
$\tilde{b}_n$, for the wave function of the $(-\,-)$ KK mode, is
obtained by the replacement $\alpha-1\rightarrow \alpha$, and the replacement
$c\rightarrow -c$ should also be made.
The  coefficient $b_n^\prime(m_n)$ for the $(-+)$ KK mode is instead given
by:
\begin{equation}
-b^{\prime}_n=\frac{J_{\alpha }(m_n /k)}{Y_{\alpha }(m_n /k)}
=\frac{J_{\alpha -1}(m_ne^{k\pi R}/k)}{Y_{\alpha -1}(m_ne^{k\pi R}
/k)}\, ,\label{KKCoeff2}\end{equation}
while the coefficient
$\tilde{b}^{\prime}_n(m_n)$, for the wave function of the $(+-)$
KK mode, is obtained by the replacement $\alpha\leftrightarrow
\alpha -1$.
The normalization factor $N_n$, for $(++)$ modes, is as follows
\begin{eqnarray}
(N_n^2)_{(++)}&=&\frac{1}{2k\pi R}\left[e^{2k\pi
R}\left(J_\alpha(m_n\frac{e^{k\pi
R}}{k})+b_\alpha(m_n)Y_\alpha(m_n\frac{e^{k\pi
R}}{k})\right)^2\right.\nonumber\\&&\left.-\left(J_\alpha(\frac{m_n}{k})
+b_\alpha(m_n)Y_\alpha(\frac{m_n}{k})\right)^2\right]\, ,\label{N++}\end{eqnarray}
and the one for $(--)$ KK modes is given by the replacement
$\alpha\rightarrow \alpha-1$. The normalization factor $N_n$, for
$(-+)$ KK modes, is instead
\begin{eqnarray}
(N_n^2)_{(-+)}&=&\frac{1}{2k\pi R}\left[e^{2k\pi
R}\left(J_\alpha(m_n\frac{e^{k\pi
R}}{k})+b_\alpha(m_n)Y_\alpha(m_n\frac{e^{k\pi
R}}{k})\right)^2\nonumber\right.\\&&\left.
-\left(J_{\alpha-1}(\frac{m_n}{k})
+b_\alpha(m_n)Y_{\alpha-1}(\frac{m_n}{k})\right)^2\right]\, ,\label{N-+}\end{eqnarray}
and the one for $(+-)$ KK modes is given by the replacement
$\alpha\leftrightarrow\alpha-1$.
 For all KK modes, in the limit $m_n<<k$ and $kR>>1$, the normalization
factor is well approximated by
\begin{equation}
N_n\simeq\frac{e^{k\pi R}}{\sqrt{2k\pi R}}J_\alpha
\left(m_n\frac{e^{k\pi R}}{k}\right)\simeq \frac{e^{k\pi
R/2}}{\sqrt{\pi^2Rm_n}}\, .\label{KK:Norm}
\end{equation}
In this way the value of all KK modes on the IR brane is
approximately $\sqrt{2k}$, as also in \cite{IsidoriPLB} and others.
The above definitions of the fermionic KK normalization constants
are needed when writing the 4D Lagrangian of Eq.~(\ref{lagrangian}) in terms of
  the Yukawa couplings in Eqs.~(\ref{LOYukawa}) and~(\ref{NLOYukawa}).

\noindent The overlap correction factors for the KK modes in Eq.~(\ref{lagrangian})
are thus defined as follows
\begin{equation}
r_{nm}^{H\Phi}(c_n,c_m,y)=\frac{\int_{0}^{\pi R}
dy\sqrt{-g}\chi_n(c_n,y)\chi_m(c_m,y)(H_0\Phi_0/k^2)e^{(4+\beta_H+\beta_\Phi)k(|y|-\pi
R)}}{\chi_n(c_n,\pi R)\chi_m(c_m,\pi R)(vv_\Phi^{4D})/(k^2e^{2k\pi
R})}\, , \label{Def:Overlap}\end{equation}
where $n,m=0, 1, 1^{-+},
1^{-}, 1^{+-}, 2,...$ denote the KK states. In the following, we
will only consider the effects of the first KK level, thus taking
$n,m=0,1,1^{-+},1^{-},1^{+-}$.
Notice also that  the overlap integral of Eq.~(\ref{Def:Overlap}) with two bulk scalar fields is
equivalent to the overlap integral of a single bulk Higgs field with
$\beta=2+\beta_{H}+\beta_{\Phi}$, rescaled by a $\mathcal{O}(1)$
correction factor, $R_{H\Phi}$
\begin{equation}
R_{H\Phi}=\frac{H_0\Phi_0/k^2}{(vv_\Phi^{4D}e^{2k\pi
R}/k)\sqrt{2(3+\beta_H+\beta_\Phi)}}
\simeq\frac{2\sqrt{(1+\beta_H)}\sqrt{(1+\beta_\Phi)}}{\sqrt{2(1+\beta_H
+\beta_\Phi)}}\simeq 1.6\, . \label{HiggsRatio}
\end{equation}
All overlap factors can eventually be rewritten in terms of $R_{H\Phi}$, in order
to make a direct comparison with the case of a single bulk scalar field, the Higgs,
and no flavon fields.

\noindent Since $c_q^L$ is strongly constrained by electroweak
precision measurements \cite{A4Warped}, and $H$ and $\Phi$ are
exponentially peaked towards the IR brane, the $c_{u_i,d_i}$ dependence
of the various overlap corrections in Eq.~(\ref{Def:Overlap}) is mild. In addition, the
continuous $(-\,-)$ and $(+\,-)$ wave functions vanish at the IR
brane, thus further suppressing the corresponding overlap corrections.
 It is also important to notice that Eqs.~(\ref{KKCoeff1})\,--\,(\ref{N-+}) imply
 that the $(-+)$ modes imitate the $(++)$ modes, while the
$(+-)$ modes imitate the $(-\,-)$ modes. The same behavior should be reflected in
the corresponding overlap correction factors.
\subsection{Numerical results for the overlap correction factors}
\label{app:A1}
We calculate the overlap
integrals in Eq.~(\ref{Def:Overlap}) numerically for the first level KK
modes and for the bulk masses assignments in \cite{A4Warped} and
Eq.~(\ref{BulkAssignments}). In the following $n,m=0,1$ and
we define: \begin{equation} r_{n(n^-)m(m^-)}^{u_i,d_i}\equiv
r_{n(n^-)m(m^-)}^{H\Phi}(c_q^L,c_{u_i,d_i},\beta) \qquad
r_{nm^{-+}}^{u_i,d_i}\equiv
r_{nm^{-+}}^{H\Phi}(c_q^L,c_{d_i,u_i},\beta)\label{indices1}.\end{equation}
Similarly we also define: \begin{equation}
r_{n^-m^{+-}}^{u_i,d_i}\equiv
r_{n^-m^{+-}}^{H\Phi}(c_{d_i,u_i},c_q^L,\beta).\label{indices2}\end{equation}
The $(++)$ and $(-+)$ KK-KK overlap corrections are given by:
\begin{eqnarray}
r_{11}^{u,d}\simeq r_{11^{-+}}^{u,d}\simeq 0.747 &
r_{11}^{s}\simeq r_{11^{-+}}^s\simeq 0.744 & r_{11}^{c}\simeq
r_{11^{-+}}^c\simeq 0.740 \nonumber\\r_{11}^{b}\simeq
r_{11^{-+}}^b\simeq0.738 & r_{11}^{t}\simeq 0.645 &
r_{11^{-+}}^t\simeq 0.708.\label{r11Results}\end{eqnarray}
The $(-\,-)$ and $(+\,-)$ KK-KK overlap corrections are given by:
\begin{eqnarray}
r_{1^-1^-}^{u,d}\simeq 0.096 & r_{1^-1^{+-}}^{u,d}\simeq 0.070 &
r_{1^-1^-}^s\simeq 0.093 \nonumber\\
r_{1^-1^{+-}}^{s}\simeq 0.075
& r_{1^-1^-}^{c}\simeq 0.090 & r_{1^-1^{-+}}^c\simeq 0.080  \nonumber\\
r_{1^-1^-}^{b}\simeq 0.087 & r_{1^-1^{-+}}^b\simeq 0.082 &
r_{1^-1^-}^t\simeq 0.112 \nonumber\\
r_{1^-1^{+-}}^t\simeq 0.048\,.\label{r222Results}\end{eqnarray}
The $(++)$ KK-zero and $(-\,+)$ KK-zero  overlap corrections
$r_{01}^{u_i,d_i}$ are given by: \begin{eqnarray}
r_{01}^{u,d}\simeq r_{01^{-+}}^{u,d}\simeq 0.800 &
r_{01}^{s}\simeq r_{01^{-+}}^s\simeq 0.794 & r_{01}^{c}\simeq
r_{01^{-+}}^c\simeq 0.790 \nonumber\\
r_{01}^{b}\simeq r_{01^{-+}}^b\simeq0.780 & r_{01}^{t}\simeq 0.670 &
r_{01^{-+}}^t\simeq 0.755\, .\label{r01Results}\end{eqnarray}
The zero-KK $(++)$ and zero-KK $(-\,+)$  overlap corrections
$r_{10}^{u_i,d_i}$ are given by:
\begin{eqnarray}
r_{10}^{u,d}\simeq 0.806 & r_{1^{-+}0}^{u,d}\simeq 0.822&
r_{10}^{s}\simeq 0.795 \,\,\,\, r_{1^{-+}0}^s\simeq 0.811 \nonumber\\
r_{10}^{c}\simeq 0.790 & r_{1^{-+}0}^c\simeq 0.803 &
r_{10}^{b}\simeq 0.784 \nonumber\\ r_{1^{-+}0}^b\simeq0.798&
r_{10}^{t}\simeq 0.720 & r_{1^{-+}0}^t\simeq 0.730\, .
\label{r10Results}\end{eqnarray}
Using Eqs.~(\ref{r00result})  and
(\ref{r11Results})\,--\,(\ref{r10Results}) we obtain the
coefficients $B_P^{u,d}$ to be used in the  spurion-overlap formula in Eq.~(\ref{BPdef}),
\begin{equation}
B_P^u=B_P^d\simeq 1.5.\label{BPResult}
\end{equation}
Notice that, while $B_P^{u,d}$ is larger than one, each
independent overlap correction factor is always smaller than one in
magnitude and approaches one for IR localized $H$ and $\Phi$ fields.
\section{Diagonalization of the KK mass matrices}
\label{app:B}
We provide more details of the diagonalization
procedure described in Sec.~\ref{Sec:MassDiag}, starting from the
one-generation case, and then considering three generations.
We first specify the KK mass spectrum
corresponding to the bulk parameters assignment in
Eq.~(\ref{BulkAssignments}).  Masses are obtained by solving
Eqs.~(\ref{KKCoeff1}) and (\ref{KKCoeff2}) numerically. The common
left-handed bulk mass parameter $c_q^L$ determines the mass of all LH
$(++)$  KK modes, $Q_L^{(1)_{u_i,d_i}}$, providing $M^{KK}_{Q_L}\equiv
M_{KK}\simeq2.55\,ke^{-k\pi R}$. The rest of the KK mass spectrum
for the down-sector, in units of $R'^{-1}=ke^{-k\pi R}$ and omitting the label KK
to ease the notation, is
\begin{equation}\begin{array}{ccc}
M_{d_R^{(1)}}=2.8 & M_{\tilde{d}_R^{(1)^{-+}}}=2.8 &
M_{s_R^{(1)}}=2.75 \\ M_{\tilde{s}_R^{(1)^{-+}}}=2.55 &
M_{b_R^{(1)}}=2.5 & M_{\tilde{b}_R^{(1)^{-+}}}=1.23\, ,
\end{array}\label{KKDspectrum}\end{equation}
while the mass
spectrum in the up sector is given by:
\begin{equation}\begin{array}{ccc}
M_{u_R^{(1)}}=2.8 & M_{\tilde{u}_R^{(1)^{-+}}}=2.8 &
M_{c_R^{(1)}}=2.56 \\ M_{\tilde{c}_R^{(1)^{-+}}}=2.7 &
M_{t_R^{(1)}}=3.4 & M_{\tilde{t}_R^{(1)^{-+}}}=2.5\,.
\end{array}\label{KKUspectrum}\end{equation}
Considering the above numerical values, we are going to  treat two of the
three KK modes for each generation as almost degenerate, when
diagonalizing the one-generation mass matrices to $\mathcal{O}(x)$. Naively, this
approximation may cease to be a good one for differences in KK masses of
$\mathcal{O}(0.2ke^{-k\pi R})$, which happens to be characteristic of the first two
generations since their masses are only three times larger than
$x=v/M_{KK}$. For this reason, we kept track of $\mathcal{O}(x^2)$
terms in the perturbative diagonalization process, and verified a posteriori that
they are sufficiently suppressed and can be neglected within the
$\mathcal{O}(x)$ approximation.

\subsection{Diagonalization of the one-generation mass matrices}
\label{app:B1}
\noindent Starting from the down sector KK mass matrix in Eq.~(\ref{M4KK}),
and using Eq.~(\ref{KKDspectrum}) we realize that we have to
perform a $\pi/4$ rotation in the $(2,4)$ plane when obtaining
$O_{L}^{d_{KK}}$ and a $\pi/4$ rotation in the $(2,3)$ plane, when
obtaining $O_{R}^{d_{KK}}$. These rotations act on
$\hat{\bf{M}}_d^{KK}(\hat{\bf{M}}_d^{KK})^\dagger$ for
$O_{L}^{d_{KK}}$ and
$(\hat{\bf{M}}_d^{KK})^{\dagger}\hat{\bf{M}}_d^{KK}$ for
$O_{R}^{d_{KK}}$. Once these rotations are performed and the
corresponding degenerate subspaces have been diagonalized, the
resulting matrix can be diagonalized using non degenerate
perturbation theory. The same holds for all the one-generation mass matrices, where
the main differences reside in
the degenerate subspaces. The resulting diagonalization matrix for
the down sector of the first generation is
\begin{equation}\noindent \footnotesize{O_L^{d_{KK}}=\phantom{bbbbbbbbbbbbbbbbbbbbbbbbbbbbbbb
bbbbbbbbbbbbbbbbbbbbbbbbbbbbbbbbbbbbbbbbbbbbbbbbbbbbbbbbb}}
\nonumber\end{equation}
\begin{equation}
\scriptsize{\left(\begin{array}{cccc} 1&
0.64f_Q^{-1}(r_{01}\breve{y}_d-r_{101}\breve{y}_u)x &
\mathcal{O}(x^2) &
0.64f_Q^{-1}(r_{01}\breve{y}_d+r_{101}\breve{y}_u)x\\
-0.91f_Q^{-1}r_{01}\breve{y}_d^*x & \frac{1}{\sqrt{2}} &
(-5.24r_{11}-4.76r_{111})\breve{y}_d^*x & \frac{1}{\sqrt{2}}\\
\mathcal{O}(x^2)& (3.7(r_{11}-r_{111})\breve{y}_u
+3.37(r_{22}-r_{222})\breve{y}_d)x & 1 &
(3.7(r_{11}+r_{111})\breve{y}_u
+3.37(r_{22}+r_{222})\breve{y}_d)x\\
-0.91f_Q^{-1}r_{101}\breve{y}_u^*x & \frac{-1}{\sqrt{2}} &
(-5.24r_{11}-4.76r_{111})\breve{y}_u^*x & \frac{1}{\sqrt{2}}\\
\end{array}\right)}\label{OLdKK}
\end{equation}
Since the KK mass spectrum in the up sector of the first generation is substantially
identical to the one in the down sector $(c_d=0.77,c_u=0.79)$, the
matrix $O_L^{u_{KK}}$ can be obtained from the above equation by
the replacement $\breve{y}_u\leftrightarrow \breve{y}_d$. For the right
diagonalization matrix, we obtain:
\begin{equation}\noindent \footnotesize{O_R^{d_{KK}}=\phantom{bbbbbbbbbbbbbbbbbbbbbbbbbbbbbbb
bbbbbbbbbbbbbbbbbbbbbbbbbbbbbbbbbbbbbbbbbbbbbbbbbbbbbbbbb}}
\nonumber\end{equation}
\begin{equation}
\scriptsize{\left(\begin{array}{cccc} 1&
f_d^{-1}r_{10}\breve{y}_dx & \mathcal{O}(x^2) & \mathcal{O}(x^2)
\\
-f_d^{-1}r_{10}\breve{y}_dx & 1 & 3.37(r_{11}+r_{22})\breve{y}_dx
-3.7(r_{111}+r_{222})\breve{y}_ux&
3.37(r_{11}+r_{22})\breve{y}_dx +(3.7(r_{111}+r_{222})\breve{y}_ux\\
\mathcal{O}(x^2) & -(4.76r_{11}+5.24r_{22})\breve{y}_d^*x &
\frac{1}{\sqrt{2}} & \frac{1}{\sqrt{2}}\\ \mathcal{O}(x^2)&
-(4.76r_{111}+5.24r_{222})\breve{y}_u^*x & \frac{-1}{\sqrt{2}} &
\frac{1}{\sqrt{2}}
\end{array}\right)}\label{ORdKK}
\end{equation}
where, once again, $O_R^{u_{KK}}$ is obtained with the replacement
$\breve{y}_u\leftrightarrow\breve{y}_d$.
 Next, we turn to the $s$-quark sector, which due to the ``fake"
$SU(2)_R$ partner of the $c_R$ quark, $\tilde{s}_R^{(1)^{-+}}$,
shows the same KK mass degeneracy pattern as the one in the $t$-quark sector. Then $O_L^{s_{KK}}$ is given by:
\begin{equation}\noindent \footnotesize{O_L^{s_{KK}}=\phantom{bbbbbbbbbbbbbbbbbbbbbbbbbbbbbbb
bbbbbbbbbbbbbbbbbbbbbbbbbbbbbbbbbbbbbbbbbbbbbbbbbbbbbbbbb}}
\nonumber\end{equation}
\begin{equation}
\footnotesize{\left(\begin{array}{cccc} 1&
0.94f_Q^{-1}r_{01}\breve{y}_sx&
\frac{f_Q^{-1}}{\sqrt{2}}r_{101}\breve{y}_cx &
\frac{f_Q^{-1}}{\sqrt{2}}r_{101}\breve{y}_cx\\
-0.94f_Q^{-1}r_{01}\breve{y}_sx & 1 &
(6.06r_{11}+4.72r_{22})e^{i\theta_c}\breve{y}_s^*x & -(6.06r_{11}+4.72r_{22})
e^{i\theta_c}\breve{y}_s^*x\\
\mathcal{O}(x^2)& (8.6r_{11}+8.1r_{22})\breve{y}_sx
 &
-\frac{e^{i\theta_c}}{\sqrt{2}}+\frac{(r_{111}^2-r_{222}^2)\breve{y}_c+r_{11}^2
(|\breve{y}_s|^2/\breve{y}_c^*)
} {4\sqrt{2}(r_{111}+r_{222})}x &
\frac{e^{i\theta_c}}{\sqrt{2}}+\frac{(r_{111}^2-r_{222}^2)
\breve{y}_c+r_{11}^2(|\breve{y}_s|^2/\breve{y}_c^*)
} {4\sqrt{2}(r_{111}+r_{222})}x\\
-f_Q^{-1}r_{101}\breve{y}_c^*x & \mathcal{O}(x^2) &
\frac{1}{\sqrt{2}}+\frac{(r_{111}^2-r_{222}^2)\breve{y}_c+r_{11}^2
(|\breve{y}_s|^2/|\breve{y}_c|)
} {4\sqrt{2}(r_{111}+r_{222})}x &
\frac{1}{\sqrt{2}}-\frac{(r_{111}^2-r_{222}^2)\breve{y}_c+r_{11}^2
(|\breve{y}_s|^2/|\breve{y}_c|)
} {4\sqrt{2}(r_{111}+r_{222})}x
\end{array}\right)}\label{OLsKK}
\end{equation}
where $\theta_c$ denotes the phase of $\breve{y}_c$. The right
diagonalization matrix, $O_R^{s_{KK}}$, is instead
\begin{equation}\noindent \footnotesize{O_R^{s_{KK}}=\phantom{bbbbbbbbbbbbbbbbbbbbbbbbbbbbbbb
bbbbbbbbbbbbbbbbbbbbbbbbbbbbbbbbbbbbbbbbbbbbbbbbbbbbbbbbb}}
\nonumber\end{equation}
\begin{equation}
\footnotesize{\left(\begin{array}{cccc} 1&
-\frac{f_s^{-1}}{\sqrt{2}}r_{10}e^{i\theta_c}\breve{y}_s^*x &
\mathcal{O}(x^2)
&\frac{f_s^{-1}}{\sqrt{2}}r_{10}e^{i\theta_c}\breve{y}_s^*x
\\
-f_s^{-1}r_{10}\breve{y}_sx &
-\frac{e^{i\theta_c}}{\sqrt{2}}+\frac{(r_{222}^2-r_{111}^2)
\breve{y}_c+r_{22}^2(|\breve{y}_s|^2/\breve{y}_c^*)
} {4\sqrt{2}(r_{111}+r_{222})}x & (8r_{11}+8.5r_{22})\breve{y}_sx&
\frac{e^{i\theta_c}}{\sqrt{2}}+\frac{(r_{222}^2-r_{111}^2)
\breve{y}_c+r_{22}^2(|\breve{y}_s|^2/\breve{y}_c^*)
} {4\sqrt{2}(r_{111}+r_{222})}x\\
\mathcal{O}(x^2) &(5.7r_{11}+6r_{22})e^{i\theta_c}\breve{y}_s^*x &
1 &
 -(5.7r_{11}+6r_{22})e^{i\theta_c}\breve{y}_s^*x\\ \mathcal{O}(x^2)&
\frac{1}{\sqrt{2}}+\frac{(r_{222}^2-r_{111}^2)\breve{y}_c+r_{22}^2
(|\breve{y}_s|^2/|\breve{y}_c|)
} {4\sqrt{2}(r_{111}+r_{222})}\,x& \mathcal{O}(x^2) &
\frac{1}{\sqrt{2}}-\frac{(r_{222}^2-r_{111}^2)\breve{y}_c+r_{22}^2
(|\breve{y}_s|^2/|\breve{y}_c|)
} {4\sqrt{2}(r_{111}+r_{222})}x
\end{array}\right)}\label{ORsKK}
\end{equation}
The diagonalization matrices for the $t$-quark sector, similar
in structure to the matrices in the $s$-sector, are as follows
\begin{equation}\noindent \footnotesize{O_L^{t_{KK}}=\phantom{bbbbbbbbbbbbbbbbbbbbbbbbbbbbbbb
bbbbbbbbbbbbbbbbbbbbbbbbbbbbbbbbbbbbbbbbbbbbbbbbbbbbbbbbb}}
\nonumber\end{equation}
\begin{equation}
\tiny{\left(\begin{array}{cccc} 1& 0.75f_Q^{-1}r_{01}\breve{y}_tx&
\frac{f_Q^{-1}}{\sqrt{2}}r_{101}\breve{y}_bx &
\frac{f_Q^{-1}}{\sqrt{2}}r_{101}\breve{y}_bx\\
-0.75188f_Q^{-1}r_{01}\breve{y}_tx & 1 &
(1.22r_{11}+0.92r_{22})e^{i\theta_b}\breve{y}_t^*x & -
(1.22r_{11}+0.92r_{22})e^{i\theta_b}\breve{y}_t^*x\\
\mathcal{O}(x^2)& (1.73(r_{11}+1.3r_{22})\breve{y}_tx
 &
-\frac{e^{i\theta_b}}{\sqrt{2}}+\frac{(r_{111}^2-r_{222}^2)\breve{y}_b+
(f_t^2r_{10}^2+r_{11}^2)(|\breve{y}_t|^2/\breve{y}_b^*)
} {4\sqrt{2}(r_{111}+r_{222})}x &
\frac{e^{i\theta_b}}{\sqrt{2}}+\frac{(r_{111}^2-r_{222}^2)\breve{y}_b+
(f_t^2r_{10}^2+r_{11}^2)(|\breve{y}_t|^2/\breve{y}_b^*)
} {4\sqrt{2}(r_{111}+r_{222})}x\\
-f_Q^{-1}r_{101}\breve{y}_b^*x & \mathcal{O}(x^2) &
\frac{1}{\sqrt{2}}+\frac{(r_{111}^2-r_{222}^2)\breve{y}_b+(f_t^2r_{10}^2+
r_{11}^2)(|\breve{y}_t|^2/|\breve{y}_b|)
} {4\sqrt{2}(r_{111}+r_{222})}x &
\frac{1}{\sqrt{2}}-\frac{(r_{111}^2-r_{222}^2)\breve{y}_b+(f_t^2r_{10}^2+
r_{11}^2)(|\breve{y}_t|^2/|\breve{y}_b|)
} {4\sqrt{2}(r_{111}+r_{222})}x
\end{array}\right)}\label{OLtKK}
\end{equation}
where $\theta_b$ denotes the phase of $\breve{y}_b$, and
\begin{equation}\noindent \footnotesize{O_R^{t_{KK}}=\phantom{bbbbbbbbbbbbbbbbbbbbbbbbbbbbbbb
bbbbbbbbbbbbbbbbbbbbbbbbbbbbbbbbbbbbbbbbbbbbbbbbbbbbbbbbb}}
\nonumber\end{equation}
\begin{equation}
\footnotesize{\left(\begin{array}{cccc} 1&
-\frac{f_t^{-1}}{\sqrt{2}}r_{10}e^{i\theta_b}\breve{y}_t^*x &
\mathcal{O}(x^2)
&\frac{f_t^{-1}}{\sqrt{2}}r_{10}e^{i\theta_b}\breve{y}_t^*x
\\
-f_t^{-1}r_{10}\breve{y}_tx &
-\frac{e^{i\theta_b}}{\sqrt{2}}+\frac{(r_{222}^2-r_{111}^2)
\breve{y}_b+r_{22}^2(|\breve{y}_t|^2/\breve{y}_b^*)
} {4\sqrt{2}(r_{111}+r_{222})}x &
(1.33r_{11}+1.8r_{22})\breve{y}_tx&
\frac{e^{i\theta_b}}{\sqrt{2}}+\frac{(r_{222}^2-r_{111}^2)
\breve{y}_b+r_{22}^2(|\breve{y}_t|^2/\breve{y}_b^*)
} {4\sqrt{2}(r_{111}+r_{222})}x\\
\mathcal{O}(x^2)
&(0.92r_{11}+1.22r_{22})e^{i\theta_b}\breve{y}_t^*x & 1 &
 -(0.92r_{11}+1.22r_{22})e^{i\theta_b}\breve{y}_t^*x\\ \mathcal{O}(x^2)&
\frac{1}{\sqrt{2}}+\frac{(r_{222}^2-r_{111}^2)\breve{y}_b+r_{22}^2
(|\breve{y}_t|^2/|\breve{y}_b|)
} {4\sqrt{2}(r_{111}+r_{222})}\,x& \mathcal{O}(x^2) &
\frac{1}{\sqrt{2}}-\frac{(r_{222}^2-r_{111}^2)\breve{y}_b+r_{22}^2
(|\breve{y}_t|^2/|\breve{y}_b|)
} {4\sqrt{2}(r_{111}+r_{222})}x
\end{array}\right)}\label{ORtKK}
\end{equation}
Finally, we consider the $b$-quark sector and the analogous $c$-quark sector.
Starting with $O_L^{b_{KK}}$, we obtain:
\begin{equation}\noindent \footnotesize{O_L^{b_{KK}}=\phantom{bbbbbbbbbbbbbbbbbbbbbbbbbbbbbbb
bbbbbbbbbbbbbbbbbbbbbbbbbbbbbbbbbbbbbbbbbbbbbbbbbbbbbbbbb}}
\nonumber\end{equation}
\begin{equation}
\footnotesize{\left(\begin{array}{cccc} 1&
-\frac{f_Q^{-1}}{\sqrt{2}}r_{01}|\breve{y}_b|x&
\frac{f_Q^{-1}}{\sqrt{2}}r_{01}|\breve{y}_b|x &
2f_Q^{-1}r_{101}\breve{y}_tx\\
-f_Q^{-1}r_{01}\breve{y}_b^*x &
-\frac{e^{-i\theta_b}}{\sqrt{2}}-\frac{(r_{11}^2-r_{22}^2)
\breve{y}_b^*+r_{111}^2(|\breve{y}_t|^2/\breve{y}_b)
} {4\sqrt{2}(r_{111}+r_{222})}x  &
\frac{e^{-i\theta_b}}{\sqrt{2}}-\frac{(r_{11}^2-r_{22}^2)
\breve{y}_b^*+r_{111}^2(|\breve{y}_t|^2/\breve{y}_b)
} {4\sqrt{2}(r_{111}+r_{222})}x & \mathcal{O}(x^2)\\
\mathcal{O}(x^2)&
\frac{1}{\sqrt{2}}-\frac{(r_{11}^2-r_{22}^2)|\breve{y}_b|+r_{111}^2
(|\breve{y}_t|^2/|\breve{y}_b|)
} {4\sqrt{2}(r_{111}+r_{222})}x
 &
\frac{1}{\sqrt{2}}+\frac{(r_{11}^2-r_{22}^2)|\breve{y}_b|+r_{111}^2
(|\breve{y}_t|^2/|\breve{y}_b|)
} {4\sqrt{2}(r_{111}+r_{222})}x &
-\frac{2}{3}(r_{111}y_t+2r_{222}y_t)x\\
-2f_Q^{-1}r_{101}\breve{y}_t^*x & 0.47(r_{111}+2r_{222})y_t^*x &
0.47(r_{111}+2r_{222})y_t^*x  & 1
\end{array}\right)}\label{OLbKK}
\end{equation}
and for the right-handed diagonalization matrix
\begin{equation}\noindent \footnotesize{O_R^{b_{KK}}=\phantom{bbbbbbbbbbbbbbbbbbbbbbbbbbbbbbb
bbbbbbbbbbbbbbbbbbbbbbbbbbbbbbbbbbbbbbbbbbbbbbbbbbbbbbbbb}}
\nonumber\end{equation}
\begin{equation}
\footnotesize{\left(\begin{array}{cccc} 1&
-\frac{f_b^{-1}}{\sqrt{2}}r_{10}|\breve{y}_b|x&
\frac{f_b^{-1}}{\sqrt{2}}r_{10}|\breve{y}_b|x &
\mathcal{O}(x^2)\\
-f_b^{-1}r_{10}\breve{y}_bx &
-\frac{e^{i\theta_b}}{\sqrt{2}}+\frac{(r_{22}^2-r_{11}^2)\breve{y}_b^*
+r_{222}^2(|\breve{y}_t|^2/\breve{y}_b)
} {4\sqrt{2}(r_{11}+r_{22})}x  &
\frac{e^{i\theta_b}}{\sqrt{2}}+\frac{(r_{22}^2-r_{11}^2)\breve{y}_b^*
+r_{222}^2(|\breve{y}_t|^2/\breve{y}_b)
} {4\sqrt{2}(r_{11}+r_{22})}x & -\frac{2}{3}(2r_{111}+r_{222})y_tx\\
\mathcal{O}(x^2)&
\frac{1}{\sqrt{2}}+\frac{(r_{22}^2-r_{11}^2)|\breve{y}_b|+r_{222}^2(|\breve{y}_t|^2/|\breve{y}_b|)
} {4\sqrt{2}(r_{11}+r_{22})}x
 &
\frac{1}{\sqrt{2}}-\frac{(r_{22}^2-r_{11}^2)|\breve{y}_b|+r_{222}^2(|\breve{y}_t|^2/|\breve{y}_b|)
} {4\sqrt{2}(r_{11}+r_{22})}x &
\mathcal{O}(x^2)\\
\mathcal{O}(x^2) &
-\frac{\sqrt{2}}{3}(r_{111}+2r_{222})e^{i\theta_b}\breve{y}_t^*x &
\frac{\sqrt{2}}{3}(r_{111}+2r_{222})e^{i\theta_b}\breve{y}_t^*x  & 1
\end{array}\right)}\label{ORbKK}
\end{equation}
Analogously, $O_L^{c_{KK}}$ is given by:
\begin{equation}\noindent \footnotesize{O_L^{c_{KK}}=\phantom{bbbbbbbbbbbbbbbbbbbbbbbbbbbbbbb
bbbbbbbbbbbbbbbbbbbbbbbbbbbbbbbbbbbbbbbbbbbbbbbbbbbbbbbbb}}
\nonumber\end{equation}
\begin{equation}
\footnotesize{\left(\begin{array}{cccc} 1&
-\frac{f_Q^{-1}}{\sqrt{2}}r_{01}|\breve{y}_c|x&
\frac{f_Q^{-1}}{\sqrt{2}}r_{01}|\breve{y}_c|x &
0.94f_Q^{-1}r_{101}\breve{y}_sx\\
-f_Q^{-1}r_{01}\breve{y}_c^*x &
-\frac{e^{-i\theta_c}}{\sqrt{2}}-\frac{(r_{11}^2-r_{22}^2)\breve{y}_c^*+r_{111}^2(|\breve{y}_s|^2/\breve{y}_c)
} {4\sqrt{2}(r_{111}+r_{222})}x  &
\frac{e^{-i\theta_c}}{\sqrt{2}}-\frac{(r_{11}^2-r_{22}^2)\breve{y}_c^*+r_{111}^2(|\breve{y}_s|^2/\breve{y}_c)
} {4\sqrt{2}(r_{111}+r_{222})}x & \mathcal{O}(x^2)\\
\mathcal{O}(x^2)&
\frac{1}{\sqrt{2}}-\frac{(r_{11}^2-r_{22}^2)|\breve{y}_c|+r_{111}^2(|\breve{y}_s|^2/|\breve{y}_c|)
} {4\sqrt{2}(r_{111}+r_{222})}x
 &
\frac{1}{\sqrt{2}}+\frac{(r_{11}^2-r_{22}^2)|\breve{y}_c|+r_{111}^2(|\breve{y}_s|^2/|\breve{y}_c|)
} {4\sqrt{2}(r_{111}+r_{222})}x &
(8.6r_{111}+8.1r_{222})y_sx\\
-0.94f_Q^{-1}r_{101}\breve{y}_s^*x &
-(6.1r_{111}+5.7r_{222})y_s^*x & -(6.1r_{111}+5.7r_{222})y_s^*x  &
1
\end{array}\right)}\label{OLcKK}
\end{equation}
while $O_R^{c_{KK}}$ is
\begin{equation}\noindent \footnotesize{O_R^{c_{KK}}=\phantom{bbbbbbbbbbbbbbbbbbbbbbbbbbbbbbb
bbbbbbbbbbbbbbbbbbbbbbbbbbbbbbbbbbbbbbbbbbbbbbbbbbbbbbbbb}}
\nonumber\end{equation}
\begin{equation}
\small{\left(\begin{array}{cccc} 1&
-\frac{f_c^{-1}}{\sqrt{2}}r_{10}|\breve{y}_c|x&
\frac{f_c^{-1}}{\sqrt{2}}r_{10}|\breve{y}_c|x &
\mathcal{O}(x^2)\\
-f_c^{-1}r_{10}\breve{y}_cx &
-\frac{e^{i\theta_c}}{\sqrt{2}}+\frac{(r_{22}^2-r_{11}^2)\breve{y}_c^*+r_{222}^2(|\breve{y}_s|^2/\breve{y}_c)
} {4\sqrt{2}(r_{11}+r_{22})}x  &
\frac{e^{i\theta_c}}{\sqrt{2}}+\frac{(r_{22}^2-r_{11}^2)\breve{y}_c^*+r_{222}^2(|\breve{y}_s|^2/\breve{y}_c)
} {4\sqrt{2}(r_{11}+r_{22})}x & (8r_{111}+8.5r_{222})y_sx\\
\mathcal{O}(x^2)&
\frac{1}{\sqrt{2}}+\frac{(r_{22}^2-r_{11}^2)|\breve{y}_c|+r_{222}^2(|\breve{y}_s|^2/|\breve{y}_c|)
} {4\sqrt{2}(r_{11}+r_{22})}x
 &
\frac{1}{\sqrt{2}}-\frac{(r_{22}^2-r_{11}^2)|\breve{y}_c|+r_{222}^2(|\breve{y}_s|^2/|\breve{y}_c|)
} {4\sqrt{2}(r_{11}+r_{22})}x &
\mathcal{O}(x^2)\\
\mathcal{O}(x^2) &
(5.65r_{111}+6r_{222})e^{i\theta_c}\breve{y}_s^*x &
-(5.65r_{111}+6r_{222})e^{i\theta_c}\breve{y}_s^*x & 1
\end{array}\right)}\label{ORcKK}
\end{equation}
We also provide the KK mass spectrum obtained in the one-generation approximation.
The mass spectrum for the first generation, $u$ and $d$, KK modes remains unchanged
at $\mathcal{O}(x)$ and given by
\begin{equation}
({\bf M}_{u,d}^{KK})_{diag}=M_{KK}\,{\mbox
diag}\left(\breve{y}_{u,d}f_Q^{-1}f_{u,d}^{-1}r_{00}x, 1.1, 1,
1.1\right)\label{MKKUDspectrum}\end{equation}
The mass spectrum for the $s$ and $t$ KK modes is modified in a similar manner, as follows
\begin{eqnarray}
({\bf M}_s^{KK})_{diag}&=&M_{KK}\,{\mbox
diag}\left(\breve{y}_sf_Q^{-1}f_s^{-1}r_{00}x, 1.125,
1-(r_{111}+r_{222})|\breve{y}_c|x,
1+(r_{111}+r_{222})|\breve{y}_c|x\right)\nonumber\\
({\bf M}_t^{KK})_{diag}&=&M_{KK}\times{\mbox
diag}\left(\breve{y}_tf_Q^{-1}f_t^{-1}r_{00}x, 1.33,
1-(r_{111}+r_{222})|\breve{y}_b|x,
1+(r_{111}+r_{222})|\breve{y}_b|x\right)\nonumber\\\label{MKKstSpectrum}\end{eqnarray}
The mass spectrum for the $b$ and $c$ KK modes is also modified in a similar manner
\begin{eqnarray}
({\bf M}_b^{KK})_{diag}&=&M_{KK}\,{\mbox
diag}\left(\breve{y}_bf_Q^{-1}f_b^{-1}r_{00}x,
1-(r_{111}+r_{222})|\breve{y}_b|x,
1+(r_{11}+r_{22})|\breve{y}_b|x, 0.5\right)\nonumber\\
({\bf M}_c^{KK})_{diag}&=&M_{KK}\times{\mbox
diag}\left(\breve{y}_cf_Q^{-1}f_c^{-1}r_{00}x,
1-(r_{11}+r_{22})|\breve{y}_c|x, 1+(r_{11}+r_{22})|\breve{y}_c|x,
1.125\right)\nonumber\\\label{MKKbcSpectrum}\end{eqnarray}
\subsection{Overlap dependence of Dipole Operators in the one-generation approximation}
\label{app:B2}
In this section we evaluate the coefficients $B_P^{u,d}$ that enter the spurion-overlap
analysis of the new physics contributions to the neutron EDM, $\epsilon'/\epsilon$ and
$b\rightarrow s\gamma$, when it is combined with the direct diagonalization of the
one-generation mass matrices.  The approach is described in Sec.~\ref{Subsec:KKdiag1}
and makes use of Eqs.~(\ref{KK1genAppD}) and
(\ref{KK1genAppU}), the Yukawa matrices of
Eqs.(\ref{Y4KK})--(\ref{HDUmatrix}) and the diagonalization
matrices of appendix~\ref{app:B1}. In the evaluation of $B_P^{u,d}$ for this case,
we neglect the very moderate generational dependence of overlap
corrections and use their maximal value, leading to
the most conservative estimate.

\noindent We start with the down-type contribution to the neutron EDM. The
overlap dependence of this contribution in the one-generation KK
diagonalization scheme is encoded in $(B_P^{d})_{nEDM}^{KK
(1gen)}$, for which the most dominant contributions are as follows
\begin{equation}
(B_P^{d})_{nEDM}^{KK
(1gen)}=(r_{01}+r_{101})\left[2.6r_{10}(r_{11}+r_{111})+1.88r_{10}(r_{22}+r_{222})
+0.91f_Q^2r_{00}(r_{01}+r_{101})\right].\label{EDMOvercoeff}\end{equation}
Here and in the following $r_{111}=r_{11^{-+}}$, $r_{101}=r_{01^{-+}}$,
$r_{22}=r_{1^-1^-}$, $r_{222}=r_{1^{-}1^{+-}}$ and the notation
for the rest of the overlaps is the same as in
Eq.~(\ref{lagrangian}).
As we  naively expect
from the spurion-overlap analysis in the mass insertion approximation, each term
in the above equation is cubic in the overlap correction factors
and of the characteristic form $(0-KK)(KK-KK)(KK-0)$, or
$(0-0)(0-KK)(0-KK)$ for $f_Q^2$ proportional terms. Notice that
the latter, with $f_Q^2\approx0.1$, is suppressed compared to
other $\mathcal{O}(x)\approx\mathcal{O}(0.1)$ terms.
For this reason we can safely neglect the overlap correction terms proportional to
$\mathcal{O}(0.5)f_Q^2$ in the expressions below.

\noindent The overlap dependence of the up-type contribution to the neutron EDM is encoded in
$(B_P^u)_{nEDM}^{KK(1gen)}$, for which the most dominant
contributions are as follows
\begin{equation}
(B_P^u)_{nEDM}^{KK(1gen)}=(r_{10}+r_{101})\left[2.6r_{10}(r_{11}+r_{111})
+2.9r_{10}(r_{22}+r_{222})+0.46f_Q^2r_{00}(r_{01}+r_{101})\right]\, . \label{EDMUpOverCoeff}\end{equation}
 The overlap dependence of the up- and down-type contributions within the first generation is
almost identical, due to the similarity of the corresponding
one-generation KK diagonalization matrices of Eqs.~(\ref{OLdKK})
and (\ref{ORdKK}) and given that $c_d=0.77$ and $c_u=0.79$. More substantial
differences between the overlap dependence of up- and down-type
contributions to $\epsilon'/\epsilon$ and $b\rightarrow s\gamma$ are to be expected,
 since they involve less degenerate bulk parameters from the second and third generation.
 In addition, to account conservatively for the generational dependence within the
one-generation approximation, we will take the maximum over all generations for the following $B_P^{u,d}$ factors.

\noindent The overlap dependence of the down-type (neutral Higgs) contribution to
$\epsilon'/\epsilon$  is thus encoded
in $(B_P^{d})_{\epsilon'/\epsilon}^{KK(1gen)}$, for which the
dominant terms are as follows
\begin{equation} (B_P^{d})_{\epsilon'/\epsilon}^{KK(1gen)}={\mbox max}\left [
r_{01}r_{10}\left(\frac{1}{4}r_{111}-\frac{5}{4}r_{222}+\frac{r_{11}^2}{4(r_{111}+r_{222})}+
f_Q^2r_{00}r_{101}\right)\right ]\, .\label{(e'/e)OverCoeffD}\end{equation}
For the up-type (charged Higgs) contribution to $\epsilon'/\epsilon$
 we obtain
\begin{eqnarray}(B_P^{u})_{\epsilon'/\epsilon}^{KK(1
gen)}&=&{\mbox max}\left [
\frac{r_{01}r_{10}}{r_{11}+r_{22}}\left(\phantom{\frac{1}{4}}\!\!\!\!\!3.24r_{11}^2-6.06r_{111}(r_{11}+r_{22})
-0.125r_{111}^2+5.93r_{11}r_{22}\right.\right.\nonumber\\&&\left.+
2.68r_{22}^2-6.43r_{222}(r_{11}+r_{22})\phantom{\frac{1}{4}}\!\!\!\!\!\right)
+\frac{r_{10}r_{101}}{r_{11}+r_{22}}\left(\frac{1}{8}
(r_{11}^2+r_{111}^2)+0.5r_{11}r_{22}\right.\nonumber\\&&\left.\left.-2.7r_{111}(r_{11}+r_{22})+0.375r_{22}^2
-3.37r_{222}(r_{11}+r_{22})\phantom{\frac{1}{4}}\!\!\!\!\!\right)\right ]\, .
\label{(e'/e)OverCoeffU}
\end{eqnarray}
For the down-type (neutral Higgs) contribution to $b\rightarrow s\gamma$, the
dominant overlap dependence is as follows
\begin{eqnarray}(B_P^{d})_{b\rightarrow s\gamma}^{KK(1
gen)}&=&{\mbox max}\left\{  \frac{r_{10}}{r_{11}+r_{22}}
\left[\phantom{\frac{1}{4}}\!\!\!\!r_{01}\left(0.2\,r_{11}^2+\sqrt{2}\,r_{111}^2
-(\sqrt{2}/2)\,r_{11}r_{22}
-0.9r_{22}^2\right)\right.\right.\nonumber\\&&\left.+\,r_{101}\left(
-0.125\,r_{11}^2-r_{111}^2+0.5\,r_{11}r_{22}+0.625\,r_{22}^2\right)\phantom{\frac{1}{4}}\!\!\!\!\!
\right]\nonumber\\&&+\frac{r_{10}}{r_{111}+r_{222}}\left[\phantom{\frac{1}{4}}
\!\!\!\!r_{01}\left(4r_{11}^{\phantom{2}}r_{111}
+4.8r_{111}r_{22}+4.04r_{11}r_{222}+4.76r_{22}r_{222}\right)\right.\nonumber\\
&&\left.+\,r_{101}
\left(-0.1r_{111}^2+0.125r_{22}^2+0.54r_{111}r_{222}+0.625r_{222}^2\right)
\phantom{\frac{1}{4}}\!\!\!\!\right]\nonumber
\\&&\left. -\,r_{101}\left(1.32r_{10}r_{111}+4.62r_{10}r_{222}\right)\right\}
\phantom{\frac{1}{4}}\!\!\!\!\!\, ,\label{(BsGammaN)OverCoeff}\end{eqnarray}
while the dominant up-type (charged
Higgs) contribution is
\begin{eqnarray}(B_P^{u})_{b\rightarrow s\gamma}^{KK(1
gen)}&=&{\mbox max}\left\{  \frac{r_{10}}{r_{11}+r_{22}}\left[
\phantom{\frac{1}{4}}\!\!\!\!r_{01}\left(-0.18\,r_{11}^2
+\sqrt{2}\,r_{111}^2-2\,r_{11}r_{22}
-1.8r_{22}^2\right)\right.\right.\nonumber\\&&\left.+\,r_{101}\left(
0.125\,r_{11}^2+r_{111}^2+\sqrt{2}\,r_{11}r_{22}+1.275\,r_{22}^2\right)
\phantom{\frac{1}{4}}\!\!\!\!\!
\right]\nonumber\\&&+\frac{r_{10}}{r_{111}+r_{222}}\left[\phantom{\frac{1}{4}}\!\!\!\!r_{01}
\left(-4r_{11}^{\phantom{2}}r_{111}
-3.8r_{111}r_{22}-4.04r_{11}r_{222}-3.8r_{22}r_{222}\right)\right.\nonumber\\&&\left.+\,r_{101}
\left(0.125r_{111}^2-0.125r_{22}^2+0.47r_{111}r_{222}+0.347r_{222}^2\right)
\phantom{\frac{1}{4}}\!\!\!\!\right]\nonumber
\\&&\left. +\,r_{101}\left(1.32r_{10}r_{111}-2.9r_{10}r_{222}\right)\right\}
\phantom{\frac{1}{4}}\!\!\!\!\!\, .\label{(BsGammaH+)OverCoeff}\end{eqnarray}
Using the numerical results for the overlap corrections in the RS-A${}_4$
setup, as given in appendix~\ref{app:A1}, we obtain the following conservative estimate for the
modified $B_P^{u,d}$ overlap factors
\begin{eqnarray}
(B_P^d)_{nEDM}^{KK(1gen)}\simeq5.4&
(B_P^u)_{nEDM}^{KK(1gen)}\simeq 5.3 &
(B_P^d)_{\epsilon'/\epsilon}^{KK(1 gen)}\simeq 0.2, \nonumber\\
(B_P^u)_{\epsilon'/\epsilon}^{KK(1 gen)}\simeq-2.63, &
(B_P^d)^{KK(1 gen)}_{b\rightarrow s\gamma}\simeq 0.98,&
(B_P^u)^{KK(1 gen)}_{b\rightarrow s\gamma}\simeq -1.54\, ,
\label{KK1genFactros}\end{eqnarray} 
to be used in the combined spurion-overlap analysis with the diagonalization of
the one-generations mass matrices -- see Sec.~\ref{Subsec:KKdiag1}.
\subsection{Approximate analytical diagonalization of the three-ge\-ne\-ration mass matrices}
\label{app:B3}
This section collects more details of the approximate
analytical diagonalization scheme described in
Sec.~\ref{Subsec:KKFulldiag}. In particular, we are going to
inspect the structure of the down-type mass matrix, $\hat{{\bf
M}}_{Full}^D$, once it is rotated by the block KK diagonalization
matrices, ${\bf O}_{L,R}^{D_{KK}}$. This allows to understand why an additional
rotation by the A$_4$ $12\times12$ matrices, $\hat{{\bf
O}}^{D_{A_4}}_{L,R}$ may provide an almost complete diagonalization.
Similar arguments hold for the up-type
mass matrix, $\hat{{\bf M}}^U_{Full}$, which is of an analogous
structure. Using the $4\times 4$ block notation, we write:
\begin{eqnarray}
\frac{({\bf O}_{L}^{D_{KK}})^{\dagger}\hat{{\bf M}}_{Full}^D{\bf
O}_{R}^{D_{KK}}}{M_{KK}}&\equiv&\tilde{{\bf M}}_{(KK)}^D\nonumber\\
&&\hspace{-2.0truecm}=\left(\begin{array}{ccc} (\hat{{\bf M}}^d_{KK})_{diag}/M_{KK} &
(O_L^{d_{KK}})^\dagger \hat{Y}_s^{\prime_{KK}}O_R^{s_{KK}}x &
(O_L^{d_{KK}})^\dagger
\hat{Y}_b^{\prime_{KK}}O_R^{b_{KK}}x\\(O_L^{s_{KK}})^\dagger
\hat{Y}_d^{\prime_{KK}}O_R^{d_{KK}}x &
(\hat{{\bf M}}^s_{KK})_{diag}/M_{KK} & (O_L^{s_{KK}})^\dagger
\hat{Y}_b^{\prime\prime_{KK}}O_R^{b_{KK}}x\\
(O_L^{b_{KK}})^\dagger \hat{Y}_d^{\prime\prime_{KK}}O_R^{d_{KK}}x
& (O_L^{b_{KK}})^\dagger \
\hat{Y}_s^{\prime\prime_{KK}}O_R^{s_{KK}}x & (\hat{{\bf
M}}^b_{KK})_{diag}/M_{KK}
\end{array}\right)\nonumber\\\label{KKBlock}\end{eqnarray}
In the above equation, $\hat{Y}_{d,s,b}^{(\prime,\prime\prime)_{KK}}$
is a shorthand notation for the off-diagonal entries in
Eq.~(\ref{MFullD}), namely
$\hat{Y}_d^{\prime_{KK}}=\hat{Y}_d^{KK}((\hat{y}_{{21}}^{LO}+\hat{y}_{{21}}^{NLO},f_d)$
and so on. Using this notation, the matrix
$\hat{Y}_d^{\prime_{KK}}$ has the same structure as
$\hat{Y}_d^{KK}$ in Eq.~(\ref{Y4KK}), up to the replacement
$\breve{y}_{u,d}\rightarrow
\breve{y}_{u,d}^{\,\prime}$. At LO, from Eq.~\ref{LOYukawa}, one also has
 $ \breve{y}_{u,d}^{\,\prime}\equiv(\hat{Y}_{12}^{u,d})_{LO}=2
(\hat{y}^{u,d}_{12})_{LO} v_\Phi^{4D}e^{k\pi
R}/k$, while Eq.~\ref{NLOYukawa} provides the NLO contribution. Since
we want to perform the diagonalization to
$\mathcal{O}(x)$, the slight generational dependence of the overlap
corrections is numerically negligible at this order. The same goes
for the zero-zero interaction terms in each of the off-diagonal
blocks.
 It is important to notice that the latter are proportional to $x$, hence it
is clear that only the $\mathcal{O}(1)$ terms in
$O_{L,R}^{(d,s,b)_{KK}}$ will supplement us with off-diagonal
$\mathcal{O}(x)$ entries in the ``KK-rotated" mass matrix of
Eq.~(\ref{KKBlock}). The $\mathcal{O}(1)$ terms in the KK
diagonalization matrices arise from the rotations that
diagonalize the corresponding nearly degenerate subspaces.

\noindent To illustrate it, we first focus on the $(12)$ block in
Eq.~(\ref{KKBlock}) and write it explicitly. From
Eqs.~(\ref{OLdKK}) and (\ref{ORdKK}), we learn that the
$\mathcal{O}(1)$ terms in $O_{L,R}^{d_{KK}}$ correspond to $\pi/4$
rotations in the $(24)$ and $(34)$ subspaces, respectively.
Similarly the $\mathcal{O}(1)$ terms in $O_{L,R}^{s_{KK}}$ of
Eqs.~(\ref{OLsKK})--~(\ref{ORsKK}), correspond to $\pi/4$
rotations (plus a phase $\theta_c\equiv{\mbox Arg}(y_c)$) in the
$(34)$ and $(24)$ subspaces, respectively.
Using $(\hat{y}^{u,d}_{12})_{LO}=(\hat{y}^{u,d}_{11})_{LO}$, we
have
$\breve{y}_{c,s}^{\,\prime}\simeq\breve{y}_{c,s}$\,\footnote{Notice
that we also have $\breve{y}_c\simeq\breve{y}_s$, which is an
exact equality for the LO Yukawa interactions.} and the $(12)$
block of $\tilde{{\bf M}}_{(KK)}^D$ turns out to be:
\begin{equation}
\left(\tilde{{\bf
M}}_{(KK)}^D\right)_{12}=\left(\begin{array}{cccc}
f_Q^{-1}f_s^{-1}r_{00}\breve{y}_sx &f_Q^{-1}r_{101} \breve{y}_cx
&f_Q^{-1}r_{10}\breve{y}_sx &
\frac{f_Q^{-1}r_{101}}{\sqrt{2}}\breve{y}_cx\\ 0 &
\frac{e^{i\theta_c}}{2}\left(r_{222}\breve{y}_c^*-r_{22}\breve{y}_s^*\right)x
&0
&\frac{e^{i\theta_c}}{2}\left(r_{222}\breve{y}_c^*-r_{22}\breve{y}_s^*\right)x\\
f_s^{-1}r_{10}\breve{y}_sx & \frac{1}{2}r_{111}\breve{y}_cx &
r_{11}\breve{y}_sx & \frac{1}{2}r_{111}\breve{y}_cx\\0 &
-\frac{e^{i\theta_c}}{2}\left(r_{222}\breve{y}_c^*+r_{22}\breve{y}_s^*\right)x
&0 &
-\frac{e^{i\theta_c}}{2}\left(r_{222}\breve{y}_c^*+r_{22}\breve{y}_s^*\right)x
\end{array}\right)\label{Block12}
\end{equation}
Because of the near degeneracy of $r_{22}$ and $r_{222}$, one concludes
that the second row is approximately vanishing. A similar pattern
will emerge for the $(21)$ block, where the third column will be
the one that approximately vanishes at this order.
Considering the $(13)$ block, we recall that
$(\hat{y}^{u,d}_{13})_{LO}=(\hat{y}^{u,d}_{11})_{LO}$, which
implies $\breve{y}_{t,b}^{\,\prime}\simeq\breve{y}_{t,b}$, thus leading to
\begin{eqnarray}
\left(\tilde{{\bf
M}}_{(KK)}^D\right)_{13}&=&\nonumber\\
&&\hspace{-2.0truecm}\left(\begin{array}{cccc}
f_Q^{-1}f_b^{-1}r_{00}\breve{y}_bx
&\frac{f_Q^{-1}}{\sqrt{2}}r_{01} \breve{y}_bx
&\frac{f_Q^{-1}}{\sqrt{2}}r_{01} \breve{y}_bx &
f_Q^{-1}r_{101}\breve{y}_tx\\ 0 &
\frac{e^{i\theta_b}}{2}\left(r_{222}\breve{y}_t^*-r_{22}\breve{y}_b^*\right)x
&\frac{e^{i\theta_c}}{2}\left(r_{222}\breve{y}_t^*-r_{22}\breve{y}_b^*\right)x
&0\\
f_b^{-1}r_{10}\breve{y}_bx & \frac{r_{11}}{\sqrt{2}}\breve{y}_bx &
 \frac{r_{11}}{\sqrt{2}}\breve{y}_bx & r_{111}\breve{y}_tx\\0 &
-\frac{e^{i\theta_b}}{2}\left(r_{222}\breve{y}_t^*+r_{22}\breve{y}_b^*\right)x
&-\frac{e^{i\theta_b}}{2}\left(r_{222}\breve{y}_t^*+r_{22}\breve{y}_b^*\right)x
&0
\end{array}\right)\nonumber\\\label{Block13}
\end{eqnarray}
It is already at this level that we notice the modifications
induced by $y_t\simeq2.8$, as compared to $y_{u,c,d,s,b}=1$. The
difference of Yukawa couplings now spoils the vanishing of the second
row for the (13) block,  differently from the (12) block.  An analogous situation arises in the
third column of the $(31)$ block.This will be the main obstacle in obtaining a fully analytical
diagonalization of all blocks involving the third generation.

\noindent Finally, we focus on the $(23)$ block of $\tilde{{\bf
M}}_{(KK)}^D$. From Eqs.~(\ref{OLbKK}) and ~(\ref{ORbKK}) it is
clear that the $\mathcal{O}(1)$ terms in $O_{L,R}^{b_{KK}}$
correspond to a $\pi/4$ rotations in the $(23)$ subspace of the
$(\tilde{{\bf M}}_{(KK)}^D)_{23}$ block plus a phase
$\theta_b={\mbox Arg}(y_b)$. Using
$(\hat{y}^{u,d}_{23})_{LO}=\omega^2(\hat{y}^{u,d}_{11})_{LO}$, we
get $\breve{y}_{t,b}^{\,\prime\prime}\simeq\omega^2\breve{y}_{t,b}$ and
the $(23)$ block turns out to be:
\begin{eqnarray}
\left(\tilde{{\bf
M}}_{(KK)}^D\right)_{23}&=&\nonumber\\
&&\hspace{-3.0truecm}\footnotesize{
\left(\begin{array}{cccc}
\omega^2f_Q^{-1}f_b^{-1}r_{00}\breve{y}_bx
&\omega^2\frac{f_Q^{-1}}{\sqrt{2}}r_{01} \breve{y}_bx
&\omega^2\frac{f_Q^{-1}}{\sqrt{2}}r_{01} \breve{y}_bx &
\omega^2f_Q^{-1}r_{101}\breve{y}_tx\\0 &\frac{-\omega
e^{i\theta_b}}{\sqrt{2}}r_{22}\breve{y}_b^*x &\frac{\omega
e^{i\theta_b}}{\sqrt{2}}r_{22}\breve{y}_b^*x &
0\\\frac{-\omega^2e^{i\theta_c}}{\sqrt{2}}f_Q^{-1}r_{10}\breve{y}_b
& -(\frac{\omega^2e^{i\theta_c}}{2}\breve{y}_br_{11}+\frac{\omega
e^{i\theta_b}}{2}\breve{y}_t^*r_{222})x &
(\frac{-\omega^2e^{i\theta_c}}{2}\breve{y}_br_{11}+\frac{\omega
e^{i\theta_b}}{2}\breve{y}_t^*r_{222})x
&\frac{-\omega^2e^{i\theta_c}}{\sqrt{2}}r_{111}\breve{y}_tx\\
\frac{\omega^2e^{i\theta_c}}{\sqrt{2}}f_Q^{-1}r_{10}\breve{y}_b &
(\frac{\omega^2e^{i\theta_c}}{2}\breve{y}_br_{11}-\frac{\omega
e^{i\theta_b}}{2}\breve{y}_t^*r_{222})x &
(\frac{-\omega^2e^{i\theta_c}}{2}\breve{y}_br_{11}+\frac{\omega
e^{i\theta_b}}{2}\breve{y}_t^*r_{222})x
&\frac{\omega^2e^{i\theta_c}}{\sqrt{2}}r_{111}\breve{y}_tx
\end{array}\right)  } \nonumber\\\label{Block23}
\end{eqnarray}
As expected,  the cancellation pattern we encountered in the
$(12)$ block gets modified even further than in the $(13)$ block,
due to the different rotations involved in the $(23)$ block of
$\tilde{{\bf M}}_{(KK)}^D$. Observing
Eqs.~(\ref{Block12})--~(\ref{Block23}), we realize that the
distribution of the 1, $\omega$ and $\omega^2$ factors in the off-diagonal
 blocks of $\tilde{{\bf M}}_{(KK)}^D$ approximately
corresponds to the one of the leading order Yukawa couplings
 $(\hat{y}^{u,d}_{ij})_{LO}$, up to complex
conjugations of terms proportional to $\breve{y}_{u,c,t,d,s,b}^*$.
For this reason, we expect the A$_4$ $12\times 12$ rotation
matrices defined as ${\bf O}^{(U,D)_{A_4}}_{L,R}\equiv
V_{L,R}^{u,d}\otimes\tilde{\mathbbm{1}}_{4\times4}$ to induce some
cancellations among the various  blocks of $\tilde{{\bf
M}}_{(KK)}^D$. However, due to the difference of $y_t$ from the rest of the
LO Yukawa couplings and due to the different rotations in each of the
off-diagonal blocks of $\tilde{{\bf M}}_{(KK)}^D$, it is clear
that the cancellations induced by ${\bf O}^{(U,D)_{A_4}}_{L,R}$
can never be exact, even if we only consider the LO Yukawa
interactions. Thus, the above diagonalization scheme will still
fail to fully diagonalize the degenerate subspace. On the other
hand, off-diagonal elements in the non degenerate subspaces like the
zero-KK and a few KK-KK entries can be treated using non
degenerate perturbation theory.

\noindent At this level, the exact structure of the approximate $12\times 12$
diagonalization matrices composed of the ${\bf O}_{L,R}^{D_{KK}}$,
${\bf O}^{(U,D)_{A_4}}_{L,R}$ and the perturbative rotation
matrices, used in attempting the analytical diagonalization, is very
complicated and impossible to write in a compact way. Instead, to better estimate  the
inaccuracy of this diagonalization scheme, we assign
$y_{u,c,d,s,b}=1$, $y_t\simeq2.8$ and set the bulk masses
according to Eq.~(\ref{BulkAssignments}), which yield the physical
running quark masses at the scale $R'^{-1}\simeq1.8$ TeV  and
fix all the overlap correction factors. The NLO Yukawas are
assigned according to Eq.~(\ref{CKMAssignment}), to provide a realistic CKM
matrix in the ZMA. The $x$ parameter is left
unassigned. We then write the magnitude of the
off-diagonal elements in the degenerate subspace of $\tilde{{\bf
M}}_{(KK)}^D\tilde{{\bf M}}_{(KK)}^{D\dagger}$ to gain an insight
into the ``contaminations", which can not be treated using non
degenerate perturbation theory and can only be partly reduced
by the ${\bf O}^{(D)_{A_4}}_{L}$ rotation matrices. A similar
procedure is followed for $\tilde{{\bf
M}}_{(KK)}^{D\dagger}\tilde{{\bf M}}_{(KK)}^D$ and the right-handed rotations.

\noindent From Eqs.~(\ref{MKKUDspectrum})--\,(\ref{MKKbcSpectrum}), we learn
that the degenerate subspace approximately decomposes in two blocks, one
corresponding to the $(2,4,6)$ subspace, where all KK masses are
approximately $1.1M_{KK}$ and the $(3,7,8,10,11)$ subspace\footnote{Since the
contaminations in the off-diagonal elements of the degenerate subspace are of
$\mathcal{O}(x)$, the $\mathcal{O}(x)$ corrections to the KK
masses will have a minor effect on this estimation.}, where
all KK masses are approximately
$M_{KK}\simeq2.55\, R'^{-1}$. In
$\tilde{{\bf M}}_{(KK)}^D\tilde{{\bf M}}_{(KK)}^{D\dagger}$ these
masses appear squared, since every diagonal block is
proportional to a diagonalized one-generation KK mass matrix squared
plus additional $\mathcal{O}(x^2)$ terms. Hence in total, we have the
squared spectrum $(1.21,1,1.21,1.21,1,1,1,1)$ on the diagonal of the $(2,3,4,6,7,8,10,11)$
subspace of $\tilde{{\bf M}}_{(KK)}^D\tilde{{\bf
M}}_{(KK)}^{D\dagger}$, and the contaminations in the same subspace amount to
\begin{eqnarray} (\tilde{{\bf
M}}^{D}\tilde{{\bf
M}}^{D\dagger})_{(KK_{Deg.})}&\approx&\nonumber\\
&&\hspace{-4.0truecm}\footnotesize{
\left(\begin{array}{cccccccc}
1.21&\mathcal{O}(x^2)&\mathcal{O}(x^2)&0.03x &0.02x &0.02x
&-0.1x&-0.15x\\
\mathcal{O}(x^2)&1&\mathcal{O}(x^2)&0.2x&0.25x&0.2x&0.25x&0.25x\\
\mathcal{O}(x^2)&\mathcal{O}(x^2)&1.21&-0.03x&-0.25x&0.25x&0.05x&0.2x\\0.03x
&0.2x
&-0.03x&1.21&\mathcal{O}(x^2)&\mathcal{O}(x^2)&(-0.05+0.15i)x&-0.05+0.1ix\\
0.02x&0.25x&-0.25x&\mathcal{O}(x^2)&1 &\mathcal{O}(x^2)&
0.25ix&0.2ix\\
0.02x&0.2x&0.25x&\mathcal{O}(x^2)&\mathcal{O}(x^2)&1&-0.1x&-0.14x\\
-0.1x&0.3x&0.05x&(-0.05+0.15i)x&0.25ix&-0.1x&1&\mathcal{O}(x^2)\\
-0.15x&0.3x&0.2x&(-0.05+0.1i)x&0.2ix&-0.14x&\mathcal{O}(x^2)&1
\end{array}\right) }\nonumber\\\label{Mcontaminate}\end{eqnarray}
\normalsize  We realize that the largest contaminations of
$\mathcal{O}(0.25x)$ are numerically suppressed by three orders of
magnitude compared to the diagonal entries for $x\simeq0.04$,
which corresponds to $R'^{-1} = 1.8$ TeV. For this reason, the
results of the approximate analytical diagonalization scheme can
still provide an order of magnitude estimate for the
physical couplings between zero modes and KK modes. From the
above matrix, one can also infer the way  ${\bf
O}_L^{D_{KK}}$ deviates from the ``true" diagonalization matrix; in
particular it is evident that the first row of ${\bf
O}_L^{D_{KK}}$ is the least contaminated. This qualitatively
explains why the semianalytical estimation for the neutron EDM
is in better agreement with the numerical result than in the case of
$\epsilon'/\epsilon$ and $b\rightarrow s\gamma$. We have failed to
find a better analytical method which would allow us to further
diagonalize the contaminated subspace of Eq.~(\ref{Mcontaminate}).
Nevertheless, the adopted scheme allows to qualitatively understand
the way some of the cancellation mechanisms still act in the
full $12\times 12$ case.
An analogous situation occurs for  $(\tilde{{\bf
M}}^{D\dagger}\tilde{{\bf M}}^{D})_{(KK_{Deg.})}$ and the right-handed
diagonalization matrix ${\bf O}_R^{D_{KK}}$.


\begin{thebibliography}{99}
\bibitem{A4Warped}
  A.~Kadosh and E.~Pallante,
  JHEP {\bf 1008}, 115 (2010)
  [arXiv:1004.0321 [hep-ph]].
\bibitem{a4} E.~Ma and G.~Rajasekaran, Phys.\ Rev.\ {\bf D64}, 113012 (2001)
K.~S.~Babu, E.~Ma and J.~W.~F.~~Valle, Phys.\ Lett.\ {\bf B552}
207 {2003}; E.~Ma, Phys.\ Rev.\ {\bf D70}, 031901(2004) Talk at
SI2004, Fuji-Yoshida, Japan, hep-ph/0409075.
\bibitem{RS}
  L.~Randall and R.~Sundrum,
  Phys.\ Rev.\ Lett.\  {\bf 83}, 3370 (1999)
  [arXiv:hep-ph/9905221];
  L.~Randall and R.~Sundrum,
  Phys.\ Rev.\ Lett.\  {\bf 83}, 4690 (1999)
  [arXiv:hep-th/9906064].
\bibitem{Csaki:2008qq}
  C.~Csaki, C.~Delaunay, C.~Grojean and Y.~Grossman,
  JHEP {\bf 0810}, 055 (2008)
  [arXiv:0806.0356 [hep-ph]].
\bibitem{Volkas}
  X.~G.~He, Y.~Y.~Keum and R.~R.~Volkas,
  JHEP {\bf 0604}, 039 (2006)
  [arXiv:hep-ph/0601001].
  \bibitem{Agashe:2003zs}
  K.~Agashe, A.~Delgado, M.~J.~May and R.~Sundrum,
  JHEP {\bf 0308}, 050 (2003)
  [arXiv:hep-ph/0308036].
\bibitem{Carena:2007}
 M.~Carena, E.~Ponton, J. Santiago and C.E.M.~Wagner, arXiv:hep-ph/0701055.
  \bibitem{rsgim1}
  K.~Agashe, G.~Perez and A.~Soni,
  Phys.\ Rev.\ Lett.\  {\bf 93}, 201804 (2004)
  [arXiv:hep-ph/0406101];
  \bibitem{Cacciapaglia:2007fw}
  G.~Cacciapaglia, C.~Csaki, J.~Galloway, G.~Marandella, J.~Terning and A.~Weiler,
  JHEP {\bf 0804}, 006 (2008)
  [arXiv:0709.1714 [hep-ph]].
\bibitem{Agashe:2004cp}
  K.~Agashe, G.~Perez and A.~Soni,
  Phys.\ Rev.\  D {\bf 71}, 016002 (2005)
  [arXiv:hep-ph/0408134];
\bibitem{Fitzpatrick:2007sa}
  K.~Agashe, M.~Papucci, G.~Perez and D.~Pirjol,
  arXiv:hep-ph/0509117;
  Z.~Ligeti, M.~Papucci and G.~Perez,
  Phys.\ Rev.\ Lett.\  {\bf 97}, 101801 (2006)
  [arXiv:hep-ph/0604112].
  A.~L.~Fitzpatrick, G.~Perez and L.~Randall,
  arXiv:0710.1869 [hep-ph].
\bibitem{Chen:2008qg}
  M.~C.~Chen and H.~B.~Yu,
  Phys.\ Lett.\  B {\bf 672}, 253 (2009)
  [arXiv:0804.2503 [hep-ph]].
\bibitem{Perez:2008ee}
  G.~Perez and L.~Randall,
  JHEP {\bf 0901}, 077 (2009)
  [arXiv:0805.4652 [hep-ph]].
\bibitem{Csaki:2009wc}
  C.~Csaki, G.~Perez, Z.~Surujon and A.~Weiler,
  arXiv:0907.0474 [hep-ph].
\bibitem{Azatov}
  K.~Agashe, A.~Azatov and L.~Zhu,
  Phys.\ Rev.\  D {\bf 79}, 056006 (2009)
  [arXiv:0810.1016 [hep-ph]].
  \bibitem{IsidoriPLB}
O. Gedalia, G. Isidori and G. Perez, Phys.\ Lett.\  {\bf B 682},
200 (2009). 
\bibitem{TPrime}
F.~Feruglio, C.~Hagedorn, Y.~Lin and L.~Merlo, Nucl.\ Phys.\ B {\bf 775}, 120 (2007)
[arXiv:hep-ph/0702194];
M.~C.~Chen, K.~T.~Mahanthappa and F.~Yu,
  arXiv:0909.5472 [hep-ph];
P.~H.~Frampton and T.~W.~Kephart,
  Int.\ J.\ Mod.\ Phys.\ A {\bf 10}, 4689 (1995)
  [arXiv:hep-ph/9409330];
A.~Aranda, C.~D.~Carone and R.~F.~Lebed,
  Phys.\ Lett.\ B {\bf 474}, 170 (2000) [arXiv:hep-ph/9910392];
A.~Aranda, C.~D.~Carone and R.~F.~Lebed,
  Phys.\ Rev.\ D {\bf 62}, 016009 (2000)
  [arXiv:hep-ph/0002044].
\bibitem{HiggsContino}
  K.~Agashe and R.~Contino,
  Phys.\ Rev.\  D {\bf 80}, 075016 (2009)
  [arXiv:0906.1542 [hep-ph]].
  \bibitem{HiggsFCNC}
  A.~Azatov, M.~Toharia and L.~Zhu,
  Phys.\ Rev.\  D {\bf 80}, 035016 (2009)
  [arXiv:0906.1990 [hep-ph]].
  \bibitem{PDG}
  C.~Amsler {\it et al.}  [Particle Data Group],
  Phys.\ Lett.\ {\bf B667} 1 (2008).
\bibitem{Agashe:2006at}
  K.~Agashe, R.~Contino, L.~Da Rold and A.~Pomarol,
  Phys.\ Lett.\  B {\bf 641}, 62 (2006)
  [arXiv:hep-ph/0605341].
\bibitem{WiseScalar}
  W.~D.~Goldberger and M.~B.~Wise,
  Phys.\ Rev.\ Lett.\  {\bf 83}, 4922 (1999)
  [arXiv:hep-ph/9907447];
G.~Cacciapaglia, C.~Csaki, G.~Marandella and J.~Terning,
  JHEP {\bf 0702}, 036 (2007)
  [arXiv:hep-ph/0611358].
\bibitem{Buras}
  G.~Buchalla, A.~J.~Buras and M.~E.~Lautenbacher,
  Rev.\ Mod.\ Phys.\  {\bf 68}, 1125 (1996)
  [arXiv:hep-ph/9512380].
\bibitem{EDMBound}
  C.~A.~Baker {\it et al.},
  Phys.\ Rev.\ Lett.\  {\bf 97}, 131801 (2006)
  [arXiv:hep-ex/0602020].
  \bibitem{EpsilonUncertainty}
  V.~Cirigliano,
  Eur.\ Phys.\ J.\  C {\bf 33}, S333 (2004);
  W.~Lee,
  PoS {\bf LAT2006}, 015 (2006)
  [arXiv:hep-lat/0610058];
  S.~Bertolini, J.~O.~Eeg and M.~Fabbrichesi,
  Phys.\ Rev.\  D {\bf 63}, 056009 (2001)
  [arXiv:hep-ph/0002234];
  A.~Pich,
  [arXiv:hep-ph/0410215].
  A.~J.~Buras and J.~M.~Gerard,
  Phys.\ Lett.\  B {\bf 517}, 129 (2001)
  [arXiv:hep-ph/0106104].
  \bibitem{HadronBag}
  A.~J.~Buras, G.~Colangelo, G.~Isidori, A.~Romanino and L.~Silvestrini,
  Nucl.\ Phys.\  B {\bf 566}, 3 (2000)
  [arXiv:hep-ph/9908371].
\bibitem{BurasHiggs}
  A.~J.~Buras, B.~Duling and S.~Gori,
  arXiv:0905.2318 [hep-ph];
  M.~Blanke, A.~J.~Buras, B.~Duling, S.~Gori and A.~Weiler,
  JHEP {\bf 0903}, 001 (2009)
  [arXiv:0809.1073 [hep-ph]].
  \bibitem{UTfit}
  M.~Bona {\it et al.}  [UTfit Collaboration],
  JHEP {\bf 0803}, 049 (2008)
  [arXiv:0707.0636 [hep-ph]].
\bibitem{Casagrande_ECust}
S. Casagrande et al., arXiv:1005.4315 [hep-ph]
  \bibitem{bulkfermion1}
Y.~Grossman and M.~Neubert,
  Phys.\ Lett.\  B {\bf 474}, 361 (2000)
  [arXiv:hep-ph/9912408];
  T.~Gherghetta and A.~Pomarol,
  Nucl.\ Phys.\  B {\bf 586}, 141 (2000)
  [arXiv:hep-ph/0003129];
  S.~J.~Huber and Q.~Shafi,
  Phys.\ Lett.\  B {\bf 498}, 256 (2001)
  [arXiv:hep-ph/0010195];
  S.~J.~Huber and Q.~Shafi,
  Phys.\ Lett.\  B {\bf 512}, 365 (2001)
  [arXiv:hep-ph/0104293].
\end{thebibliography}
\end{document}